\RequirePackage{fix-cm}
\documentclass[twocolumn]{svjour3}          
\smartqed  
\usepackage{graphicx}
\usepackage{natbib}
\usepackage{amsmath}
\usepackage{mathrsfs}
\usepackage{amsfonts}
\usepackage{diagbox}
\usepackage{makecell}

\newcommand{\de}{\mathrm{d}}

\newcommand{\x}{{\mathbf x}}
\newcommand{\R}{\mathbb R}

\newcommand{\1}{{\mathbf 1}}
\newcommand{\V}{{\mathcal V}}

\def\P{\mathbb P}
\def\E{\mathbb E}
\DeclareMathOperator{\e}{e}
\DeclareMathOperator{\Var}{Var}
\DeclareMathOperator{\Cov}{Cov}
\DeclareMathOperator{\argmin}{arg \, min}

\newcommand{\beann}{\begin{eqnarray*}}
\newcommand{\eeann}{\end{eqnarray*}}

\newtheorem{conj}{Conjecture}

\journalname{}
\begin{document}

\title{Resample-smoothing of Voronoi intensity estimators
}


\author{M. Mehdi Moradi   \and
        Ottmar Cronie \and 
        Ege Rubak \and 
        Raphael Lachieze-Rey \and 
        Jorge Mateu \and 
        Adrian Baddeley
}


\institute{M. Mehdi Moradi \at
              Institute of New Imaging Technologies (INIT), University Jaume I, Castellon, Spain 
              \\
           \and
           Ottmar Cronie (Corresponding author) \at
              Department of Mathematics and Mathematical Statistics, Ume{\aa} University, Sweden\\
              \email{ottmar.cronie@umu.se}\\
              Tel.: +46-90-7865742\\
              \and 
              Ege Rubak \at
              Department of Mathematical Sciences, Aalborg University, Denmark
              \\
              \and 
              Raphael Lachieze-Rey \at
              Universit\'e Paris Descartes, Sorbonne Paris Cit\'e, France
              \and 
              Jorge Mateu \at 
              Department of Mathematics, University Jaume I, Castellon, Spain
               \\
              \and 
              Adrian Baddeley \at
              Department of Mathematics \& Statistics, Curtin University, Perth, Australia
              \\
}

\date{Received: date / Accepted: date}

\maketitle

\begin{abstract}
Voronoi intensity estimators, which are non-parametric estimators for intensity functions of point processes, are both parameter-free and adaptive; the intensity estimate at a given location is given by the reciprocal size of the Voronoi/Dirichlet cell containing that location. 
Their major drawback, however, is that they tend to under-smooth the data in regions where the point density of the observed point pattern is high and over-smooth in regions where the point density is low. 
To remedy this problem, i.e.\ to find some middle-ground between over- and under-smoothing, 
we propose an additional smoothing technique for  Voronoi intensity estimators for point processes in arbitrary metric spaces, which is based on repeated independent thinnings of the point process/pattern. Through a simulation study we show that our resample-smoothing technique improves the estimation significantly. In addition,  we study statistical properties such as unbiasedness and variance, and propose a rule-of-thumb and a data-driven cross-validation approach to choose the amount of thinning/smoothing to apply. We finally apply our proposed intensity estimation scheme to two datasets: locations of pine saplings (planar point pattern) and motor vehicle traffic accidents (linear network point pattern).
\end{abstract}

\begin{acknowledgements}
M.M. Moradi  gratefully acknowledges funding from the European union through the GEO-C project (H2020-MSCA-ITN-2014, Grant Agreement Number 642332, http://www.geo-c.eu/); J. Mateu is partially funded by grants MTM2016-78917-R and P1-1B2015-40.
Ege Rubak was supported
by The Danish Council for Independent Research $\mid$ Natural Sciences,
grant DFF -- 7014--00074 ``Statistics for point processes in space and beyond'';
and by the Centre for Stochastic Geometry and Advanced Bioimaging,
funded by grant 8721 from the Villum Foundation.
We thank Ned Levine for kindly providing us with the dataset on Houston vehicle traffic accidents as well as helpful discussions on such data. We also thank David Cohen for fruitful discussions. 
\end{acknowledgements}

\pagebreak

\section{Introduction}
\label{S:Intro}
In point pattern analysis \citep{VanLieshoutBook,Diggle14Book,BRT15}, exploratory data analyses often start with a non-parametric analysis of the spatial intensity of events/data points. The {\em intensity function}, which is a first order moment characterisation of the point process assumed to have generated the data, 
reflects the abundance of points in different regions and may be seen as a ``heat map" for the events. For most datasets, assuming that the underlying point process is homogeneous, i.e.\ that its intensity function is constant, is rarely a realistic assumption. Hence, the natural way to start is to assume inhomogeneity for the underlying point process. 

The most prominent approach to non-parametric intensity estimation is undoubtedly kernel estimation \citep{Diggle14Book,BRT15,cronie2018bandwidth}. 
A key point with kernel intensity estimation, and kernel-based estimation in general, is that equally much smoothing is applied to the whole dataset. The degree of smoothing is controlled by a smoothing parameter, the so-called bandwidth, and the resulting estimates heavily depend on the choice of bandwidth. A small bandwidth may result in under-smoothing whereas a large bandwidth might result in over-smoothing of the intensity. Regarding the bandwidth selection, for point processes/patterns in Euclidean spaces some progress has been made \citep{cronie2018bandwidth}. 

Concerning other spatial domains, recently there has been an increasing interest in point patterns on linear networks \citep{OS12,BRT15,RTMMNMB}; examples of data include street crimes or traffic accidents on a road network (of a city).
Here the matter of kernel estimation is more delicate due to the geometry of the underlying network and the methodology is still under development. \citet{B03,B05,B08} proposed several methods for kernel smoothing of network data without discussing statistical properties.  \citet{XZY08} defined a kernel-based intensity estimator for network point patterns without taking the topography of the network into consideration and as a result the estimation errors tended to be large, thus making the estimator heavily biased. 
\cite{OSS09} further introduced a class of so-called equal-split network kernel density estimators which support both continuous and discontinuous schemes.  
By exploiting properties of diffusion on networks, \cite{mcswiggan2017} developed a kernel estimation method based on the heat kernel, which is the appropriate linear network analogue of the Gaussian kernel. In addition, \cite{MFJ17} extended the classical spatial edge corrected kernel intensity estimator to point patterns on linear networks. 

As a consequence of e.g.\ covariates, such as demography and human mobility, 
it is quite common to encounter situations where there are notable abrupt spatial changes in the distribution of the events, with a large number of events in particular parts of the study region and nearly empty parts close by. E.g., 
street crimes or traffic accidents tend to happen in particular streets/roads/junctions and they are often surrounded by empty neighbourhoods. 
The classical kernel estimation approach does often not fit such types of data. 

We argue, similarly to \cite{BSV10}, that kernel-based approaches may be unsatisfactory when there are sharp boundaries between parts with high and low intensity. Indeed, a fixed kernel smoothing bandwidth results in over-smoothing in parts with high intensity and under-smoothing in low-intensity parts \citep{BRT15}. In addition, choosing a fixed bandwidth is itself a well studied and challenging problem \citep{BRT15,cronie2018bandwidth}. 
By considering an adaptive estimator, i.e. an estimator that adapts locally to the distribution of events, we may reduce such problems when estimating the intensity function \citep{BRT15,Diggle14Book,SilvermanBook}.

A first idea would be to consider adaptive kernel estimators, which use an individual bandwidth for each point of the point pattern or a spatially varying bandwidth function. 
In the planar case, i.e.\ in the 2-dimensional Euclidean setting, some efforts have been made \citep{davies2010adaptive,Diggle14Book,davies2016symmetric,davies2018fast}. 
The issue with adaptive kernel estimation, however, is that optimal bandwidth selection becomes even more challenging and important \citep{Diggle14Book,SilvermanBook}. 

As an alternative, one could consider an approach without any choice of tuning parameters, e.g.\ a tessellation-based approach \citep{vanIntensity,schaap2007delaunay}. One such approach is provided by Voronoi intensity estimation \citep{Ord78,BSV10,OS12}, defined such that within a given Voronoi cell of the point pattern the intensity estimate is set to the reciprocal of the size of that Voronoi cell \citep{Okabe}. When employing the Voronoi intensity estimator, one thing that quickly becomes evident is that it often accentuates local features too much, in particular in regions with high event density. This reflects a previously observed phenomenon: adaptive estimators, such as the Voronoi intensity estimator, may smooth too little whereas kernel estimators may smooth too much in dense regions \citep[Section 6.5.2]{BRT15}. Hence, one should be able to find some middle ground and we here aim at providing a contribution to that. 

Our idea is simple. In dense parts surrounded by empty neighbourhoods, Voronoi intensity estimators tend to smooth too little, thus generating excessive peaks in the intensity estimate in those parts. By removing points in such a dense part we reduce the peaks, which results in a smoother intensity estimate, with a shape more similar to the true intensity function. However, the problem of doing this ``manually" is twofold: 1) it is not clear which specific points we should remove, and 2) we need to compensate for the reduced total mass. To solve these issues, we propose to independently thin the original point pattern some $m\geq1$ times, according to some retention probability, in order to obtain $m$ different point patterns and thereby $m$ different Voronoi intensity estimates. In order to compensate for the reduced mass, we then scale each of the $m$ estimates by the applied retention probability and use the corresponding average as final estimate of the intensity function. We propose this technique for point patterns in arbitrary metric spaces.

The paper is structured as follows. In Section \ref{s:Preliminaries} we give a short background on point processes and intensity estimation. In Section \ref{s:Smoothing} we introduce our resample-smoothing technique, study its statistical properties and discuss ways to choose the amount of smoothing, i.e.\ thinning, to apply. In Section \ref{s:Evaluations} we evaluate our approach numerically for a few different planar point processes and in Section \ref{SectionDataAnalysis} we apply our methodology to two datasets: a planar point pattern and a linear network point pattern. 
Section \ref{s:Discussion} contains a discussion and some directions for future work and in the Appendix we provide the proofs of the theoretical results in the paper as well as bias and variance plots for the simulation study in Section \ref{s:Evaluations}.

\section{Preliminaries}
\label{s:Preliminaries}

Let $X$ be a simple point process in an arbitrary space $S$, by which we here mean a complete separable metric space with associated metric/distance $d(\cdot,\cdot)$ \citep{DVJ2}. 
Throughout, all subsets $A\subseteq S$ considered are Borel sets and we endow $S$ with some suitable locally finite Borel reference measure $A\mapsto|A|\geq0$, $A\subseteq S$; we denote integration with respect to this measure by $\int \de u$. 

A realisation $\x=\{x_1,\ldots,x_n\}\subseteq S$, $n\geq0$, of $X$, i.e.\ an almost surely (a.s.) finite over bounded Borel sets (locally finite) collection of distinct points in $S$, will be referred to as a point pattern.

The cardinality of the set $X\cap A$, $A\subseteq S$, will be denoted by $N(X\cap A)\in\{0,1,\ldots\}$ and we note that by definition we a.s.\ have $N(X\cap A)<\infty$ for bounded $A\subseteq S$ and $N(X\cap\{u\})\in\{0,1\}$ for any $u\in S$. 

This setup is the usual one in the general study of point processes and common examples include:
\begin{itemize}
\item The points of $X$ are located in $d$-dimensional Euclidean space $S=\R^d$, $d\geq1$ \citep{VanLieshoutBook,Diggle14Book,BRT15}. Here $d(u,v)=\|u-v\|$, $u,v\in\R^d$, where $\|\cdot\|=\|\cdot\|_d$ denotes the Euclidean norm, and $|\cdot|$ is Lebesgue measure (volume).

\item The underlying space is given by a linear network, i.e., a union
$$
S=L=\bigcup_{i=1}^{k}l_i
$$ 
of $k\in\{1,2,\ldots\}$ line segments $l_i=[u_i,v_i]=\{tu_i + (1-t)v_i:0\leq t\leq 1\}\subseteq\R^d$, $d\geq1$. A common choice for $d(u,v)$ is the shortest path distance, which gives the shortest length of any path joining $u,v\in L$ \citep{OS12,rakshit17}. Treated as a graph with vertices given by the intersections and endpoints of the line segments, one assumes that $L$ is connected. 
The measure $|\cdot|$ here corresponds to integration with respect to arc length. 

\item The point process $X$ generates collections of points on the sphere $S=\alpha\mathbb{S}^{d-1}=\{x\in\R^d:\|x\|_d=\alpha\}$, $\alpha>0$, $d\geq1$, where $d(\cdot,\cdot)$ is the great circle distance and $|\cdot|$ is the spherical measure \citep{lawrence2016sphere,moller2016sphere}. 

\end{itemize}
We emphasise that in each of the above cases there exist other metrics and measures which may be more suited for a particular context. 

At times, we will assume that $X$ is stationary, or invariant. More specifically, there is a family of transformations/shifts $\{\theta_s:s\in S\}$, $\theta_s:S\to S$, along $S$, which induces a so-called flow, under which the distribution of $\theta_s X=\{\theta_s(x):x\in X\}$ coincides with that of $X$ for any $s\in S$. The underlying assumption will be that $S$ is a so-called (unimodular) homogeneous space with a fixed origin $o\in S$, with $d(\cdot,\cdot)$ chosen such that it metrizes $S$ and $|\cdot|$ chosen to be the associated (left) Haar measure \citep{Last2010, SchneiderWeil}. To exemplify, in Euclidean spaces with $|\cdot|$ chosen to be Lebesgue measure, we let $\theta_s(u)=u+s\in\R^d$, $u,s\in\R^d$, which yields the classical notion of stationarity, and on a sphere with the corresponding spherical measure we consider the orthogonal group of rotations. 
Note that a more general setting is also possible \citep[Chapter 7]{Kallenberg2017}. 

\subsection{Intensity functions
}

To characterise the first moment of $X$, i.e.\ the marginal distributional properties of its points, we consider its intensity function $\rho:S\to[0,\infty)$. It may be defined through the Campbell formula \citep{DVJ2} which states that for any measurable function $f\geq0$ on $S$,
$$
\E\left[\sum_{x\in X}f(x)\right] = \int_S f(u) \rho(u) \de u.
$$
In particular, 
$$
\E[N(X\cap A)]=\int_{A}\rho(u)\de u 
$$
for any $A\subseteq S$ and if $X$ is stationary then $\rho(u)\equiv\rho\in(0,\infty)$ for any $u\in S$. 
Since for any infinitesimal neighbourhood $du\subseteq S$ of $u\in S$ with size/measure $\de u$, we have that $\E[N(X\cap du)]$ coincides with the probability of finding a point of $X$ in $du$, this probability is given by $\rho(u)\de u$.

\subsection{Independent thinning}
\label{sec:Thinning}

A key ingredient in our smoothing technique is the notion of independent thinning \citep[Section 5.1]{CSKWM13}: given some measurable retention probability function $p(u)\in(0,1]$, $u\in S$, we run through the points of $X$ and delete a point $x\in X$ with probability $1-p(x)$, independently of the deletions carried out for the other points of $X$. It follows that the resulting thinned process has intensity
$$
\rho_{th}(u)
=
p(u)\rho(u), \quad u\in S,
$$
where $\rho(\cdot)$ is the intensity of the original process $X$ \cite[Section 5.1]{CSKWM13}. 
For further details on the thinning of point processes, see e.g.\ \citet{MollerSchoenberg} and \citet[Section 11.3.]{DVJ2}. 

It is worth mentioning that a Poisson process stays Poissonian after independent thinning \citep[Exercise 11.3.1]{DVJ2} and, in addition, the independent thinning of an arbitrary point process $X$
with low retention probability
results in a point process which, from a distributional point of view, is approximately a Poisson process \citep[Section 9.2.2]{BRT15}.

\subsection{Voronoi tessellations}
The next key ingredient in our estimation scheme is the Voronoi/Dirichlet tessellation of a point pattern $\x=\{x_1,\ldots,x_n\}$ contained in some subset $W\subseteq S$ \citep{CSKWM13,Okabe}.  
Generally speaking, a tessellation of $W$ is a tiling such that i) the union of all tiles constitutes all of $W$, and ii) the interiors of any two tiles have empty intersections. 

The Voronoi/Dirichlet cell $\V_{x}$ associated with $x\in\x$ consists of all $u\in S$ which are closer to $x$ than any $y\in\x\setminus\{x\}$, i.e.
\begin{align}
\label{VorCell}
\V_{x}&=\V_{x}(\x,W)
\\
&= \{ u \in W: d(x,u) \leq d(y,u) \text{ for all } y\in X\setminus\{x\}\}.\nonumber
\end{align}
The tiling  $\{\V_{x}\}_{x\in\x}$ is referred to as the Voronoi/Dirichlet tessellation generated by $\x$. 
Clearly, the shape of each $\V_{x}$ depends on the distance $d(\cdot,\cdot)$ chosen for $S$ and its size, $|\V_{x}|$, depends on the chosen reference measure $|\cdot|$. 

\subsection{Intensity estimation}
Given a point pattern $\x=\{x_1,\ldots,x_n\}$ in some study region $W\subseteq S$, $|W|>0$, we next set out to estimate $\rho(u)$, $u\in W$, under the assumption that $\x$ is a realisation of $X\cap W$.

Before going into details about specific estimators, we briefly mention how different estimators' performances may be evaluated and compared. 
To evaluate the performance of an estimator $\widehat\rho(\cdot)=\widehat\rho(\cdot;X,W)$ of $\rho(u)$, $u\in W$, it is common practice to employ 
the {\emph{Mean Integrated Square Error} ({\rm MISE})}: 
\begin{align}
\label{e:MISEtrue}
{\rm MISE} 
&=\E\left[\int_W\left(\widehat\rho(u) - \rho(u)\right)^2\de u\right]
\nonumber
\\
&= \int_W\Var(\widehat\rho(u))\de u + \int_W{\rm bias}(\widehat\rho(u))^2\de u
\nonumber
\\ 
&=
{\rm IV} + {\rm ISB}
,
\end{align}
where ${\rm bias}(\widehat\rho(u))=\E[\widehat{\rho}(u)]-\rho(u)$. 
Given $k\geq1$ realisations of $X\cap W$, to obtain an estimate of {\rm MISE} we average over the integrated square errors generated by each of the $k$ realisations. 

Alternatively, we may find estimates of the functions $\Var(\widehat\rho(u))$ and ${\rm bias}(\widehat\rho(u))$, $u\in W$, based on the $k$ patterns and integrate these over $W$.
This is the setup chosen for the numerical evaluations presented in Section \ref{s:Evaluations}.

\subsubsection{Voronoi intensity estimation}

In practice, it is often the case that events mainly occur in specific parts of the study region, e.g.\ that accidents often happen in more crowded streets or on specific parts of a highway, or that trees tend to grow mainly in specific parts of a forest. In other words, there are sharp boundaries between parts with high and low intensities. 
We argue, similarly to \citet{BSV10}, that in order not to blur such boundaries, it is preferable to employ an adaptive intensity estimation scheme, which adapts locally to changes in the spatial distribution of the events.

We here choose to focus on a particular kind of adaptive intensity estimator, namely Voronoi intensity estimators. Recalling the Voronoi cells in expression \eqref{VorCell}, we formally define the Voronoi intensity estimator of the intensity function of $X\subseteq S$ as follows.

\begin{definition}
For a point process $X$ with intensity function $\rho(\cdot)$,  
the Voronoi intensity estimator of $\rho(u)$, $u\in W\subseteq S$, $|W|>0$, is given by
\begin{align}
\label{Vor} 
\widehat{\rho}^{V}(u) \nonumber
&=
\widehat{\rho}^{V}(u;X,W)
=
\sum_{x\in X\cap W}
\frac{\1\{u\in\V_{x}\}}{|\V_{x}|}
\\
&=
\sum_{x\in X\cap W}
\frac{\1\{u\in\V_{x}(X,W)\}}{|\V_{x}(X,W)|}
,
\qquad u\in W,
\end{align}
where, for $u\in \V_x\cap\V_y\neq\emptyset$, $x,y\in X\cap W$, we either let $\widehat{\rho}^{V}(u)=1/|\V_x|$ or $\widehat{\rho}^{V}(u)=1/|\V_y|$ according to some arbitrary rule. 
Note that $\widehat{\rho}^{V}(u)=0$ if $X\cap W=\emptyset$.
\end{definition}
It should be noted that the points of $X$ which lie outside $W$ may interact with those inside $W$. Indeed, due to the way we define the Voronoi cells in expression \eqref{VorCell}, the Voronoi intensity estimator neglects possible edge effects.

The Voronoi intensity estimator, which was introduced by \citet{brown1965} and \citet{Ord78} in the context of Euclidean spaces, has been considered by e.g.\ \citet{baddeley2007validation,ogata11,BSV10,vanIntensity}. \citet{ebeling93} have used it to study local spatial concentration of photons, \citet{duyckaerts1994} and \citet{duyckaerts2000} have employed it to estimate neuronal density, and it has been applied in the setting of statistical seismology by \citet{ogata11} and \citet{BRT15}. 
In the context of linear networks, \citet{OS12} briefly discussed a Voronoi based density estimator, the network Voronoi cell histogram, for the purpose of non-parametric density estimation on linear networks. They further discussed geometric properties of Voronoi tessellations on linear networks. 
\citet{BSV10} focused on the planar case and particular statistical properties. 

\section{Resample-smoothing of intensity estimators}
\label{s:Smoothing}

\citet{BSV10} pointed out that when there are abrupt changes in the intensity, kernel-based estimators may yield substantial bias and high variance, and they showed that the Voronoi estimator can alleviate these problems. Unfortunately, they tend to under-smooth in very dense areas surrounded by nearly empty neighbourhoods. 
This may be said about adaptive estimators in general; there is a tendency of adapting too much to the particular features of the observed point pattern $\x$, rather than reflecting the features of the intensity function of the underlying point process $X$. 
To see how the under-smoothing, i.e.\ the over accentuating of local features of the Voronoi intensity estimator occurs, note that for a pattern $\x$, if $x\in\x$ is located in a very dense part then its Voronoi cell becomes small and, consequently, $\widehat{\rho}^{V}(u)=1/|\V_x|$ becomes very large for $u\in\V_x$. 
A further issue with the Voronoi intensity estimator is that its variance tends to be quite large, thus resulting in quite unreliable estimates. 

One may further ask the adequate question whether there are other tessellations $\{\mathcal{C}_i\}$, $\bigcup_i \mathcal{C}_i=W$, giving rise to estimators $\widehat\rho(u)=\sum_i\beta_i\1\{u\in\mathcal{C}_i\}$, $\beta_i>0$, which perform better than the Voronoi intensity estimator. Even so, the question then still remains how to explicitly generate a better one. 
In addition, an advantage of the kernel estimation approach is arguably in that it generates a smoothly varying intensity estimate, at least when using certain kernels, as opposed to the possibly unnatural ``jumps" generated by the Voronoi estimator. 

As a remedy for these issues, one suggestion is to follow \cite{BSV10} by considering the so-called centroidal Voronoi intensity estimator.
A further idea, which seems appealing, is to introduce some smoothing procedure for $\widehat{\rho}^{V}(\cdot)$, which would reduce the unnaturally extreme peaks while smoothing out the ``jumps". 
We next propose such a smoothing procedure, which we refer to as \emph{resampling-smoothing}. 

Recall the independent thinning operation in Section \ref{sec:Thinning}. 
We will here focus on the simple case where $p(u)\equiv p\in(0,1]$, $u\in W$, which is referred to as $p$-thinning \cite[Section 5.1]{CSKWM13}; we identify the case $p=1$ with the unthinned process $X$. 
From Section \ref{sec:Thinning} we have that 
$$
\rho(u)=\frac{\rho_{th}(u)}{p},
\qquad u\in S,
$$
where we recall the intensity $\rho_{th}(\cdot)$ of the thinned process $X_p$. Hence, dividing by $p$ is exactly what is needed to compensate for the reduced intensity caused by removing points. 
We exploit this relationship in the following way.  
Given a point pattern $\x$ and an estimator $\widehat\rho(\cdot)$ of $\rho(u)$, $u\in W$, fix some $p\in(0,1]$ and thin the point pattern $m\geq1$ times, each time with retention probability $p$. 
This results in the thinned patterns $\x_p^1,\ldots,\x_p^m$, each for which the intensity is estimated. We now let the average of these $m$ estimated intensity functions, divided by $p$, be reported as the final estimate; 
note the similarity with the approach considered by \cite{baddeley2007validation}. 
The resample-smoothed Voronoi intensity estimator is formally defined as follows.

\begin{definition}
Consider a point process $X\subseteq S$ with intensity function $\rho(\cdot)$. 
Given some $p\in(0,1]$ and $m\geq1$, the {\em resample-smoothed Voronoi intensity estimator} of $\rho(u)$, $u\in W\subseteq S$, $|W|>0$, is given by
\begin{align}
\label{SmoothVor}
\widehat{\rho}_{p,m}^{V}(u)
=
\widehat{\rho}_{p,m}^{V}(u;X,W)
=
\frac{1}{m}\sum_{i=1}^m
\frac{\widehat{\rho}_i^{V}(u)}{p}
,
\end{align}
where 
$$
\widehat{\rho}_i^{V}(u)
=
\widehat{\rho}^{V}(u;X_p^i,W)
=
\sum_{x\in X_p^i}
\frac{\1\{u\in\V_{x}(X_p^i,W)\}}{|\V_{x}(X_p^i,W)|}
$$
is the Voronoi intensity estimator based on the $i$th thinning $X_p^i$ of $X\cap W$.
Note that when $p=1$, $\widehat{\rho}_{p,m}^{V}(\cdot)$ reduces to $\widehat{\rho}^{V}(\cdot)$ for any $m\geq1$. 
\end{definition}

Reflecting on the effect of the thinning procedure, for each thinned version we obtain new Voronoi cells and consequently different locations of the jumps in the corresponding intensity estimate $\widehat{\rho}_i^{V}(\cdot)$. This is what results in the ``smoothing" and it is also the remedy for choosing the specific tiling in a possibly wrong/rigid way. Note also that we in fact simply are considering the average of $m$ different estimates of $\rho(\cdot)$. 

\subsection{Theoretical properties}
\label{s:Properties}

We next look closer at some statistical properties of resample-smoothed Voronoi intensity estimators. The proofs of all the results presented can be found in the Appendix.

We stress that in the case of the restriction $X\cap W$ of a point process $X$ to a (bounded) region $W\neq S$, the Voronoi cells $\V_{x}(X,W)$ are different than when $W=S$. Hereby, distributional properties of $\widehat{\rho}_{p,m}^{V}(\cdot)$ may be different depending on how $W$ is chosen.

We start by considering the asymptotic scenario where the number of thinned patterns, $m\geq1$, in the estimator \eqref{SmoothVor} tends to infinity. Note that by the result below, 
we have that the limit $\lim_{m\to\infty}\widehat{\rho}_{p,m}^{V}(u;X,W)$ a.s.\ exists for a point process $X$. 

\begin{lemma}
\label{LemmaConvergence}
Given fixed $p\in(0,1]$ and $k\geq1$, for any point pattern $\x\subseteq W\subseteq S$ 
we have that $|\widehat{\rho}_{p,m}^{V}(u;\x,W) - \widehat{\rho}_{p,m+k}^{V}(u;\x,W)|\to0$ a.s.\ as $m\to\infty$. In turn, we have that $\lim_{m\to\infty}\widehat{\rho}_{p,m}^{V}(u;\x,W)$ a.s.\ exists. 

\end{lemma}

\subsubsection{Bias}

Turning to the first order properties of $\widehat{\rho}_{p,m}^{V}(\cdot)$, we note that 
\begin{align}
\label{MassPreservation}
\int_W\widehat{\rho}_{p,m}^{V}(u)\de u \nonumber
&=
\frac{1}{mp}\sum_{i=1}^m
\sum_{x\in X_p^i}
\frac{\int_W \1\{u\in\V_{x}(X_p^i,W)\} \de u}{|\V_{x}(X_p^i,W)|}
\\
&=
\frac{1}{mp}\sum_{i=1}^m
N(X_p^i\cap W)
.
\end{align}
Hence, when $p=1$ we have preservation of mass, i.e.\ $\int_W\widehat{\rho}_{p,m}^{V}(u)\de u=N(X\cap W)$.
Taking expectations on both sides in \eqref{MassPreservation}, we obtain
$$
\E\left[\int_W\widehat{\rho}_{p,m}^{V}(u)\de u\right]
=
\frac{1}{m}\sum_{i=1}^m\frac{p\int_L\rho(u)\de u}{p}
=\int_W\rho(u)\de u,
$$ 
i.e., for any $m\geq1$ and $p\in(0,1]$, $\int_W\widehat{\rho}_{p,m}^{V}(u)\de u$ is an unbiased estimator of $\E[N(X\cap W)]$. 

Noting that $\E[\widehat{\rho}_{p,m}^{V}(u;X,W)] = \E[\widehat{\rho}^{V}(u;X_p,W)]/p$ for any $p\in(0,1]$ and $m\geq1$, we see that $\widehat{\rho}_{p,m}^{V}(u;X,W)$ is unbiased for the estimation of the intensity of $X$ if and only if the original Voronoi intensity estimator is unbiased for the estimation of the intensity of an arbitrary thinning $X_p$. 
There is unfortunately not much more to be said without explicitly assuming something about the distributional properties of $X$.

When $X$ is stationary, all Voronoi cells have the same distribution and we may speak of the typical Voronoi cell $\V_o=\V_o(X)$, which satisfies $\V_o\stackrel{d}{=}\theta_{-x}\V_x(X,S)$ for any $x\in X$; here $\theta_{-x}$ denotes the transformation/shift such that $x$ is taken to the origin $o\in S$. 
In particular, we have that $\widehat{\rho}_{p,m}^{V}(u)$ and $\widehat{\rho}_{p,m}^{V}(v)$ have the same distribution for any $u,v\in S$ and it turns out that unbiasedness holds.

\begin{theorem}
\label{thm:Unbiasedness}
For a stationary point process $X\subseteq W=S$ with intensity function $\rho>0$, the resample-smoothed Voronoi intensity estimator \eqref{SmoothVor} is unbiased for any choice of $p\in(0,1]$ and $m\geq1$. 
\end{theorem}

As our main interest lies in estimating non-constant intensity functions, stationary models are of limited practical interest.
We next turn to inhomogeneous Poisson processes in Euclidean spaces.


\begin{theorem}\label{thm:unbiasedbound}
Let $X\subseteq W=S=\R^d$, $d\geq1$, be a Poisson process with intensity function $\rho(u)$, $u\in\R^{d}$, which satisfies the Lipschitz condition that for some $\mu_u>0$, $|\rho (v)-\rho (u)| \leq \mu_u \varepsilon$ for $v\in B(u,\varepsilon)$ and $\varepsilon>0$ sufficiently small; $B(u,\varepsilon)$ denotes the Euclidean ball with centre $u$ and radius $\varepsilon>0$. 
Denoting by $C_u(X)$ the Voronoi cell containing $u\in\R^d$, 
assume further that $m^{\kappa }:=\sup_{u\in \R^{d}}\E[| C_u(X)|^{-\kappa}] < \infty$ for some $\kappa \geq 1+1/d$. 
Then, for any $u\in\R^d$, $p\in(0,1]$ and $m\geq1$, 
\begin{align*}
\left|
\rho(u)-\E\left[
\widehat\rho_{p,m}^V(u)
\right]
\right|\leq C p^{-1}(p\rho(u))^{-1/d}\log(p\rho(u))^{2/d}
\end{align*}
for some $C>0$ that depends on the intensity. 
The right hand side tends to $0$ as the intensity tends to infinity.  
\end{theorem}
 
\begin{remark}
The moment condition, and the Lipschitz assumption on $\rho $ can be relaxed to weaker versions and still have the left hand side go to $0$, but the rate would be different.

It has been conjectured that the size of the typical cell of a homogeneous Poisson process follows a (generalised) Gamma distribution (see e.g.\ \citep{CSKWM13}); note in particular Lemma \ref{LemmaHomPoisson} below. The moment condition in the statement of the above result, i.e.\ $m^{\kappa}<\infty$, would be satisfied if this is indeed the case. Under such a conjectured distribution, \citet{BSV10} showed that in the planar case the original Voronoi intensity estimator is ratio-unbiased for a given class of intensity functions. 
\end{remark}

\subsubsection{Variance} 
Regarding the variance of $\widehat{\rho}_{p,m}^{V}(u)$, 
the next result shows that in an infinite sized study region $W$, by thinning as much as possible we also obtain a variance of the resample-smoothed Voronoi estimator which is close to 0. In addition, letting $m\to\infty$ has the same effect. 
Hence, for cases where the estimator is unbiased we should, in theory, smooth as much as possible, in combination with choosing $m$ as large as possible. The problem in practice, however, is that point patterns are sampled in bounded regions $W$ and we have to resort to finite $m$. This motivates the data-driven approaches suggested in Section \ref{sec:ChoosingParameters}. 

\begin{theorem}\label{thm:Variance}
Consider a point process $X\subseteq W\subseteq S$ such that $\widehat{\rho}^{V}(u)=\widehat{\rho}^{V}(u;X,W)$, $u\in W$, has finite variance. 
Given $p\in(0,1]$ and $m\geq1$, the variance of $\widehat{\rho}_{p,m}^{V}(u)=\widehat{\rho}_{p,m}^{V}(u;X,W)$ satisfies
\begin{align*}
\Var(\widehat{\rho}_{p,1}^{V}(u))/m
&\leq 
\Var(\widehat{\rho}_{p,m}^{V}(u)) 
\leq 
\Var(\widehat{\rho}_{p,1}^{V}(u))
\\
&\leq
\Var(\widehat{\rho}_{1,m}^{V}(u))
=
\Var(\widehat{\rho}^{V}(u))
\end{align*}
and it tends to the covariance between $\widehat{\rho}^{V}(u;X_p^1,W)/p$ and $\widehat{\rho}_2^{V}(u;X_p^2,W)/p$ as $m\to\infty$. 
Moreover, for a fixed $m\geq1$ it further follows that $\lim_{p\to0}\Var(\widehat{\rho}_{p,m}^{V}(u))=1/(m|W|^2)$, which is 0 if $|W|=\infty$. 
\end{theorem}

Turning to the stationary case, from the proof of Theorem \ref{thm:Unbiasedness} (see the Appendix) we have that the $p$-thinning $X_p$ of a stationary point process $X$ with intensity $\rho>0$ is again stationary, but with intensity $p\rho$. 
For $X_p$, the distribution $\bar{P}_{p}(\cdot)$ of the size of the cell that covers $u$ is the same for any $u\in S$ and it is given by (see \citet[Section 8]{Last2010} and \citet[Theorem 10.4.1.]{SchneiderWeil})
\begin{align}
\label{e:DistributionCell}
\bar{P}_{p}(A) = p\rho \int_A t P_{|\V_o(X_p)|}(dt), 
\quad A\subseteq[0,\infty), 
\end{align}
where $P_{|\V_o(X_p)|}(\cdot)$ is the distribution of the typical cell size. 
Besides giving us the unbiasedness in Theorem \ref{thm:Unbiasedness}, i.e. 
$$
\E[\widehat{\rho}_{p,m}^{V}(u)] 
= p^{-1}p\rho \int_0^{\infty} t^{-1} t P_{|\V_o(X_p)|}(dt) = \rho,
$$
the relationship \eqref{e:DistributionCell} further yields
\begin{align*}
\E[\widehat{\rho}_{p,1}^{V}(u)^2]
&=
\frac{1}{p^2}
\int_0^{\infty}
\frac{1}{t^2}
\bar{P}_{p}(dt)
=
\frac{\rho}{p}
\int_0^{\infty}
\frac{1}{t}
P_{|\V_o(X_p)|}(dt)
\\
&=
\frac{\rho}{p}
\E[1/|\V_o(X_p)|]
,
\\
\Var(\widehat{\rho}_{p,1}^{V}(u))
&=\frac{\rho}{p}\E[1/|\V_o(X_p)|] - \rho^2
.
\end{align*}
Through the proof of Theorem \ref{thm:Variance} (see the Appendix) we obtain that the variance of $\widehat{\rho}_{p,m}^{V}(u)$ is given by
\begin{align*}
\label{e:VarianceInequality}
&\rho\left(\frac{\E[1/|\V_o(X_p)|]}{p} - \rho\right)
\times
\\
&\times\frac{
1 + (m-1){\rm Corr}(\widehat{\rho}^{V}(u;X_p^1,S), \widehat{\rho}^{V}(u;X_p^2,S))
}{m}
,
\end{align*}
where {\rm Corr} denotes correlation. 
Unfortunately, we cannot get much further in the general setup; the problem lies in that $P_{|\V_o|}(\cdot)$ typically is not known. 

There is, however, one particular case where we can say a bit more and that is for Poisson processes on $\R$. 

\begin{lemma}
\label{LemmaHomPoisson}
For a Poisson process on $\R$ with intensity $\rho>0$, for any $p\in(0,1]$ and $m\geq1$ the typical cell size of $X_p$ follows an Erlang/Gamma distribution with shape and rate parameters $2$ and $2p\rho$, respectively. Hereby, $\Var(\widehat{\rho}_{p,m}^{V}(u)) \leq \Var(\widehat{\rho}_{p,1}^{V}(u))=\rho^2$. 
\end{lemma}

Empirically, we have consistently observed that for a fixed $m$, the variance of $\widehat{\rho}_{p,m}^{V}(u)$ decreases with $p$, for $u\in W$ located a given distance from the boundary of $W\subseteq S$. As this is partly supported by Theorem \ref{thm:Variance}, we are lead to the following conjecture.

\begin{conj}
\label{ConjectureVariance}
For an arbitrary point process $X\subseteq S$ and any $m$, the variance of $\widehat{\rho}_{p,m}^{V}(u)$ is a decreasing function of $p\in(0,1]$. In particular, if $\widehat{\rho}_{p,m}^{V}(u)$ is unbiased, this means that {\rm MISE} is decreasing with $p$. 
\end{conj}

\subsection{Choosing the smoothing parameters}
\label{sec:ChoosingParameters}
When using the resample-smoothed Voronoi intensity estimator \eqref{SmoothVor} in practice, one needs to specify the smoothing parameters $m\geq1$ and $p\in(0,1]$ prior to finding the intensity estimate. We next discuss how to obtain proper choices for $m$ and $p$. 

\subsubsection{Choosing the number of thinnings}
Lemma \ref{LemmaConvergence} tells us that for a fixed $p\in(0,1]$, $k\geq1$ and any point pattern $\x\subseteq W\subseteq S$, 
we have that $|\widehat{\rho}_{p,m}^{V}(u;\x,W) - \widehat{\rho}_{p,m+k}^{V}(u;\x,W)|\to0$ a.s.\ as $m\to\infty$. The question that remains, however, is for which $m\geq1$ we have that $|\widehat{\rho}_{p,m}^{V}(u;\x,W) - \widehat{\rho}_{p,m+k}^{V}(u;\x,W)|$ is sufficiently small. 
In our numerical evaluations in Section \ref{s:Evaluations} we illustrate that the estimated bias and variance of $\widehat{\rho}_{p,m}^{V}(u)$ do not change significantly for $m\geq200$. Hence, we propose to fix $m=200$ and then proceed by finding a proper choice for $p\in(0,1]$.

\subsubsection{Choosing retention probability}

The selection of $p\in(0,1]$ is clearly the more delicate matter here; essentially we are faced with problems similar to those of choosing bandwidths in kernel estimation. 

Through our numerical evaluations (see Section \ref{s:Evaluations}) we have found that the choice $p\in[0.1,0.3]$ always seems to generate the best intensity estimates in the sense that the variance-bias-tradeoff is taken into account by keeping both the bias and variance relatively small. The lower limit $0.1$ is based on our observation that the removal of more than 90\% of the points per thinning may generate too flat estimates in some cases; from the looks of Section \ref{s:Evaluations}, Theorem \ref{thm:Variance} and Conjecture \ref{ConjectureVariance} it may seem that the smaller the $p$, the better the estimate. We refer to the choice $m=200$ and $p\in[0.1,0.3]$ as our rule-of-thumb.

If one prefers a data-driven approach over the rule-of-thumb, we also propose a cross-validation approach to select $p$. 
Recalling a previous comment in Section \ref{sec:Thinning} about independent thinnings yielding approximate Poissonian distributional properties of the resulting processes, a natural approach to choosing $p$ when the number of thinned patterns, $m$, is fixed is to consider Poisson process likelihood cross-validation. This method has a long history in the literature of point processes and has e.g. been frequently used for bandwidth selection in kernel-based estimation \citep{SilvermanBook,Load99}. More specifically, given a point pattern $\x=\{x_1,\ldots,x_n\}\subseteq W\subseteq S$ and some fixed $m\geq1$, we choose the corresponding resampling/retention probability as a maximiser of the cross-validation criterion
\begin{align*}
CV(p)
=
CV_m(p)
=&
\sum_{i=1}^n\log \widehat{\rho}_{p,m}^{V}(x_i;\x\setminus\{x_i\},W)
\\
&
-
\int_W
\widehat{\rho}_{p,m}^{V}(u;\x,W) \de u,
\quad p\in(0,1].
\end{align*}
Note that $\widehat{\rho}_{p,m}^{V}(\cdot;\x\setminus\{x_i\},W)$ is the \emph{leave-one-out} version of $\widehat{\rho}_{p,m}^{V}(\cdot;\x,W)$, i.e.\ the resample-smoothed Voronoi intensity estimator based on the reduced sample $\x\setminus\{x_i\}$. 
As the computation of $CV(p)$, $p\in(0,1]$, can be quite computationally costly, in practice we may exclude the integral term in its expression since it approximately equals the number of points in the pattern. 
Moreover, in practice we calculate $CV(p_j)$, $j=1,\ldots,k$, $0<p_{j-1}<p_j\leq1$, sequentially by first generating $X_{p_k}^i$ and then iteratively generating $X_{p_{j-1}}^i=(X_{p_j}^i)_{p_{j-1}/p_{j}}$, $i=1\ldots,m$, $j=2,\ldots,k$. 
Note that for small $m$ the graph of $CV(p)$ may not be smooth and might contain local extrema. 

Finally, if the value obtained for $p$ through the cross-validation would deviate too much from the rule-of-thumb, we advise to proceed with the rule-of-thumb; see the log-Gaussian Cox process example in Section \ref{s:Evaluations} for a situation where this occurs.

\subsection{Large scale data and sparsity}
In general, when the number of events, $n$, of an observed point pattern $\x=\{x_1,\ldots,x_n\}$ is very large, it is often natural to consider an adaptive intensity estimator as the scales of intensity likely vary a lot. 

It may not be computationally feasible to compute $\widehat{\rho}_{p,m}^{V}(\cdot)$, $p\in(0,1]$, for an arbitrary $m\geq1$ (or any other intensity estimator for that matter). 
An alternative way of exploiting the proposed setup is to consider $\widehat{\rho}_{p_0,m}^{V}(\cdot)$ for some $p_0\in[0.1,0.3]$ and $m=1$. This means that we would introduce sparsity by only having to generate Voronoi cells for 10--30\% of the original number of points. The results in Section \ref{s:Evaluations} indicate how good an estimate one would typically obtain. Moreover, if the computation of $\widehat{\rho}_{p_0,1}^{V}(\cdot)$ is reasonably quick, one could generate a further estimate $\widehat{\rho}_{p_0,1}^{V}(\cdot)$ and average over these to obtain $\widehat{\rho}_{p_0,2}^{V}(\cdot)$. One could then continue like this in a stepwise fashion, given a total computation timeframe.

\section{Numerical evaluations}
\label{s:Evaluations}
As previously pointed out, we evaluate our intensity estimation approach numerically, which we choose to do in the Euclidean setting. 

In our simulation study, we consider four different types of models with varying degrees of variation in intensity and spatial interaction; clustering, spatial randomness and regularity. For each model we use 500 realisations on $W=[0,1]^2$ to generate numerical estimates of relevant quantities such as bias, variance, Integrated Variance ({\rm IV}), Integrated Square Bias ({\rm ISB}) and Integrated Absolute Bias ({\rm IAB}) for $\widehat{\rho}_{p,m}^{V}(u)$, $u\in W$; recall that Mean Integrated Square Error ({\rm MISE}) is obtained as the sum of {\rm IV} and {\rm ISB}. 
To carry out the analysis, we make use of the \textsf{R} package \verb|spatstat| \citep{BRT15}. 
For each model considered, in the Appendix, we provide plots of the estimated bias and variance for $m=200$ and a range of values of $p\in(0,1]$, together with the estimated biases and variances obtained through kernel estimation. 

The overall conclusion is that we clearly reduce the estimation errors by resample-smoothing the Voronoi intensity estimator. Moreover, the cross-validation approach to selecting $p$ on average yields slightly poorer intensity estimates than the rule-of-thumb, in particular if the model is clustered. 

\subsection{Homogeneous Poisson process}
\label{s:HomPoi}

\begin{figure*}[!h]
\centering
\includegraphics[scale=0.27]{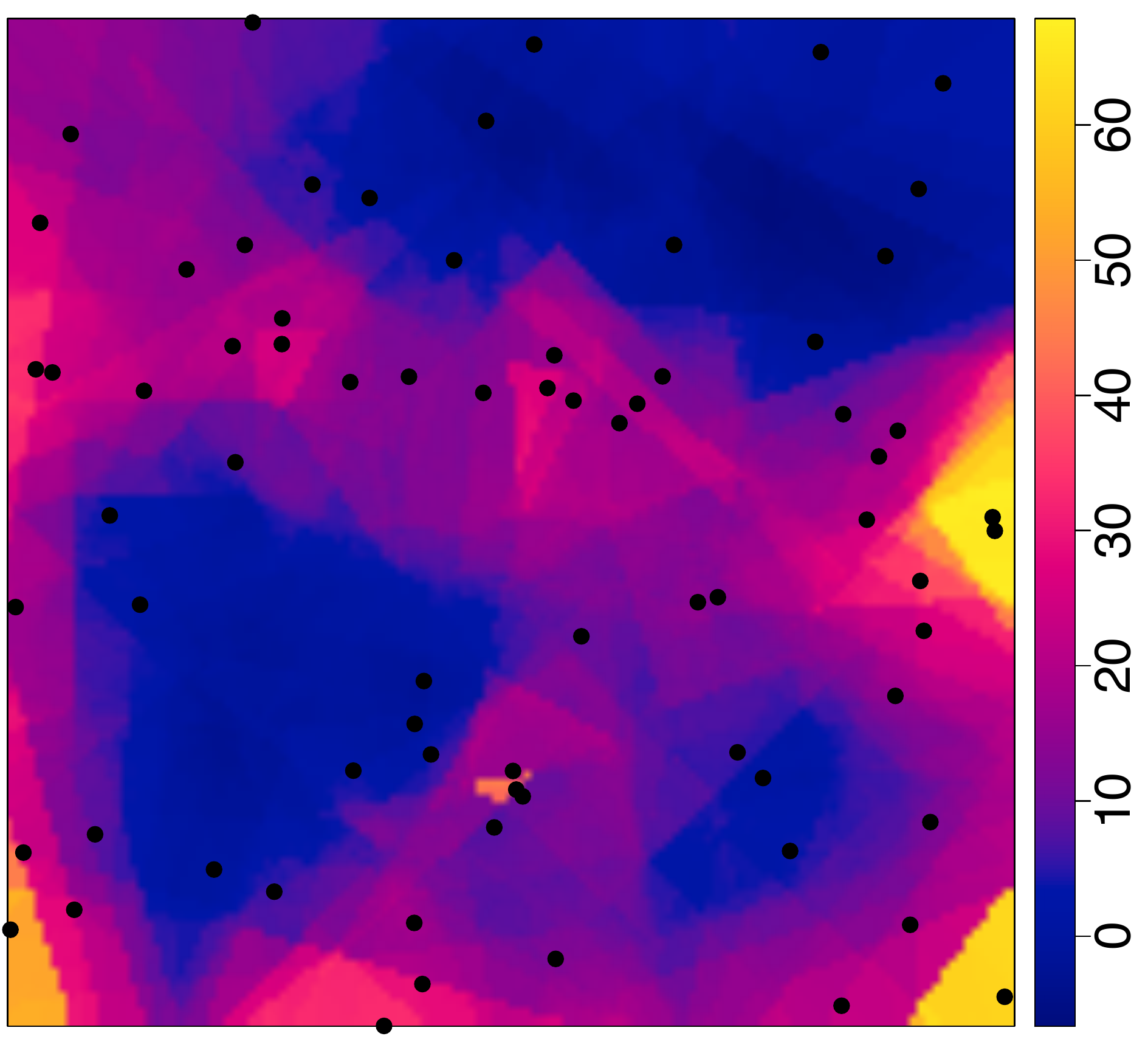}
\includegraphics[scale=0.27]{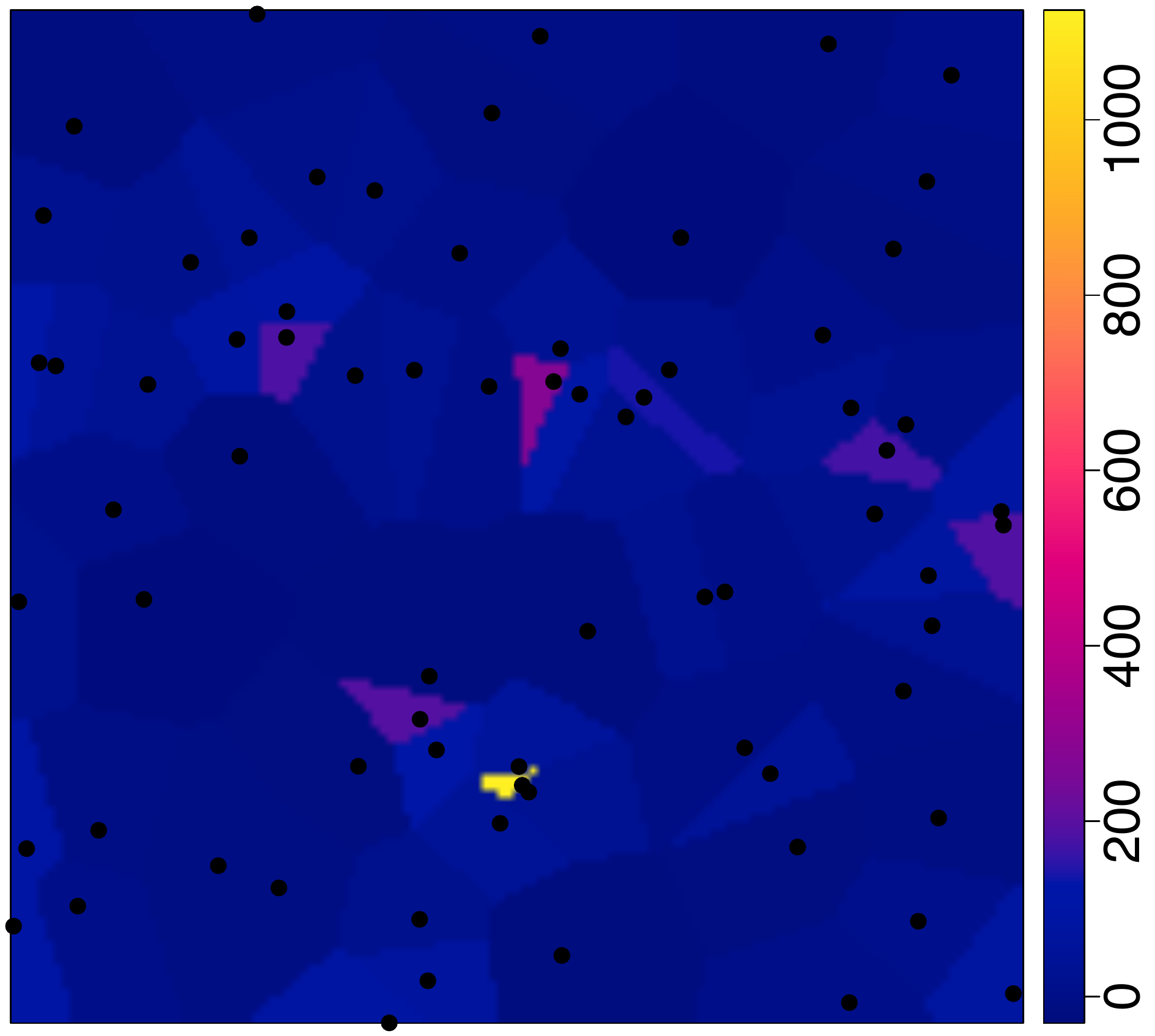}
\caption{
Estimation error plots for a realisation of a homogeneous Poisson process $X$ in $W=[0,1]^2$ with intensity $\rho=60$. \emph{Left}: $p=0.2$ and $m=200$. \emph{Right}: $p=1$.
The underlying point pattern has been superimposed in all plots.
}
\label{f:EstErrHomPoi}
\end{figure*}

We here consider a homogeneous Poisson process $X\subseteq W=[0,1]^2$ with intensity $\rho=60$. Table \ref{t:HomPoiR2} provides estimates of {\rm IAB}, {\rm ISB} and {\rm IV} for $\widehat{\rho}_{p,m}^{V}(u)$, $u\in W$, $m=200, 300, 400$, $p=0.1,\ldots,1$; recall that we use 500 realisations of $X$. Indeed, the bias seems fairly stable over the range of values for $p$ and the variance is clearly decreasing with $p$; choosing $p$ according to the rule-of-thumb keeps {\rm MISE} small. 
For illustrational purposes, in Figure \ref{f:EstErrHomPoi} we provide estimation error plots for one of the realisations, for $p=0.2$ and $p=1$ with $m=200$. One can clearly see the gain of the resample-smoothing; note that the under-estimation occurs in the empty regions. In addition,  in the Appendix we provide plots of the estimated bias and variance for $p=0.1,0.3,0.5,0.7,0.9,1$ and $m=200$, and they essentially confirm what has been observed in Table \ref{t:HomPoiR2}. 

\begin{table*}[!htbp]
\caption{Estimates of {\rm IAB}, {\rm ISB} and {\rm IV} for $\widehat{\rho}_{p,m}^{V}(u)$, $u\in W=[0,1]^2$, $m=200,300,400$, $p=0.1,\ldots,1$, based on 500 realisations of a homogeneous Poisson process in $W=[0,1]^2$ with intensity $\rho=60$.
}
\begin{center}
\begin{tabular}{|l | rrr|rrr|rrr|rr}
\hline
&  & IAB &  &  & ISB & & & IV~~ & \\
\hline
\theadfont\diagbox[width=2.5em]{$p$}{m}& 200 & 300 & 400 & 200 & 300 & 400 & 200 & 300 & 400\\
\hline
.1 & 5.7 & 5.7 & 5.7 & 43.5 & 43.2 & 43.0 & 158.4& 154.8 & 152.5\\
\hline
.2 & 4.6 & 4.6 & 4.6 & 28.4 & 28.5 & 28.4 & 264.1& 260.3& 257.9\\
\hline
.3 & 3.9 & 3.9 & 3.9 & 22.5 & 22.2 &22.2 & 375.3& 370.6 & 368.8\\
\hline
.4 & 3.5 & 35 & 3.5 & 19.7 & 19.6 &19.6 & 490.6& 488.8 & 487.8\\
\hline
.5 & 3.2 & 3.2 & 3.2 & 18.1 & 18.1 &18.1 & 672.0& 623.9 & 622.9\\
\hline
.6 & 3.0 & 3.0 & 3.0 & 17.1 & 17.1 &17.0 & 781.9& 779.4 & 779.0\\
\hline
.7 & 2.9 & 2.9 & 2.9 & 16.5 & 16.5 &16.5 & 960.0& 958.7 & 958.8\\
\hline
.8 & 2.9 & 2.9 & 2.9  & 16.0 & 16.0 &16.0 & 1172.2& 1171.8 & 1171.1\\
\hline
.9 & 2.9 & 2.9 & 2.9 & 15.8 & 15.8 &15.8 & 1422.2& 1419.6 & 1418.9\\
\hline
1 & 2.9 & 2.9 & 2.9 & 15.8 & 15.8 &15.8 & 1733.2& 1733.2 & 1733.2\\
\hline
\end{tabular}
\label{t:HomPoiR2}
\end{center}
\end{table*}

Turning to the cross-validation approach to selecting $p$, with $m=200$, based on 100 realisations of the model we obtain ${\rm IAB}=4.9$, ${\rm ISB}=30.3$ and ${\rm IV}=255$ which are in the range of what one obtains when $p$ is fixed in $(0.1,0.3)$. In Table \ref{t:PselectHomPoiR2} we further provide the 100 selected values for $p$ and we see that the majority of them fall within the range of our rule-of-thumb. Comparing with kernel estimation under uniform, or global, edge correction, using Poisson likelihood cross-validation \citep{Load99,BRT15} to select the bandwidth, we obtain ${\rm IAB}=0.24$, ${\rm ISB}=0.11$ and ${\rm IV}=126.05$. By instead employing the bandwidth selection method of \citet{cronie2018bandwidth}, we obtain ${\rm IAB}=0.87$, ${\rm ISB}=1.12$ and ${\rm IV}=688.25$.

\begin{table*}[!htbp]
\caption{Cross-validation selections of $p$ for $m=200$ in a geometric sequence, 
based on 100 realisations of a homogeneous Poisson process in $W=[0,1]^2$ with intensity $\rho=60$.
}
\begin{center}
\begin{tabular}{l | cccccccc}
\hline
$\hspace{.8cm}p$&0.10 & 0.13 & 0.18 & 0.24 & 0.33 & 0.44 & 0.59 & 0.80\\
\hline
Frequency &63 & 15 & 5 & 8 & 3 & 4 & 2&0
\\
\hline
\end{tabular}
\label{t:PselectHomPoiR2}
\end{center}
\end{table*}

\subsection{Inhomogeneous Poisson process}
\begin{figure*}[!htbp]
\centering
\includegraphics[scale=0.2]{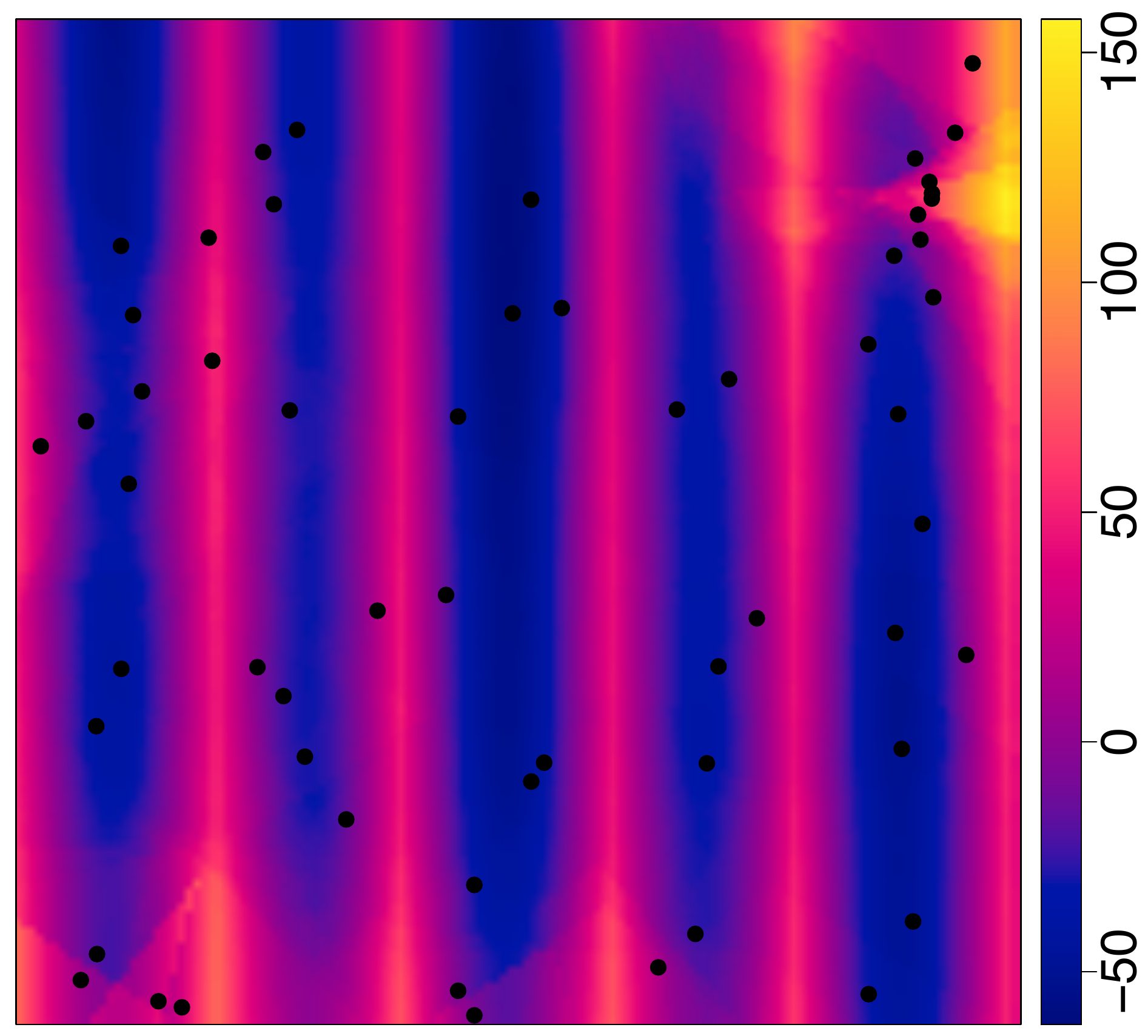}
\includegraphics[scale=0.2]{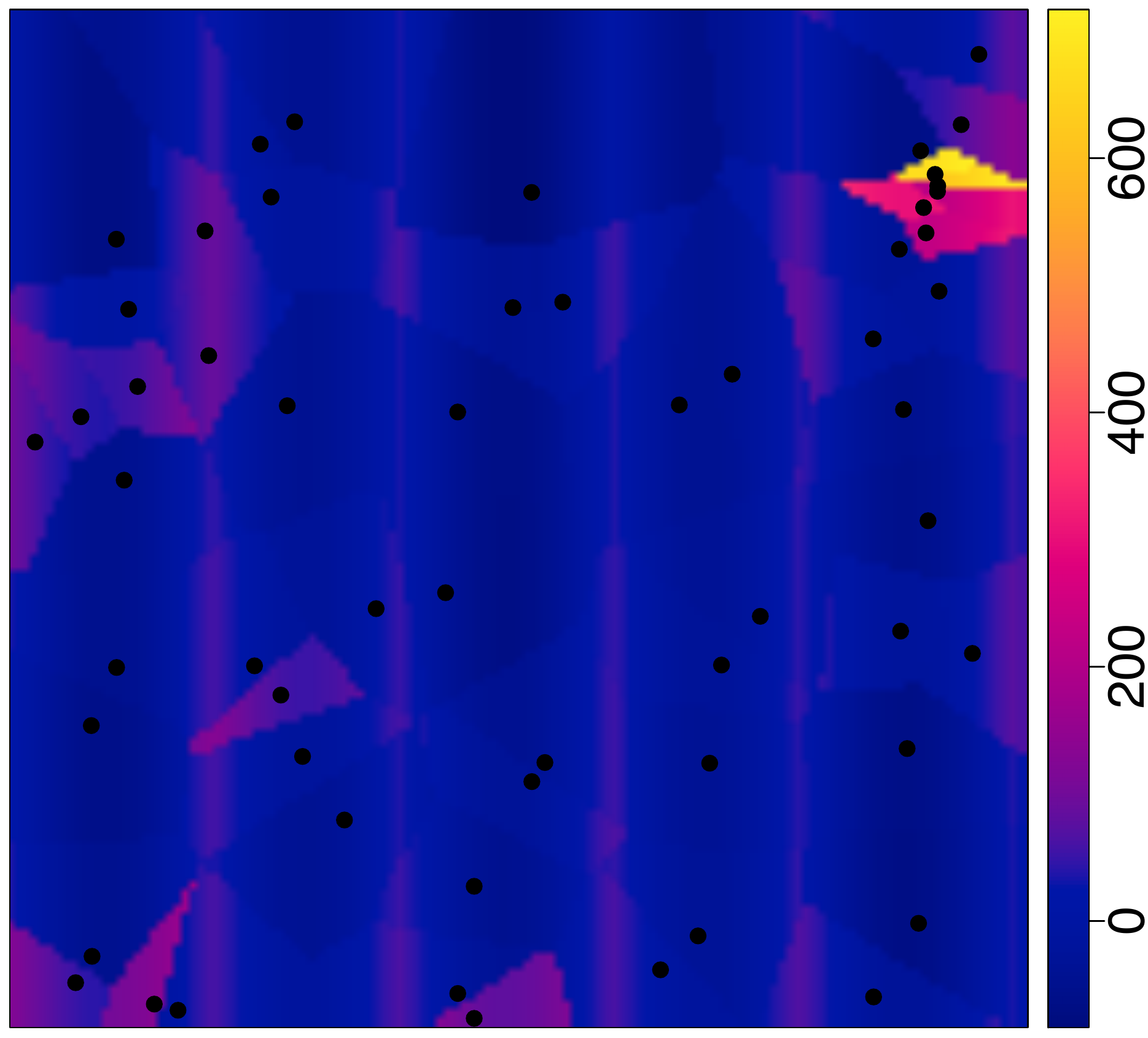}
\includegraphics[scale=0.2]{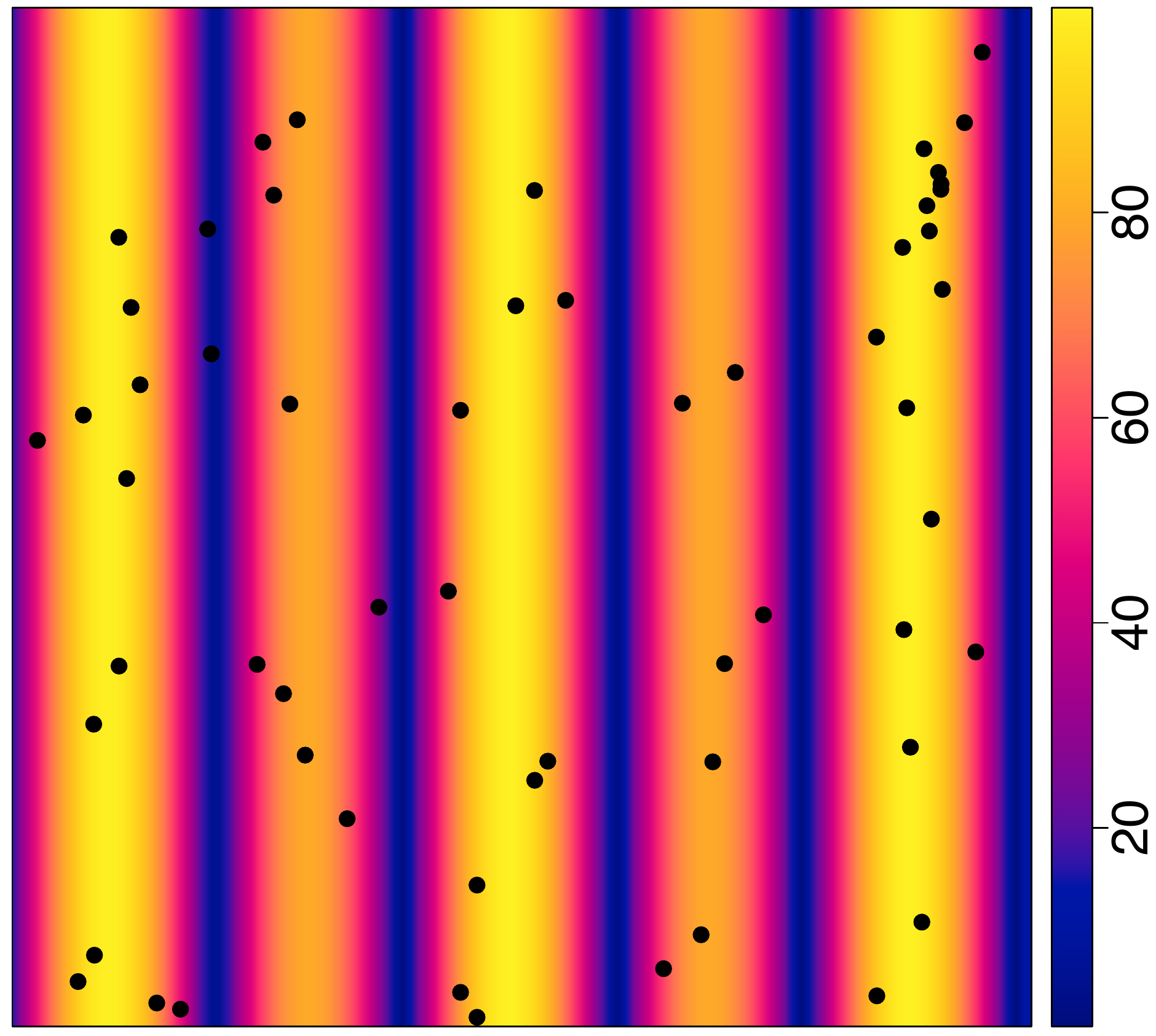}
\caption{
True intensity and estimation error plots for a realisation of an inhomogeneous Poisson process on $W=[0,1]^2$ with intensity $\rho(x,y)=|10+90\sin(16x)|$. \emph{Left}: $p=0.2$ and $m=200$. \emph{Middle}: $p=1$. \emph{Right}: True intensity.
The underlying point pattern has been superimposed in all plots.
}
\label{f:EstErrInhomPoi}
\end{figure*}

More interestingly, we next consider 500 realisations of an inhomogeneous Poisson process $X\subseteq W=[0,1]^2$ with intensity $\rho(x,y)=|10+90\sin(16x)|$; the expected total point count is $58.6$. Table \ref{t:InhomPoiR2} provides estimates of {\rm IAB}, {\rm ISB} and {\rm IV} for $\widehat{\rho}_{p,m}^{V}(u)$, $u\in W$, $m=200,300,400$, $p=0.1,\ldots,1$. 
Moreover, in Figure \ref{f:EstErrInhomPoi} we provide estimation error plots for one of the realisations, for $p=0.2$ and $p=1$ with $m=200$, and in the Appendix, we provide plots of the estimated bias and variance for $p=0.1,0.3,0.5,0.7,0.9,1$ and $m=200$. 

Turning to the cross-validation approach to selecting $p$, based on $m=200$ and 100 realisations of the model, we obtain ${\rm IAB}=25.5$, ${\rm ISB}=885.2$ and ${\rm IV}=218.5$, with the majority of the selected $p$'s coinciding with the rule-of-thumb (see Table \ref{t:PselectInhompoiR2}).

Hence, the conclusions here are essentially the same as for the homogeneous Poisson process in Section \ref{s:HomPoi}, with the main difference arguably being that inhomogeneity enforces slightly harder thinning in the cross-validation. 

Comparing with kernel estimation under uniform, or global, edge correction, using Poisson likelihood cross-validation \citep{Load99,BRT15} to select the bandwidth, we obtain ${\rm IAB}=25.16$, ${\rm ISB}=853.24$ and ${\rm IV}=158.00$. By instead employing the bandwidth selection method of \citet{cronie2018bandwidth}, we obtain ${\rm IAB}=24.43$, ${\rm ISB}=797.02$ and ${\rm IV}=636.63$.

\begin{table*}[!htbp]
\caption{Estimates of {\rm IAB}, {\rm ISB} and {\rm IV} for $\widehat{\rho}_{p,m}^{V}(u)$, $u\in W=[0,1]^2$, $m=200,300,400$, $p=0.1,\ldots,1$, based on 500 realisations of an inhomogeneous Poisson process on $W=[0,1]^2$ with intensity $\rho(x,y)=|10+90\sin(16x)|$.
}
\begin{center}
\begin{tabular}{|l | rrr|rrr|rrr|rr}
\hline
&  & IAB &  &  & ISB & & & IV~~ & \\
\hline
\theadfont\diagbox[width=2.5em]{$p$}{m}& 200 & 300 & 400 & 200 & 300 & 400 & 200 & 300 & 400\\
\hline
.1 & 25.6 & 25.6 & 25.6 & 892.3 & 891.8 &891.7 & 154.2& 150.1 & 147.6\\
\hline
.2 & 25.5 & 25.5 & 25.5 & 882.8 & 883.2 &883.3 & 249.1& 247.3& 245.6\\
\hline
.3 & 25.6 & 25.5 & 25.5 & 881.5 & 881.5 &881.5 & 360.1& 356.3 & 356.2\\
\hline
.4 & 25.5 & 25.5 & 25.5 & 878.8 & 879.0 &879.0 & 479.9& 477.2 & 475.0\\
\hline
.5 & 25.5 & 25.5 & 25.5 & 872.6 & 872.5 &872.6 & 609.8& 609.6 & 609.8\\
\hline
.6 & 25.4 & 25.4 & 25.4 & 862.7 & 862.7 &862.7 & 762.6& 764.3 & 764.1\\
\hline
.7 & 25.2 & 25.2 & 25.2 & 849.9 & 850.0 &850.0 & 952.0& 948.3 & 949.0\\
\hline
.8 & 25.0 & 25.0 & 25.0 & 835.1 & 834.8 &834.8 & 1171.9& 1172.3 & 1172.1\\
\hline
.9 & 24.7 & 24.7 & 24.7 & 817.7 & 817.6 &817.6 & 1440.1& 1440.9 & 1440.0\\
\hline
1 & 24.4 & 24.4 & 24.4 & 799.3 & 799.3 &799.3 & 1783.8& 1783.8 & 1783.8\\
\hline
\end{tabular}

\label{t:InhomPoiR2}
\end{center}
\end{table*}

\begin{table*}[!htbp]
\caption{Cross-validation selections of $p$ in a geometric sequence  for $m=200$, based on 100 realisations of an inhomogeneous Poisson process in $W=[0,1]^2$ with intensity $\rho(x,y)=|10+90\sin(16x)|$.
}
\begin{center}
\begin{tabular}{l | cccccccc}
\hline
$\hspace{.8cm}p$&0.10 & 0.13 & 0.18 & 0.24 & 0.33 & 0.44 & 0.59 & 0.80\\
\hline
Frequency &69 & 15 & 11 & 2 & 0 & 1 & 0 &2
\\
\hline
\end{tabular}
\label{t:PselectInhompoiR2}
\end{center}
\end{table*}

\subsection{Log-Gaussian Cox process}
\begin{figure*}[!htbp]
\centering
\includegraphics[scale=0.2]{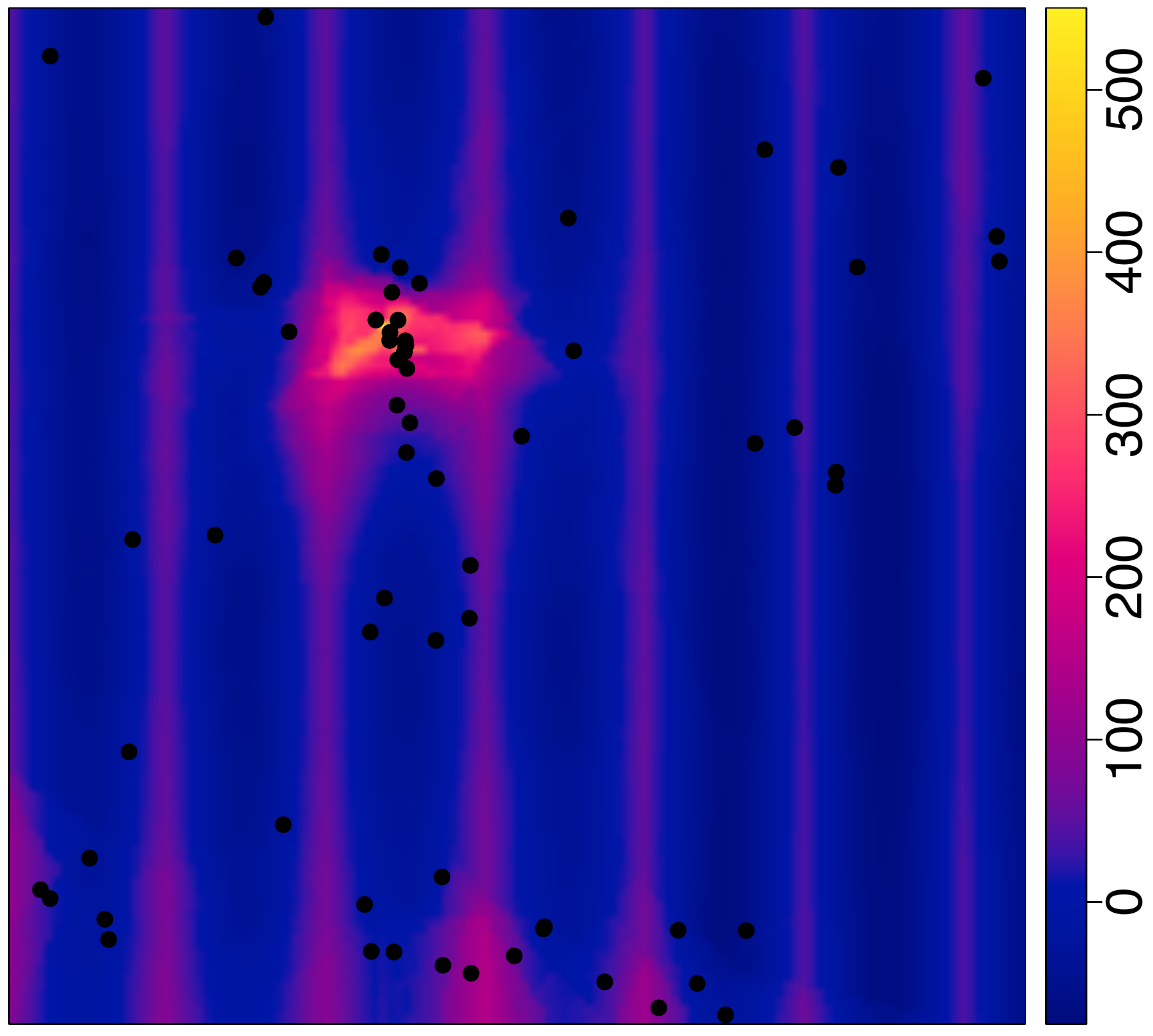}
\includegraphics[scale=0.2]{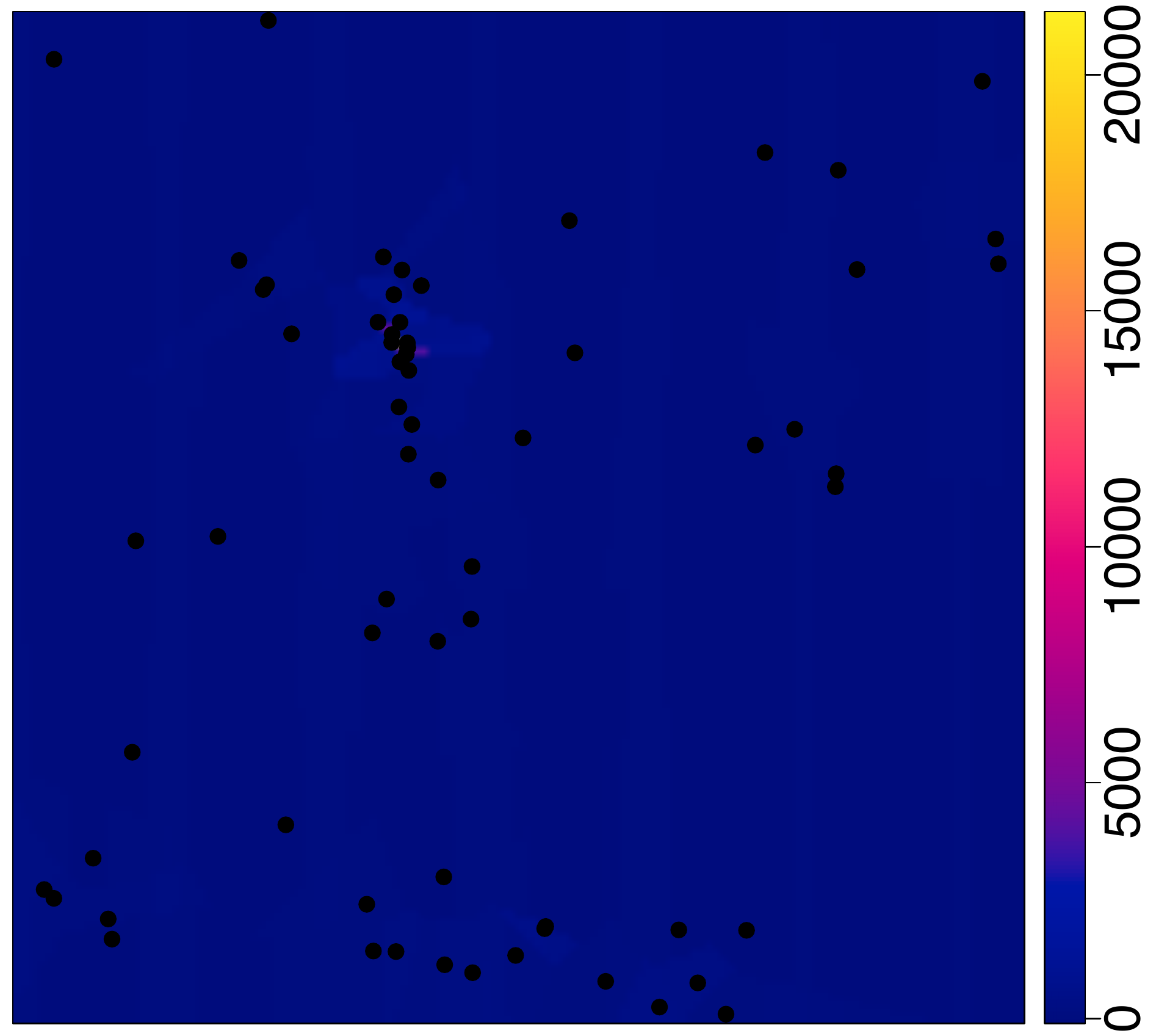}
\includegraphics[scale=0.2]{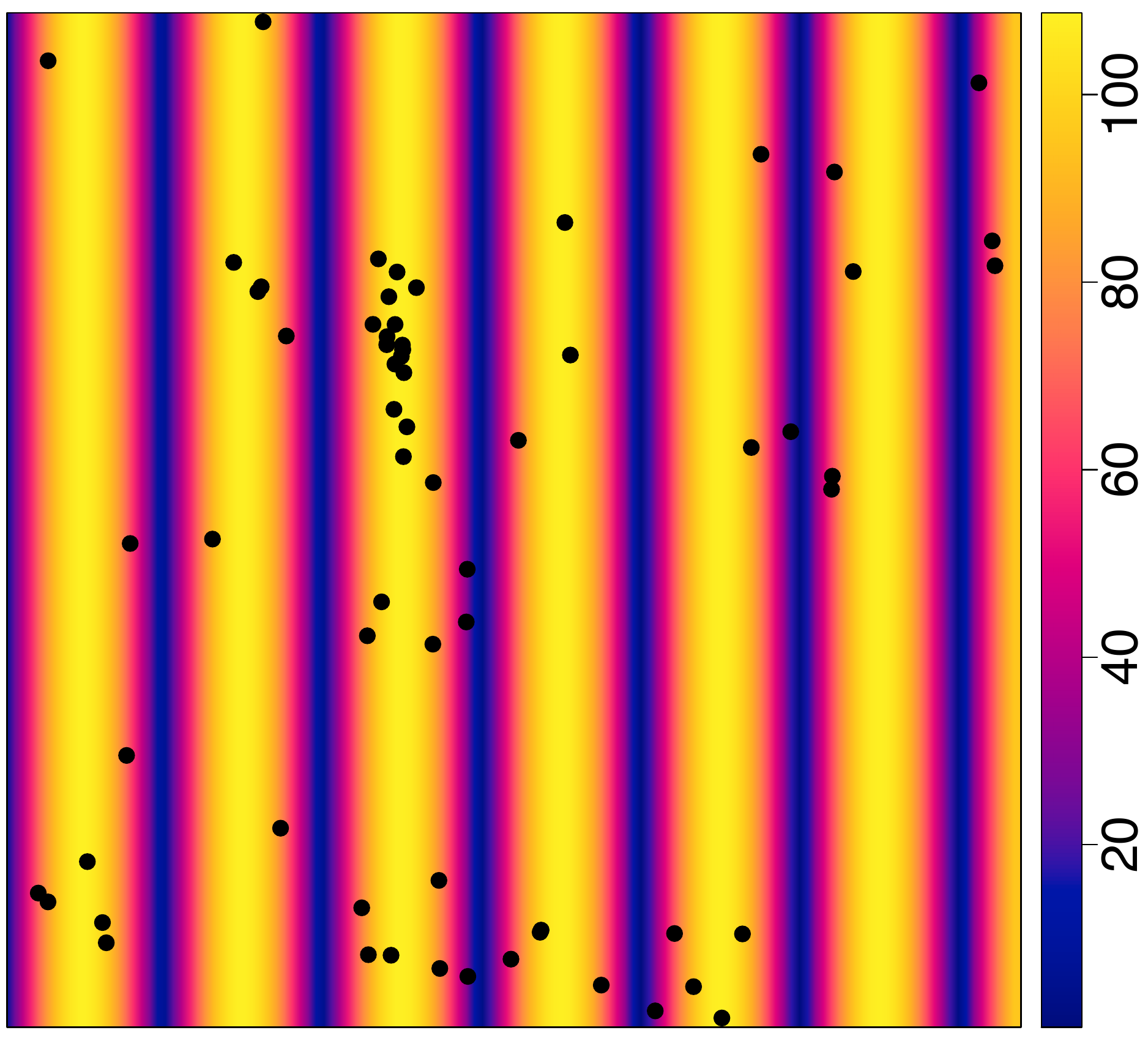}
\caption{
True intensity and estimation error plots for a realisation of a log-Gaussian Cox process in $W=[0,1]^2$ with mean function $(x,y)\mapsto\log(40|\sin(20x)|)$ and covariance function $((x_1,y_1),(x_2,y_2))\mapsto2\exp\{-\|(x_1,y_1)-(x_2,y_2)\|/0.1\}$ for the driving random field. \emph{Left}: $p=0.2$ and $m=200$. \emph{Middle}: $p=1$. \emph{Right}: True intensity.
The underlying point pattern has been superimposed in all plots.
}
\label{f:EstErrLGCP}
\end{figure*}

Turning to the scenario where the underlying point process exhibits clustering, we next consider $500$ realisations of a log-Gaussian Cox process $X\subseteq W=[0,1]^2$ where the driving Gaussian random field has the mean function $(x,y)\mapsto\log(40|\sin(20x)|)$ and covariance function $((x_1,y_1),(x_2,y_2))\mapsto2\exp\{-\|(x_1,y_1)-(x_2,y_2)\|/0.1\}$. Hereby, the intensity is given by $\rho(x,y)= 40|\sin(20x)|\e^{1}$. Table \ref{t:LGCPR2} provides estimates of {\rm IAB}, {\rm ISB} and {\rm IV} for $\widehat{\rho}_{p,m}^{V}(u)$, $u\in W$, $m=200,300,400$, $p=0.1,\ldots,1$. We see that the rule-of-thumb, i.e.\ $p\in[0.1,0.3]$, seems to be the preferable choice. 
In Figure \ref{f:EstErrLGCP} we provide estimation error plots for one of the realisations, for $p=0.2$ and $p=1$ with $m=200$, and  in the Appendix, we provide plots of the estimated bias and variance for $p=0.1,0.3,0.5,0.7,0.9,1$ and $m=200$. Here it becomes visually clear that the resample-smoothing is improving the estimation quite significantly. 

The cross-validation approach to selecting $p$, based on $m=200$ and 100 realisations of the model, yields ${\rm IAB}=28.4$, ${\rm ISB}=1118.2$ and ${\rm IV}=17207.5$, which may be comparable to the choice $p\approx0.5$. In Table \ref{t:PselectLGCPR2} we further provide the 100 selected values for $p$. 
The phenomenon that too little smoothing tends to be applied ($p$ is mainly chosen large) is not extremely surprising; as our cross-validation approach is based on a Poisson process likelihood function, it treats a realisation $\x$ of $X$ as a realisation of a Poisson process which has the corresponding realisation of the driving (random) intensity field as intensity function. In other words, it tries to perform state estimation, i.e.\ it tries to reconstruct each realisation of the driving intensity field through $\x$. This phenomenon, and that the Poisson process likelihood cross-validation approach is not performing well for clustered inhomogeneous point processes, has previously been observed in the context of kernel intensity estimation \citep{cronie2018bandwidth}. Hence, if one suspects that there is clustering in addition to inhomogeneity, or if the cross-validation generates large values for $p$, then it is wiser to stick with the proposed rule-of-thumb, $p\in[0.1,0.3]$. In fact, cross-validation-generated deviations from the rule-of-thumb  may be seen as a possible indication of clustering or inhibition. 

\begin{table*}[!htbp]
\caption{Estimates of {\rm IAB}, {\rm ISB} and {\rm IV} for $\widehat{\rho}_{p,m}^{V}(u)$, $u\in W=[0,1]^2$, $m=200,300,400$, $p=0.1,\ldots,1$, based on 500 realisations of a log-Gaussian Cox process in $W=[0,1]^2$ with mean function $(x,y)\mapsto\log(40|\sin(20x)|)$ and covariance function $((x_1,y_1),(x_2,y_2))\mapsto2\exp\{-\|(x_1,y_1)-(x_2,y_2)\|/0.1\}$ for the driving random field.
}
\begin{center}
\begin{tabular}{|l | rrr|rrr|rrr|rr}
\hline
&  & IAB &  &  & ISB & & & IV($\times10^2$) & \\
\hline
\theadfont\diagbox[width=2.5em]{$p$}{m}& 200 & 300 & 400 & 200 & 300 & 400 & 200 & 300 & 400\\
\hline
.1 & 29.5 & 29.5 & 29.5 & 1181.5 & 1181.9 &1180.9 & 48.8 & 48.8  & 48.7\\
\hline
.2 & 28.8 & 28.8 & 28.8 & 1127.3 & 1127.4 &1127.3 & 87.8& 87.2& 88.0\\
\hline
.3 & 28.2 & 28.2 & 28.2 & 1081.4 & 1081.7 &1081.6 & 123.8& 122.6 & 123.1\\
\hline
.4 & 27.6 & 27.6 & 27.6 & 1038.8 & 1039.2 &1039.4 & 153.2 & 153.0  & 152.6\\
\hline
.5 & 27.1 & 27.1 & 27.1 & 1000.1 & 999.6 &999.7 & 181.3& 182.2 & 182.0\\
\hline
.6 & 26.5  & 26.5  & 26.5  & 963.9 & 963.7 &963.5 & 212.4& 212.5 & 212.1\\
\hline
.7 & 26.0 & 26.0 & 26.0 & 930.5 & 930.4 &930.6 & 243.1& 243.0 & 243.2\\
\hline
.8 & 25.6 & 25.6 & 25.6 & 901.1 & 900.6 &900.7 & 278.8& 279.2 & 279.3\\
\hline
.9 & 25.2 & 25.2 & 25.2 & 874.4 & 874.3 &874.2 & 321.4& 321.5 & 320.9\\
\hline
1 & 24.7 & 24.7 & 24.7 & 852.3 & 852.3 &852.3 & 371.4& 371.4 & 371.4\\
\hline
\end{tabular}\label{t:LGCPR2}
\end{center}
\end{table*}


\begin{table*}[!h]
\caption{Cross-validation selections of $p$ in a geometric sequence  for $m=200$, based on 100 realisations of a log-Gaussian Cox process in $W=[0,1]^2$ with mean function $(x,y)\mapsto\log(40|\sin(20x)|)$ and covariance function $((x_1,y_1),(x_2,y_2))\mapsto2\exp\{-\|(x_1,y_1)-(x_2,y_2)\|/0.1\}$ for the driving random field.
}
\begin{center}
\begin{tabular}{l | cccccccc}
\hline
$\hspace{.8cm}p$&0.10 & 0.13 & 0.18 & 0.24 & 0.33 & 0.44 & 0.59 & 0.80\\
\hline
Frequency &4 & 4 & 0 & 1 & 8 & 14 & 34&35
\\
\hline
\end{tabular}
\label{t:PselectLGCPR2}
\end{center}
\end{table*}

Comparing with kernel estimation under uniform, or global, edge correction, using Poisson likelihood cross-validation \citep{Load99,BRT15} to select the bandwidth, we obtain ${\rm IAB}=27.75$, ${\rm ISB}=1031.03$ and ${\rm IV}=9952.85$. By instead employing the bandwidth selection method of \citet{cronie2018bandwidth}, we obtain ${\rm IAB}=28.97$, ${\rm ISB}=1117.94$ and ${\rm IV}=3856.79$.

\subsection{Thinned simple sequential inhibition point process}

To study inhomogeneity in combination with inhibition, we consider a simple sequential inhibition point process in $W=[0,1]^2$ with a total point count of $450$ and inhibition distance $0.3$, which we thin according the retention probability function $p(x,y)=\1\{x<1/3\}|x-0.02| + \1\{1/3\leq x<2/3\}|x-0.5| + \1\{x\geq2/3\}|x-0.95|$, $x,y\in W$. This results in an inhomogeneous point process with intensity $\rho(x,y)=450p(x,y)$, which yields an expected total point count of $53.6$. Table \ref{t:TSSIR2} provides estimates of {\rm IAB}, {\rm ISB} and {\rm IV} for $\widehat{\rho}_{p,m}^{V}(u)$, $u\in W$, $m=200,300,400$, $p=0.1,\ldots,1$. Just as for the previous models, we argue that $p$ should be chosen within the range of the rule-of-thumb. 

\begin{figure*}[!h]
\centering
\includegraphics[scale=0.2]{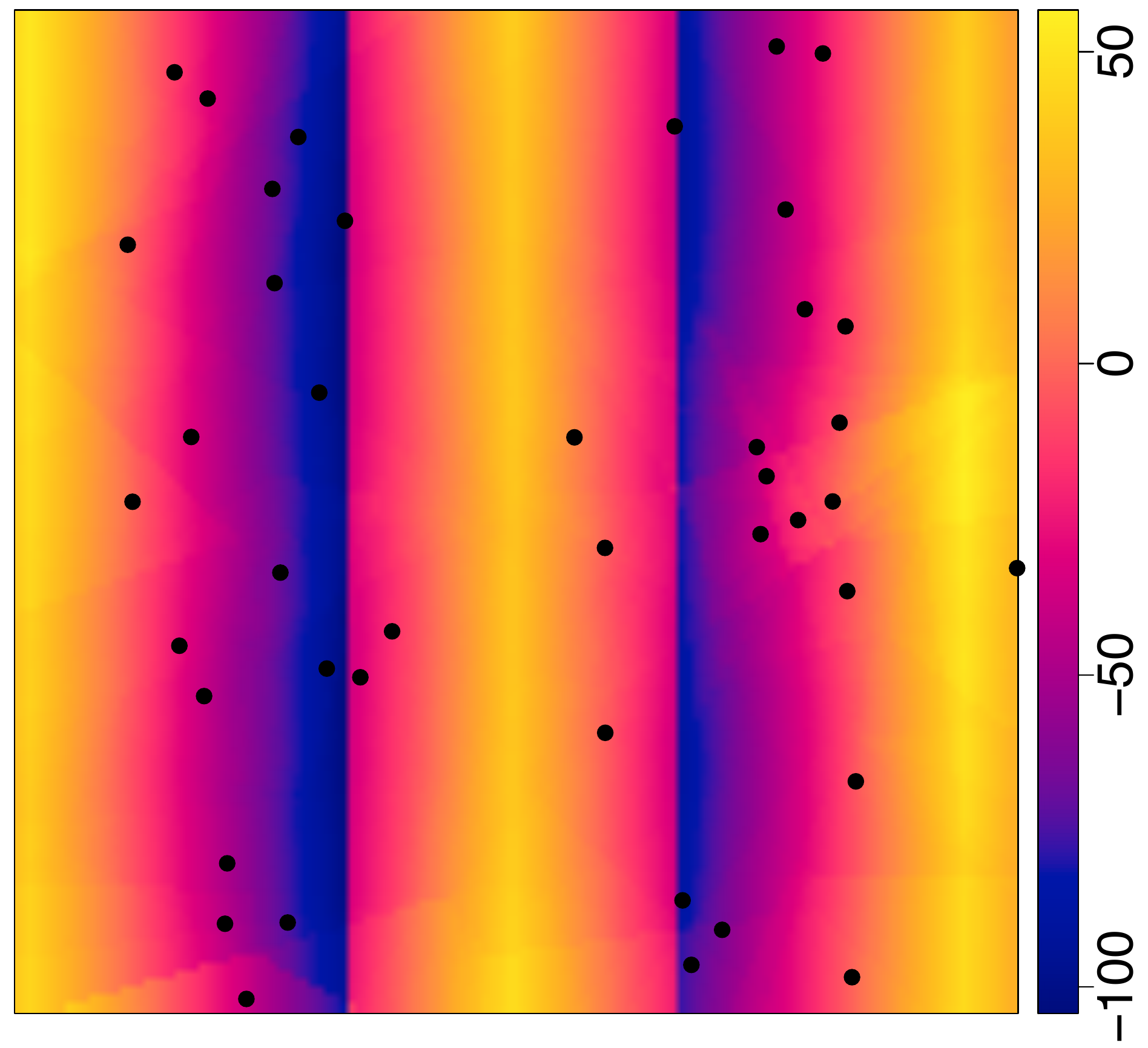}
\includegraphics[scale=0.2]{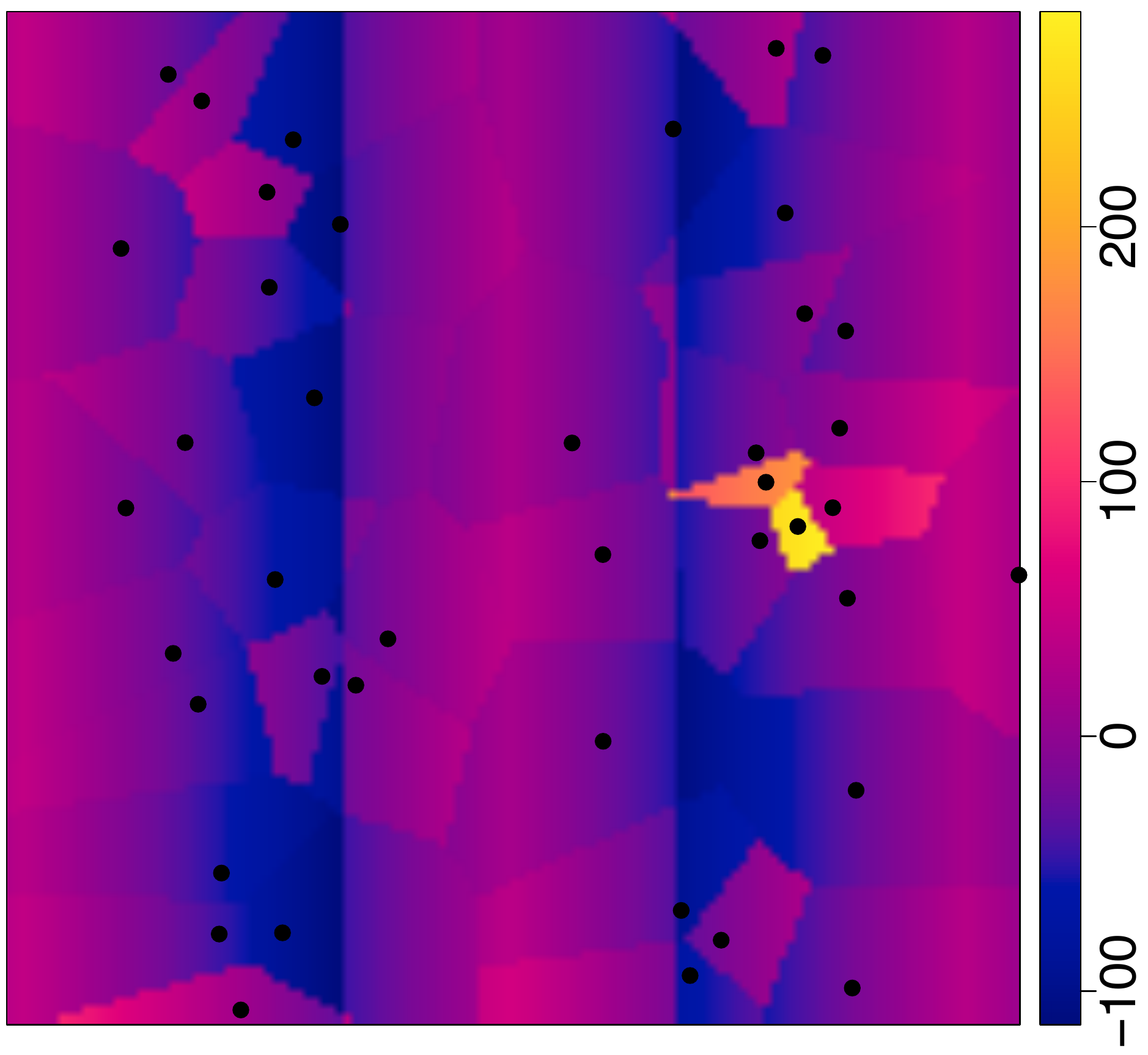}
\includegraphics[scale=0.2]{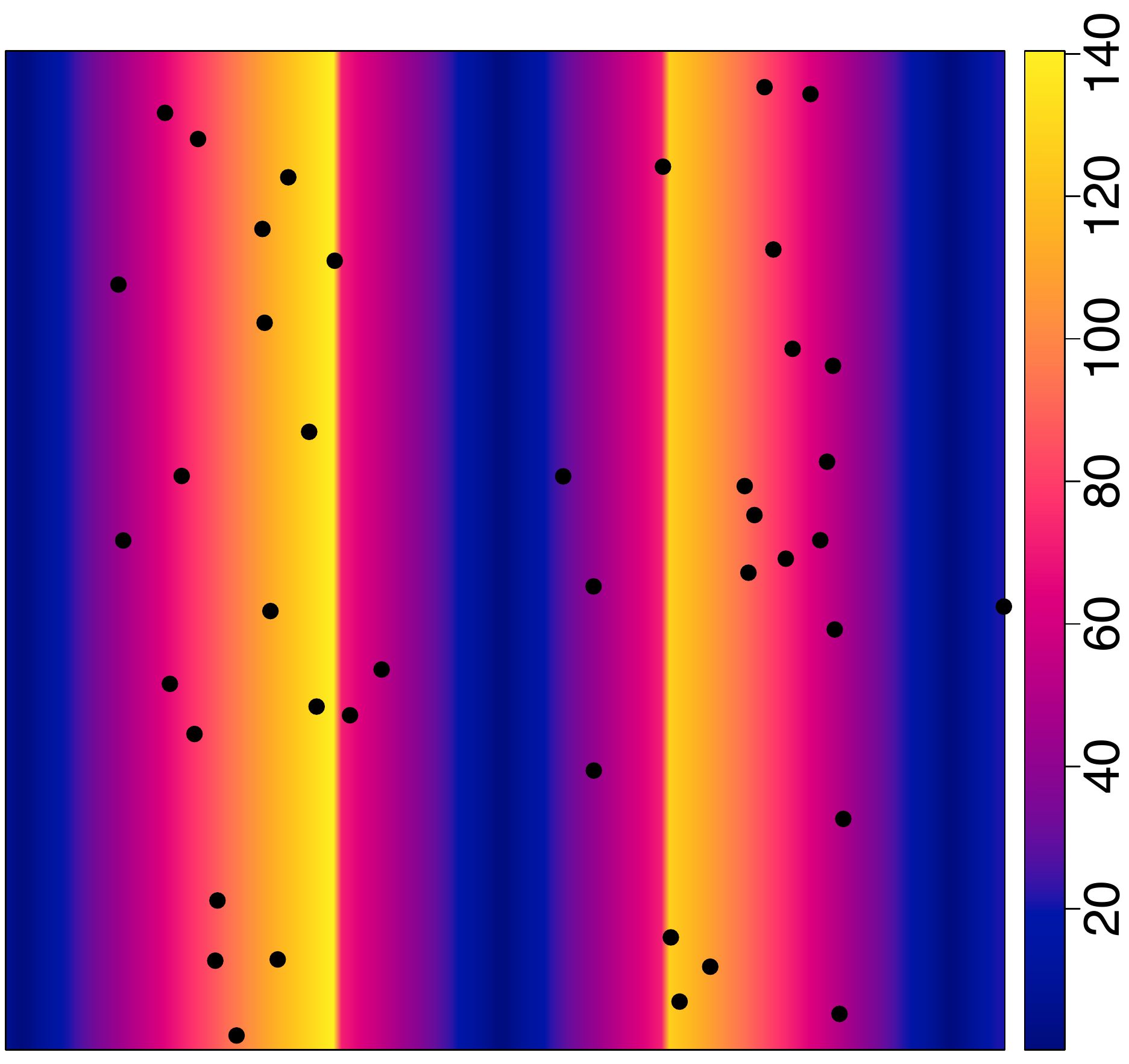}
\caption{
True intensity and estimation error plots for a realisation of an independently thinned simple sequential inhibition process in $W=[0,1]^2$ with intensity $\rho(x,y)=450p(x,y)$,  $p(x,y)=\1\{x<1/3\}|x-0.02| + \1\{1/3\leq x<2/3\}|x-0.5| + \1\{x\geq2/3\}|x-0.95|$, $x,y\in W$. \emph{Left}: $p=0.2$ and $m=200$. \emph{Middle}: $p=1$. \emph{Right}: True intensity.
The underlying point pattern has been superimposed in all plots.
}
\label{f:EstErrTSSI}
\end{figure*}

In Figure \ref{f:EstErrTSSI} we provide estimation error plots for one of the realisations, for $p=0.2$ and $p=1$ with $m=200$. Plots of the estimated bias and variance, for $p=0.1,0.3,0.5,0.7,0.9,1$ and $m=200$, can be found in the Appendix. Also here the improvements caused by the resample-smoothing are visually clear. 

The cross-validation approach to selecting $p$ based on $m=200$ and 100 realisations of the model yields ${\rm IAB}=25.1$, ${\rm ISB}=932.9$ and ${\rm IV}=595.2$, which is comparable to choosing $p\approx0.5$. Moreover, Table \ref{t:PselectTSSIR2} lists the selected values for $p$ and we see that they tend to be either very large or very small. It thus seems that approximately half of the time the cross-validation performs as it should do and approximately half of the time it chooses $p$ too large. 

Comparing with kernel estimation under uniform, or global, edge correction, using Poisson likelihood cross-validation \citep{Load99,BRT15} to select the bandwidth, we obtain ${\rm IAB}=20.5$, ${\rm ISB}=663.94$ and ${\rm IV}=485.48$. By instead employing the bandwidth selection method of \citet{cronie2018bandwidth}, we obtain ${\rm IAB}=23.97$, ${\rm ISB}=860.67$ and ${\rm IV}=308.47$.

\begin{table*}[!htbp]
\caption{Estimates of {\rm IAB}, {\rm ISB} and {\rm IV} for $\widehat{\rho}_{p,m}^{V}(u)$, $u\in W=[0,1]^2$, $m=200,300,400$, $p=0.1,\ldots,1$, based on 500 realisations of an independently thinned simple sequential inhibition process in $W=[0,1]^2$ with intensity $\rho(x,y)=450p(x,y)$,  $p(x,y)=\1\{x<1/3\}|x-0.02| + \1\{1/3\leq x<2/3\}|x-0.5| + \1\{x\geq2/3\}|x-0.95|$, $x,y\in W$.
}
\begin{center} 
\begin{tabular}{|l | rrr|rrr|rrr|rr}
\hline
&  & IAB &  &  & ISB & & & IV & \\
\hline
\theadfont\diagbox[width=2.5em]{$p$}{m}& 200 & 300 & 400 & 200 & 300 & 400 & 200 & 300 & 400\\
\hline
.1 & 32.4 & 32.4 & 32.4 & 1502.2 & 1502.7 &1502.2 & 109.4  & 105.9 & 103.4\\
\hline
.2 & 31.2 & 31.2 & 31.2 & 1385.7 & 1385.2 &1384.5 & 176.2 & 173.8  & 172.2\\
\hline
.3 & 29.2 & 29.2 & 29.2 & 1223.6 & 1223.0 &1222.8 & 253.4& 251.2  & 250.3\\
\hline
.4 & 27.0 & 27.0 & 27.0 & 1060.4 & 1060.7 &1060.3 & 348.8 & 345.3  & 345.3\\
\hline
.5 & 25.0 & 25.0 & 25.0 & 919.5 & 919.8 &920.6  & 457.3& 455.6& 454.1\\
\hline
.6 & 23.1  & 23.1  & 23.1  & 803.3 & 803.3 &803.0& 584.4& 582.7 & 581.9\\
\hline
.7 & 21.5 & 21.5 & 21.5 & 707.9  & 707.7 &707.8 & 734.2 &  733.9 & 732.8\\
\hline
.8 & 20.0  & 20.1 & 20.1 & 628.5 & 628.9 &629.1 & 916.3& 914.2 & 913.4\\
\hline
.9 & 18.9  & 18.9  & 18.9  & 567.2 & 567.5 &567.7 & 1120.5& 1118.5 & 1117.5\\
\hline
1 & 24.7 & 24.7 & 24.7 & 852.3 & 852.3 &852.3 & 1382.4& 1382.4 & 1382.4\\
\hline
\end{tabular}

\label{t:TSSIR2}
\end{center}
\end{table*}

\begin{table*}[!h]
\caption{Cross-validation selections of $p$ in a geometric sequence  for $m=200$, based on 100 realisations of an independently thinned simple sequential inhibition process in $W=[0,1]^2$ with intensity $\rho(x,y)=450p(x,y)$,  $p(x,y)=\1\{x<1/3\}|x-0.02| + \1\{1/3\leq x<2/3\}|x-0.5| + \1\{x\geq2/3\}|x-0.95|$, $x,y\in W$.
}
\begin{center}
\begin{tabular}{l | cccccccc}
\hline
$\hspace{.8cm}p$&0.10 & 0.13 & 0.18 & 0.24 & 0.33 & 0.44 & 0.59 & 0.80\\
\hline
Frequency &24 & 3 & 3 & 2 & 6 & 13 & 21&28
\\
\hline
\end{tabular}
\label{t:PselectTSSIR2}
\end{center}
\end{table*}

\section{Data analysis}
\label{SectionDataAnalysis}

We next apply our proposed intensity estimator \eqref{SmoothVor} to two real datasets, in two types of spaces. 
We first visit a linear network dataset of traffic accidents in an area of Houston, USA, and then a planar dataset of spatial locations of Finish pines.

\subsection{Houston motor vehicle traffic accidents}
\begin{figure*}[!h]
\centering
\includegraphics[scale=0.24]{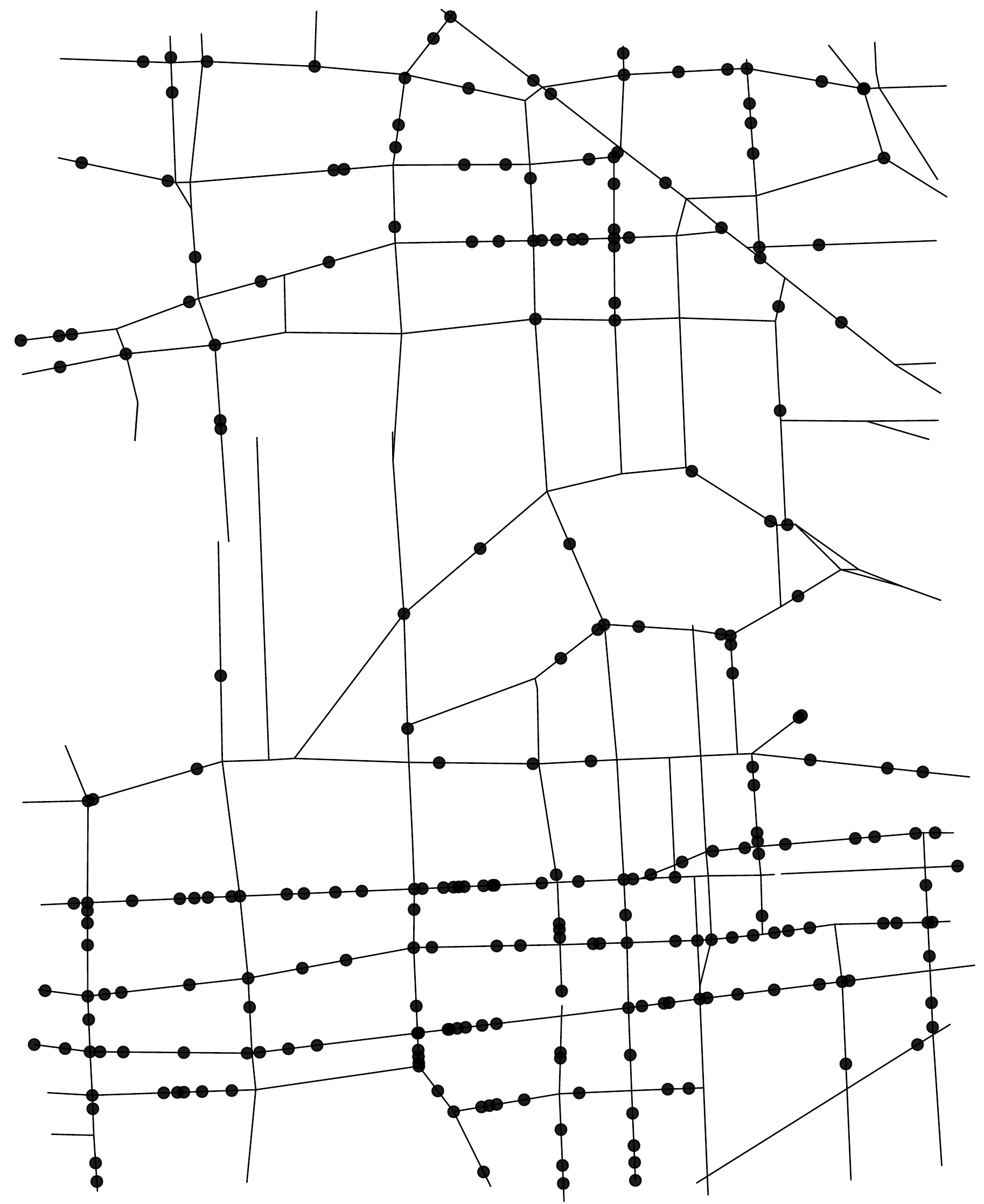}
\qquad
\includegraphics[scale=0.41]{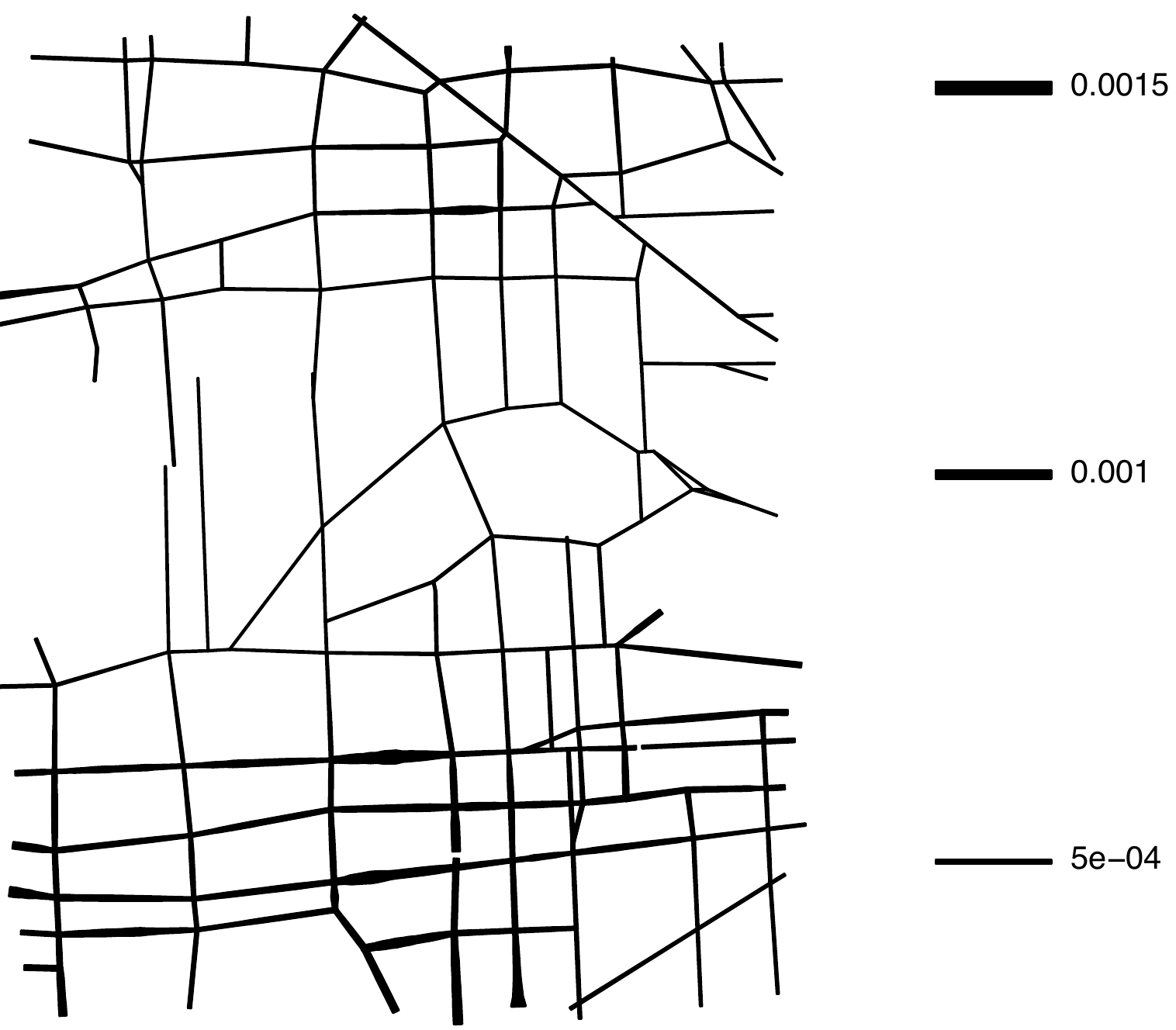}
\caption{
\emph{Left}: 
Motor vehicle traffic accidents in an area of Houston, US, during April, $1999$.
\emph{Right}: Resample-smoothed Voronoi intensity estimate for $m=200$ and $p=0.15$. 
}
\label{Houstonplot}
\end{figure*}

The dataset consists of motor vehicle traffic accident in a given area of Houston, USA, during the month of April $1999$. 
The linear network $L$ describing the road network in question (see Figure \ref{Houstonplot}) has a total length of $708,301.7$ feet, $187$ vertices, i.e.\ crossings/intersections, with a maximum vertex degree of $4$, and $253$ line segments, i.e.\ pieces of streets connecting the intersections. 

Figure \ref{Houstonplot} (left) shows the reference points of the $249$ accidents over the street network. The data have been collected by individual police departments in the Houston metropolitan area and compiled by the Texas Department of Public Safety. The compiled data have been obtained by the Houston-Galveston Area Council and then geocoded by N.\ Levine. Between $1999$ and $2001$, in the eight-county region considered, there were $252,241$ serious accidents, with an average of $84,080$ per year. From these accidents, $1,882$ were person related. See \citet{levine2006,levine2009} for details. 

In Figure \ref{Houstonplot} (right) we also provide the resample-smoothed Voronoi intensity estimate obtained for $m=200$ and $p=0.15$. 
The specific choice $p=0.15$ has been motivated by the rule-of-thumb $p\in[0.1,0.3]$ and Table \ref{Tablehous}, which shows the selected values for $p\in(0,1]$ obtained by carrying out cross-validation for the sequence $m=100,110,\ldots,200$. We see that most of the selected values for $p$ are given by $0.15$.

\begin{table*}[!htbp]
\caption{Cross-validation selected values for $p$, based on the sequence $m=100,110,\ldots,200$.}
\begin{center}
\begin{tabular}{c | ccccccccccc}
\hline
$m$& 100 &110& 120& 130& 140& 150& 160& 170& 180& 190& 200\\
\hline
$p$&0.15& 0.20& 0.20& 0.20& 0.20& 0.15& 0.15& 0.15& 0.15& 0.15& 0.15 \\
\hline
\end{tabular}
\label{Tablehous}
\end{center}
\end{table*}

Visually, there seems to be a good correspondence between the observed pattern and the obtained estimate. Note that for bigger values of $p$, in the right panel of Figure \ref{Houstonplot} we would have obtained more significant blobs in the parts corresponding to the dense parts in the left panel of Figure \ref{Houstonplot}.

\subsection{Finish pines}
The dataset, which consists of the locations of 126 pine saplings in a Finnish forest, within a rectangular window $W=[-5, 5]\times[-8, 2]$ (metres), can be found in the {\sf R} package \verb|spatstat| \citep{BRT15}. 
It has been recorded by S.\ Kellomaki, Faculty of Forestry, University of Joensuu, Finland, and further processed by A.\ Penttinen, Department of Statistics, University of Jyv\"askyl\"a, Finland.

\begin{figure*}[!htbp]
	\centering
	\includegraphics[scale=0.27]{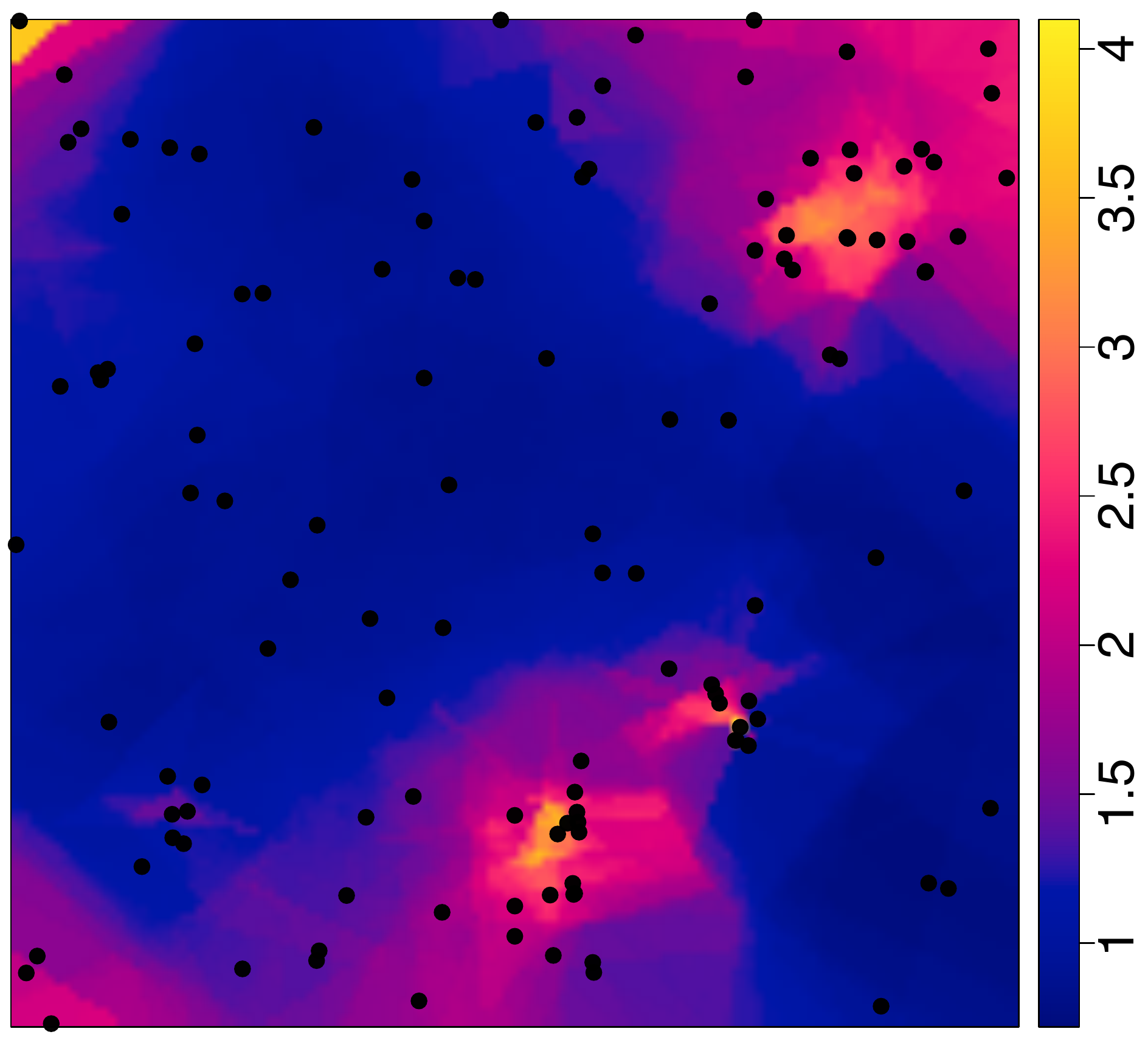}
    \qquad
	\includegraphics[scale=0.27]{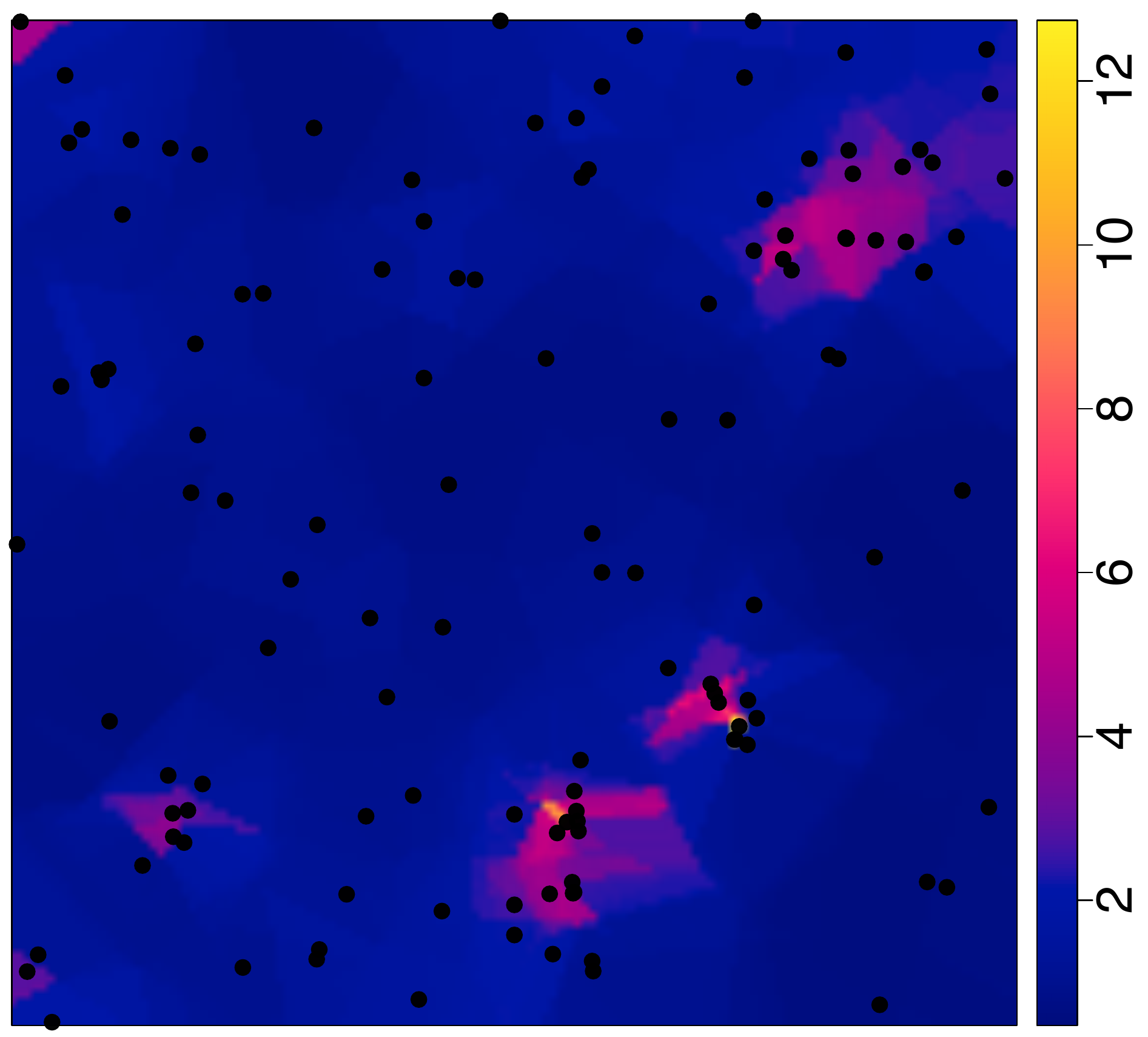}
	\caption{
		The estimate $\widehat{\rho}_{p,m}^{V}(u)$, $u\in W$, $m=200$, for $p=0.2$ (\emph{left}) and $p=0.5$ (\emph{right}), together with the locations of $126$ pine saplings in a Finnish forest, within a rectangular window $W=[-5, 5]\times[-8, 2]$ (metres).
	}
	\label{f:finpines}
\end{figure*}

In Figure \ref{f:finpines} we illustrate the estimate $\widehat{\rho}_{p,m}^{V}(u)$, $u\in W$, $m=200$, for $p=0.2$ and $p=0.5$, together with the locations of the saplings. We further provide the cross-validation results for the sequence $m=100,110,\ldots,200$ in Table \ref{Tablefinpines}; it suggests the choice $p=0.5$. We argue that $p=0.2$ is the preferable choice since it better respects the global features of the data.

\begin{table*}[!h]
	\caption{Cross-validation selected values for $p$, based on the sequence $m=100,110,\ldots,200$.}
	\begin{center}
		\begin{tabular}{c | ccccccccccc}
			\hline
			$m$& 100 &110& 120& 130& 140& 150& 160& 170& 180& 190& 200\\
			\hline
			$p$&0.65 &0.40 &0.50 &0.45& 0.40& 0.50& 0.45& 0.55& 0.50& 0.50& 0.50 \\
			\hline
		\end{tabular}
		\label{Tablefinpines}
	\end{center}
\end{table*}

\section{Discussion and future work}
\label{s:Discussion}

We have proposed a general approach for resampling, or additional smoothing, of Voronoi intensity estimators. It is based on averaging over intensity estimators generated by a set of thinned samples. 
We believe that its strength lies in that it filters out sporadic/local features in order to accentuate the structural information contained in the sample. In addition, viewing the reciprocal of a point's Voronoi cell size as a type of kernel (cf.\ \citet{vanIntensity}), centred at the point, each time we thin the pattern we change the support of that kernel. Having averaged over the thinned estimators, in essence we end up using an ``average" support for each such kernel. 

It may be noted that we alternatively may employ some retention probability function $p(u)$, $u\in W$, other than $p(u)\equiv p\in(0,1]$. 
It is, however, not clear what the benefits of such a change would be, other than possibly decreasing the computational time. Also, how to make a good choice for the function $p(\cdot)$ is not evident. 

\subsection{Future work and extensions}

Regarding future work, it would also be relevant and interesting to study the proposed setup when we replace the Voronoi tessellation by some other tessellation, generated by the point pattern in question. One such example is provided by Delaunay tessellations, as they give rise to more tractable distributional properties. 

Below follow some further possible extensions.

\subsubsection{Sequential resample-smoothing}

Since choosing the smoothing parameter $p\in(0,1]$ according to the cross-validation approach in Section \ref{sec:ChoosingParameters} can be quite computationally demanding, and thereby also time consuming, we propose an alternative and simpler version of the estimator in \eqref{SmoothVor}. 

\begin{definition}
Given some $\mathbf{p}_m=(p_1,\ldots,p_m)\in(0,1]^m$, $m\geq1$, the {\em sequentially resample-smoothed Voronoi intensity estimator} of the intensity $\rho(u)$, $u\in W\subseteq S$, $|W|>0$, of the underlying point process $X$ is defined as
\begin{align*}
\widetilde{\rho}_{\mathbf{p}_m}^{V}(u)
=
\widetilde{\rho}_{\mathbf{p}_m}^{V}(u;X,W)
=
\sum_{j=1}^m
\frac{\widehat{\rho}^{V}(u;X_{p_j},W)}{mp_j}
,\ u\in W,
\end{align*}
where $X_{p_j},\ldots,X_{p_m}$ is a sequence of independent thinnings of $X$, with the respective retention probabilities $p_j$, $j=1,\ldots,m$. 
In particular, 
$\widehat{\rho}_{p,m}^{V}(\cdot)=\widetilde{\rho}_{(p,\ldots,p)}^{V}(\cdot)$.
\end{definition}
The challenge here is clearly how to choose the sequence $\mathbf{p}_m$; we have seen that more weight clearly should be put on smaller retention probability values so an equally spaced grid over $(0,1]$ may not be the best choice. By proposing some stepwise sequencing of $(0,1]$, where we at each step $m\geq1$ obtain some $\mathbf{p}_m=(p_1,\ldots,p_m)\in(0,1]^m$, one could keep going until $\sup_{u\in W}|\widetilde{\rho}_{\mathbf{p}_m}^{V}(u)-\widetilde{\rho}_{\mathbf{p}_{m+1}}^{V}(u)|<\epsilon$ or $\sup_{u\in W}|\widetilde{\rho}_{\mathbf{p}_m}^{V}(u)-\widetilde{\rho}_{\mathbf{p}_{m+1}}^{V}(u)|/\widetilde{\rho}_{\mathbf{p}_m}^{V}(u)<\epsilon$ for some predefined $\epsilon>0$. 

\subsubsection{Edge correction in the linear network case}

Although we have neglected edge effects here, it still seems that the smoothing takes care of a significant part of the edge effects \citep{CSKWM13}. But, as noted in the data analysis, even after applying the smoothing there may be a need for edge correction \citep{BRT15,cronie2011edge}. 
In the case where $X$ is sampled on $L$, and is a subset of a process on a larger network, in which $L$ is a sub-network, edge effects come into play since the points closest to the boundary have their Voronoi cells cut off through the mapping/sampling of $L$ and the points. 
In Definition \ref{EdgeCorrection} below we propose an edge correction approach, which could be viewed as a version of Ripley's edge correction idea. 

\begin{definition}\label{EdgeCorrection} 
Given a point pattern $\x$ on a linear network $L$, 
for each boundary point $u\in\partial L$ of $L\subseteq S$, first find its closest neighbour $x_u=\argmin_{x\in \x}d(u,x)$ in terms of the shortest path distance $d(\cdot,\cdot)$. 
If $\beta_u=\min_{x\in \x\setminus\{x_u\}}d(x_u,x)/2 - d(u,x_u)>0$, 
extend $L$ by a new (set of) non-overlapping edge(s) connected to the node $u$, with total length $\beta_u$. Denote the resulting extended network by $\widetilde L(\x)$ and treat $\x$ as a linear network point pattern on/restricted to $\widetilde L(\x)$.
The edge corrected resample-smoothed Voronoi intensity estimate is given by $\widetilde{\rho}_{p,m}^{V}(u;\x,L)=\widehat{\rho}_{p,m}^{V}(u;\x,\widetilde L(\x))$ for $u\in W$. 
Note that $p=1$ results in an edge corrected version of $\widehat{\rho}^{V}(\cdot)$. 

\end{definition}

\newpage

\bibliographystyle{spbasic} 
\bibliography{NetworkRef}

\newpage
\appendix

\section{Appendix}

\subsection{Proofs}\label{sec:proofs}

\subsubsection{Proof of Lemma \ref{LemmaConvergence}}
Suppressing the dependence on $\x$ and $W$ in the notation, it follows that
\begin{align*}
&\left|\widehat{\rho}_{p,m+k}^{V}(u) - \widehat{\rho}_{p,m}^{V}(u)\right|
=\\
&=
\left|
\frac{1}{(m+k)p}
\left(
\sum_{i=1}^m
\widehat{\rho}_i^{V}(u)
+
\sum_{i=m+1}^{m+k}
\widehat{\rho}_i^{V}(u)
\right)
-
\frac{1}{mp}
\sum_{i=1}^m
\widehat{\rho}_i^{V}(u)
\right|
\\
&\leq
\left|\frac{mp}{(m+k)p} - 1\right|
\frac{1}{mp}
\sum_{i=1}^m
\widehat{\rho}_i^{V}(u)
+
\frac{1}{(m+k)p}
\sum_{i=m+1}^{m+k}
\widehat{\rho}_i^{V}(u)
\\
&\leq
\frac{\sup_{u\in W}\widehat{\rho}^{V}(u)}{p}
\left(
\left|\frac{m}{(m+k)} - 1\right|
+
\frac{k}
{(m+k)}
\right)
.
\end{align*}
Since $\sup_{u\in W}\widehat{\rho}^{V}(u)<\infty$, the right hand side tends to 0 as $m\to\infty$. Hence, $\{\widehat{\rho}_{p,m}^{V}(u)\}_{m\geq1}$ is a Cauchy sequence and by the completeness of the Euclidean space $\R$ it attains a limit as $m\to\infty$.


\subsubsection{Proof of Theorem \ref{thm:Unbiasedness}}
A $p$-thinning of $X$ is again stationary with intensity $p\rho$. By \citet[Expression (11.3.2)]{DVJ2},
for $v\in S$, 
where $G_X(\cdot)$ is the generating functional of $X$. 
Using \citet[Corollary 8.7]{Last2010} we immediately obtain that 
\begin{align*}
&\E[\widehat{\rho}_{p,m}^{V}(u)]
=\frac{\E[\widehat{\rho}^{V}(u;X_p,S)]}{p}
\\
&=\E[1/|\{\text{cell of $X_p$ containing }u\}|]/p=\frac{p\rho}{p}=\rho.
\end{align*}

\subsubsection{Proof of Theorem \ref{thm:unbiasedbound}}

We denote by $x_{u}(X) \in X$ the centre of the Voronoi cell $C_u(X)$, the cell containing $u\in\R^d$.
Let $\varepsilon>0$, $\mu=\mu_u$ and $ \rho_{-}=\min_{v\in B(u,\varepsilon)}\rho(v)$, such that $\rho(v)/2\leq\rho(v) - \mu\varepsilon \leq \rho _{-}\leq \rho(v)$ on $B(u,\varepsilon )$. Let $X_{-}$ be obtained by independently removing/adding points at rate $\rho_{-} - \rho(v)$, $v\in\R^d$. Note that $X_{-}$ is a homogeneous Poisson process with intensity $\rho _{-}$ and $X_{-}\subseteq X$ on $B(u,\varepsilon )$ a.s.. 

We call   Voronoi neighbours in some configuration $\mathbf{x} $ the centres of cells of $\mathbf{x} $ which are neighbours of $C_{u}(\mathbf{x})$. Denote by $R (\mathbf{x})$ the maximal Euclidean distance between $x_u(\mathbf{x} )$ and its  Voronoi neighbours. Remark that if $R (\mathbf{x})\leq \varepsilon$, then $C_{u}(\mathbf{x})\subseteq B(u,\varepsilon)$.
One can find a finite number of balls such that if any such ball contains a point of $\mathbf{x} $, then $R (\mathbf{x})\leq 1$. Hence, using the void probabilities of $X$, we have at the scale $\varepsilon$ for $X$ that
\begin{align*}
\P(R(X)\geq \varepsilon)\leq C_d \e^{-c_{d}\rho _-\varepsilon^{d}}
\end{align*}
for some $C_d,c_d>0$.

Now, let $\Omega $ be the event that $ X$ and $X_{-}$ coincide on $B(u,\varepsilon)$ and $R(X)\leq \varepsilon $. Conditionally on $\Omega$, $C_{u}(X) = C_{u}(X_{-})\subseteq B(u,\varepsilon)$. We obtain 
\begin{align*}
&\1_{\{\Omega^{c}\}} 
\leq 
\1_{\{R(X) > \varepsilon\}}
+
\sum_{x\in X_{-}\cap B(u,\varepsilon)}\1_{\{x\text{\rm{ eliminated at thinning}}\}},
\\
&\P(\Omega ^{c})\leq \P(R(X)>\varepsilon)
+
\int_{B(u,\varepsilon)} \mu \varepsilon \de x\leq C_{d}\e^{-c_{d}\rho _-\varepsilon^{d}}
+
c\varepsilon^{d}\mu\varepsilon.
\end{align*}
Let further $\kappa' = (1-\kappa^{-1})^{-1}\leq d+1$. By H{\"o}lder's inequality and Theorem 1 we have that
\begin{align*}
&\left|
\E\left[
\widehat\rho^V(u)
\right]
-
\rho (u)
\right|
\leq
\\
\leq& \left|
\E\left[
\1_{\{\Omega\}}\frac{1}{ | C_{u}(X) | }
\right]
-\rho (u)\right|
+
\mathbb{E}\left[
\mathbf{1}_{\{\Omega^c\}} \frac{1}{ | C_{u}(X) | }
\right]\\
\leq &
\left|
\E\left[\1_{\{\Omega\}}
\frac{1}{ | C_{u}(X _{-}) | }-\rho (u)
\right] 
\right|
+
(\E|C_{u}(X)|^{-\kappa})^{1/\kappa}
\P(\Omega^c)^{1/\kappa'}
\\ 
\leq & 
\underbrace{\E\left[
\frac{1}{| C_{u}(X_{-}) |}
-\rho_{-}  
\right]}_{=0}
+
\1_{\{\Omega^c\}}\frac{1}{ | C_{u}(X_{-}) | }
+
| \rho (u)-\rho _{-} |   
\\
&+
m (c_d\mu\varepsilon^d\varepsilon + C_d\e^{-c_{d} \rho_- \varepsilon^d})^{1/\kappa'}
\\
\leq &
\mu\varepsilon + 2m (c_d\mu\varepsilon^d\varepsilon 
+ C_d\e^{-c_{d}\rho _- \varepsilon^d})^{1/\kappa'}
.
\end{align*} 
Setting $\varepsilon = \rho_-^{-1/d}\log(\rho_{-})^{2/d}$ and recalling that $\rho(u)/2\leq \rho_-$, using that $\kappa' \leq d+1$, proves the result for the original Voronoi intensity estimator. 

As a $p$-thinning $X_p$, $p\in(0,1]$, of $X$ is a Poisson process with intensity $p\rho(\cdot)$, we finally note that
\begin{align*}
&p|\E[\widehat{\rho}_{p,m}^{V}(u)] - \rho(u)|
=
\left|
\E\left[\widehat{\rho}^{V}(u;X_p,\R^d)\right]
-
p\rho(u)
\right|
\leq
\\
&\leq
\mu p^{-1}\varepsilon + 2m(c_d\mu p^{-1}\varepsilon^d\varepsilon 
+ C_d\e^{-c_{d}p\rho(u) \varepsilon^d})^{1/\kappa'},
\end{align*}
since $\E| C_u(X_p)|^{-\kappa}\leq \E| C_u(X)|^{-\kappa}$.


\subsubsection{Proof of Theorem \ref{thm:Variance}}
Note first that 
\begin{eqnarray}
\label{e:VarianceInequality}
&&\Var(\widehat{\rho}_{p,m}^{V}(u)) 
=\\ 
&=&\frac{1}{(mp)^2}\sum_{i=1}^m\Var(\widehat{\rho}_1^{V}(u)) + \frac{1}{(mp)^2}\sum_{i\neq j}\Cov(\widehat{\rho}_i^{V}(u), \widehat{\rho}_j^{V}(u))
\nonumber
\\
&=& 
\frac{1}{m} \Var(\widehat{\rho}_1^{V}(u)/p) + \frac{m-1}{m}
\Cov(\widehat{\rho}_1^{V}(u)/p, \widehat{\rho}_2^{V}(u)/p)
\nonumber
\\
&=& 
\Var(\widehat{\rho}_{p,1}^{V}(u))
\frac{
1 + (m-1){\rm Corr}(\widehat{\rho}_1^{V}(u), \widehat{\rho}_2^{V}(u))
}{m}
,
\nonumber
\end{eqnarray}
where $\Cov$ and ${\rm Corr}$ denote covariance and correlation, respectively. 
Since the variance is non-negative, by \eqref{e:VarianceInequality} we must have that ${\rm Corr}(\widehat{\rho}_1^{V}(u), \widehat{\rho}_2^{V}(u)) \geq -1/(m-1)$ for every single $m\geq1$. Hence, the correlation must be non-negative, whereby
$
\Var(\widehat{\rho}_{p,1}^{V}(u))/m
\leq 
\Var(\widehat{\rho}_{p,m}^{V}(u)) 
\leq 
\Var(\widehat{\rho}_{p,1}^{V}(u))
$; this is obtained by setting ${\rm Corr}(\widehat{\rho}_1^{V}(u), \widehat{\rho}_2^{V}(u))=0,1$ in expression \eqref{e:VarianceInequality}. 
Also, letting $m\to\infty$ in \eqref{e:VarianceInequality}, the limit of \eqref{e:VarianceInequality} is given by $\Cov(\widehat{\rho}_1^{V}(u), \widehat{\rho}_2^{V}(u))/p^2$ since $\Var(\widehat{\rho}_1^{V}(u))<\infty$. 
For a fixed $m$, when $p=1$ it follows that ${\rm Corr}(\widehat{\rho}_1^{V}(u), \widehat{\rho}_2^{V}(u))=1$, i.e.\ the correlation is maximised, since $\widehat{\rho}_1^{V}(u) = \widehat{\rho}_2^{V}(u)$ a.s.; 
$\Var(\widehat{\rho}_{p,m}^{V}(u))\leq\Var(\widehat{\rho}_{1,m}^{V}(u))=
\Var(\widehat{\rho}^{V}(u))$.

It further follows that $\E[\widehat{\rho}_1^{V}(u) \widehat{\rho}_2^{V}(u)]/p^2$ equals to 
\begin{align*}
&
\E\Bigg[
\sum_{x\in X\cap W}
\frac{\1\{x\in X_p^1\}}{p}
\frac{\1\{u\in\V_{x}(X_p^1,W)\}}{|\V_{x}(X_p^1,W)|}
\times
\\
&\times
\sum_{x\in X\cap W}
\frac{\1\{x\in X_p^2\}}{p}
\frac{\1\{u\in\V_{x}(X_p^2,W)\}}{|\V_{x}(X_p^2,W)|}
\Bigg]
=
\\
=&
\E\Bigg[
\sum_{x_1,x_2\in X\cap W}
\frac{\1\{x_1\in X_p^1\}\1\{x_2\in X_p^2\}}{p^2}
\times
\\
&\times\frac{\1\{u\in\V_{x_1}(X_p^1,W)\cap\V_{x_2}(X_p^2,W)\}}
{|\V_{x_1}(X_p^1,W)| |\V_{x_2}(X_p^2,W)|}
\Bigg]
,
\end{align*}
which is larger than or equal to $|W|^{-2}$. 
Note that $\lim_{p\to0}\1\{u\in\V_{x_1}(X_p^1,W)\cap\V_{x_2}(X_p^2,W)\}\stackrel{a.s.}{=}1$ and $\lim_{p\to0}\1\{x_1\in X_p^1\}
\1\{x_2\in X_p^2\}/p^2\stackrel{a.s.}{=}1$, since the latter is the product of two independent Bernoulli random variables with parameter $p$. Finally we note that as $p\to0$, $(|\V_{x_1}(X_p^1,W)| |\V_{x_2}(X_p^2,W)|)^{-1}\to|W|^{-2}$ a.s., whereby $\E[\widehat{\rho}_1^{V}(u) \widehat{\rho}_2^{V}(u)]/p^2\to|W|^{-2}$, by dominated convergence. 
Applying similar arguments, it is not hard to see that also $|W|^{-1}\leq\E[\widehat{\rho}_1^{V}(u)]/p\to|W|^{-1}$ and $|W|^{-2}\leq\Var(\widehat{\rho}_1^{V}(u))/p^2\to|W|^{-2}$ as $p\to0$. This now yields $\lim_{p\to0}\Var(\widehat{\rho}_{p,m}^{V}(u))=1/(m|W|^2)$, which is 0 when $W$ is unbounded. 

\subsubsection{Proof of Lemma \ref{LemmaHomPoisson}}
Recall that $X_p$ is a homogeneous Poisson process with intensity $p\rho$.
For a typical point of $X_p$, let $\Delta_-$ and $\Delta_+$ be the distances to the point's nearest neighbours to the left and to the right, respectively; they are independent and exponentially distributed with mean $p\rho$. Since $\Delta_-/2$ and $\Delta_+/2$ are independent and exponentially distributed with mean $2p\rho$, 
the typical cell size, $\Delta_-/2 + \Delta_+/2$, follows an Erlang/Gamma distribution with shape parameter $2$ and rate $2p\rho$, whereby the density of $P_{|\V_o(X_{p})|}(\cdot)$ is given by $f_{|\V_o(X_{p})|}(t)=(2p\rho)^2 t \e^{-2p\rho t}$.
%
Through expression \eqref{e:DistributionCell}
we now obtain
\begin{align*}
\E[\widehat{\rho}_{p,1}^{V}(u)^2]
&=
\frac{\rho}{p}\E[1/|\V_o(X_p)|]
=
\frac{\rho}{p}
\int_0^{\infty}
\frac{1}{t}
(2p\rho)^2 t \e^{-2p\rho t}
\de t
\\
&=
4p\rho^3
\int_0^{\infty}
\e^{-2p\rho t}
\de t
=
\frac{4p\rho^3}{2p\rho}
=
2\rho^2
,
\end{align*}
i.e., $\Var(\widehat{\rho}_{p,m}^{V}(u)) \leq \Var(\widehat{\rho}_{p,1}^{V}(u))=2\rho^2 - \rho^2=\rho^2$ by Theorem \ref{thm:Variance}.

\subsection{Estimated bias and variance plots}\label{sec:AppendixB}
This section provides plots of the estimated bias and variance for $\widehat{\rho}_{p,m}^{V}(u)$, for each of the models described in Section 4 in the paper, when $m=200$ and $p=0.1,0.3,0.5,0.7,0.9,1$. We additionally provide kernel intensity estimates, with bandwidths selected by means of Poisson likelihood cross-validation \citep{BRT15,Load99} and the method of \citet{cronie2018bandwidth}. The estimates are generated by $500$ realisations of each of the models. 

\clearpage
\begin{figure*}[!h]
\centering
  \includegraphics[width=0.35\textwidth]{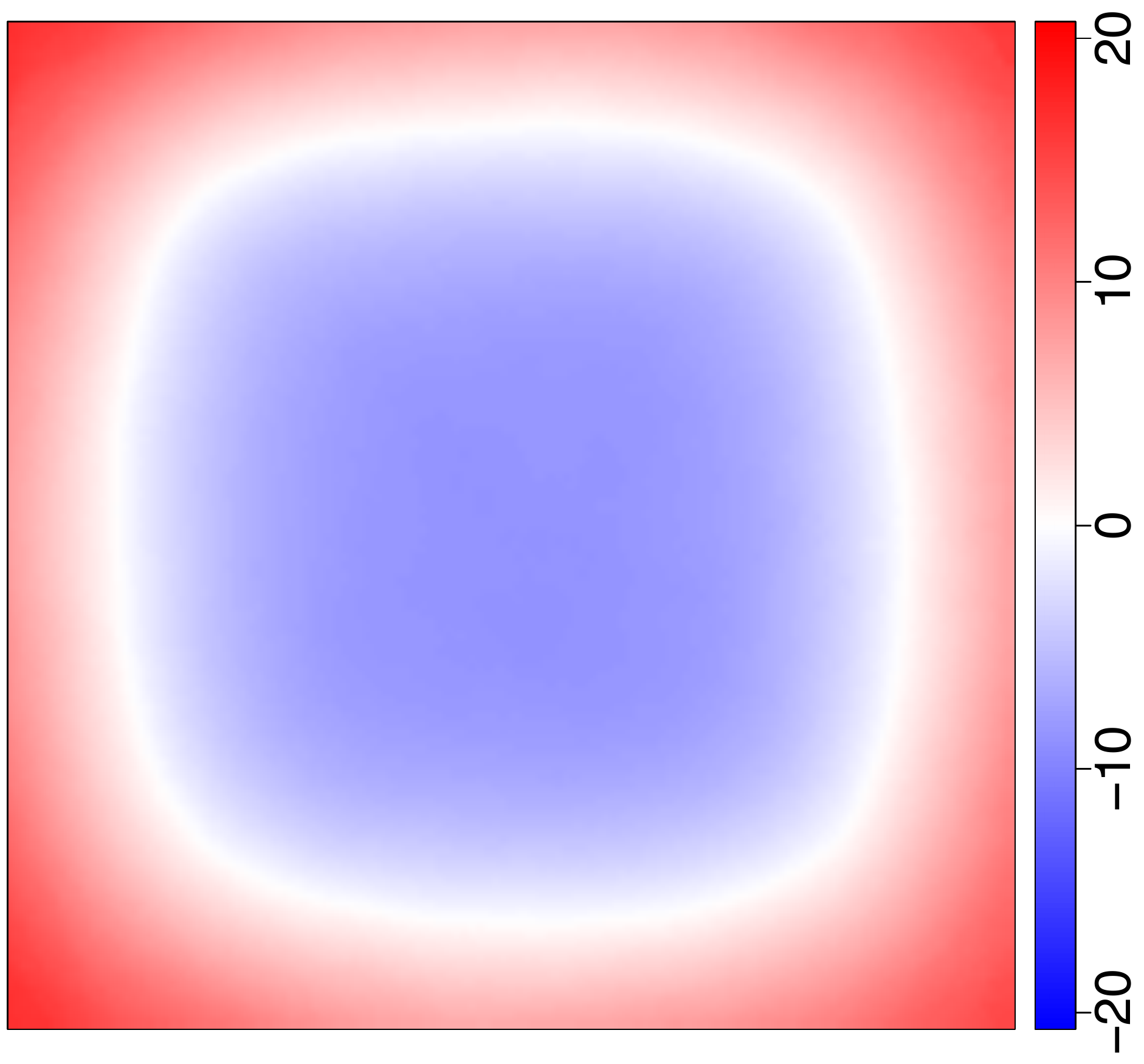}
  \qquad
 \includegraphics[width=0.35\textwidth]{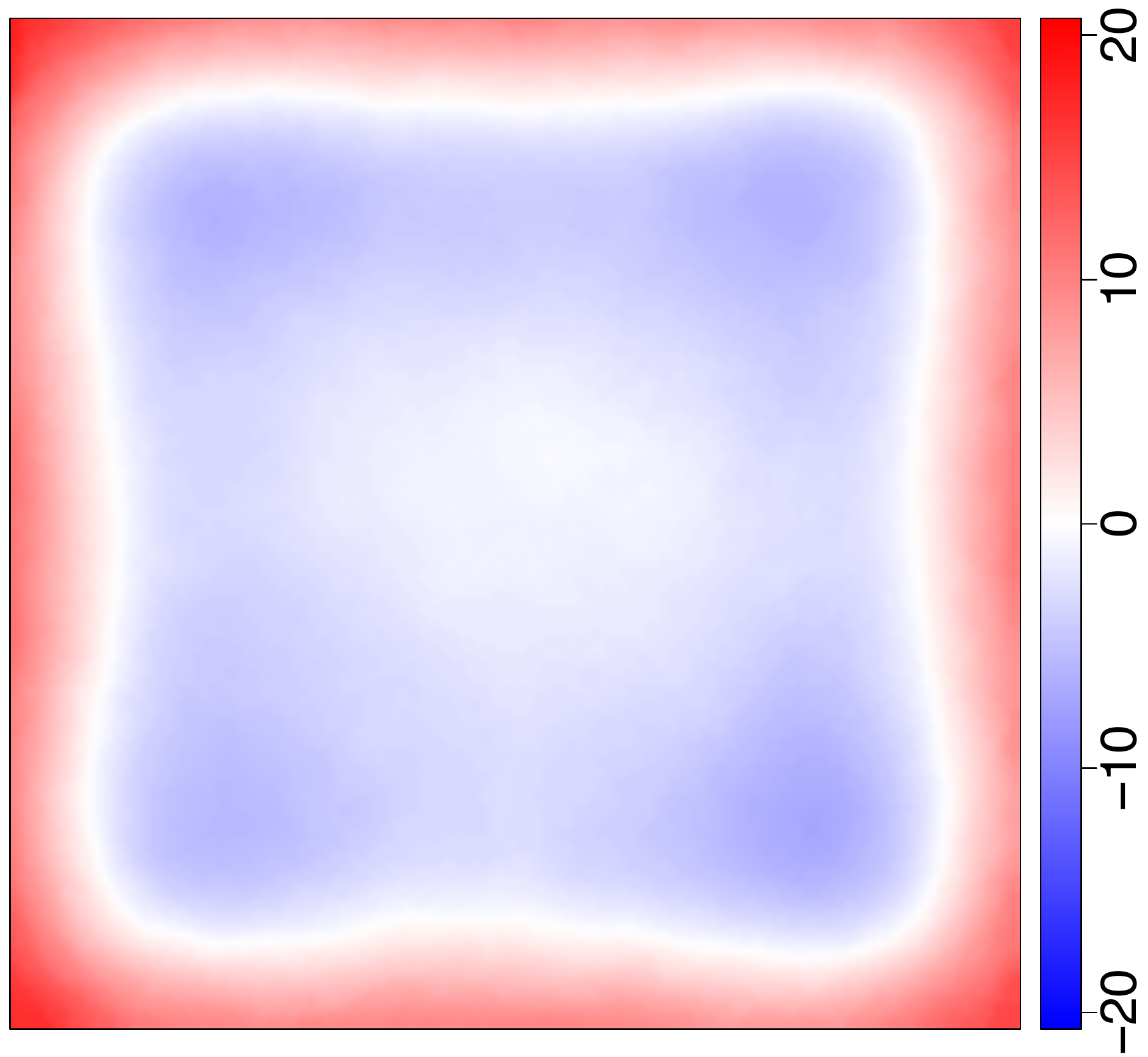}
 \qquad
  \includegraphics[width=0.35\textwidth]{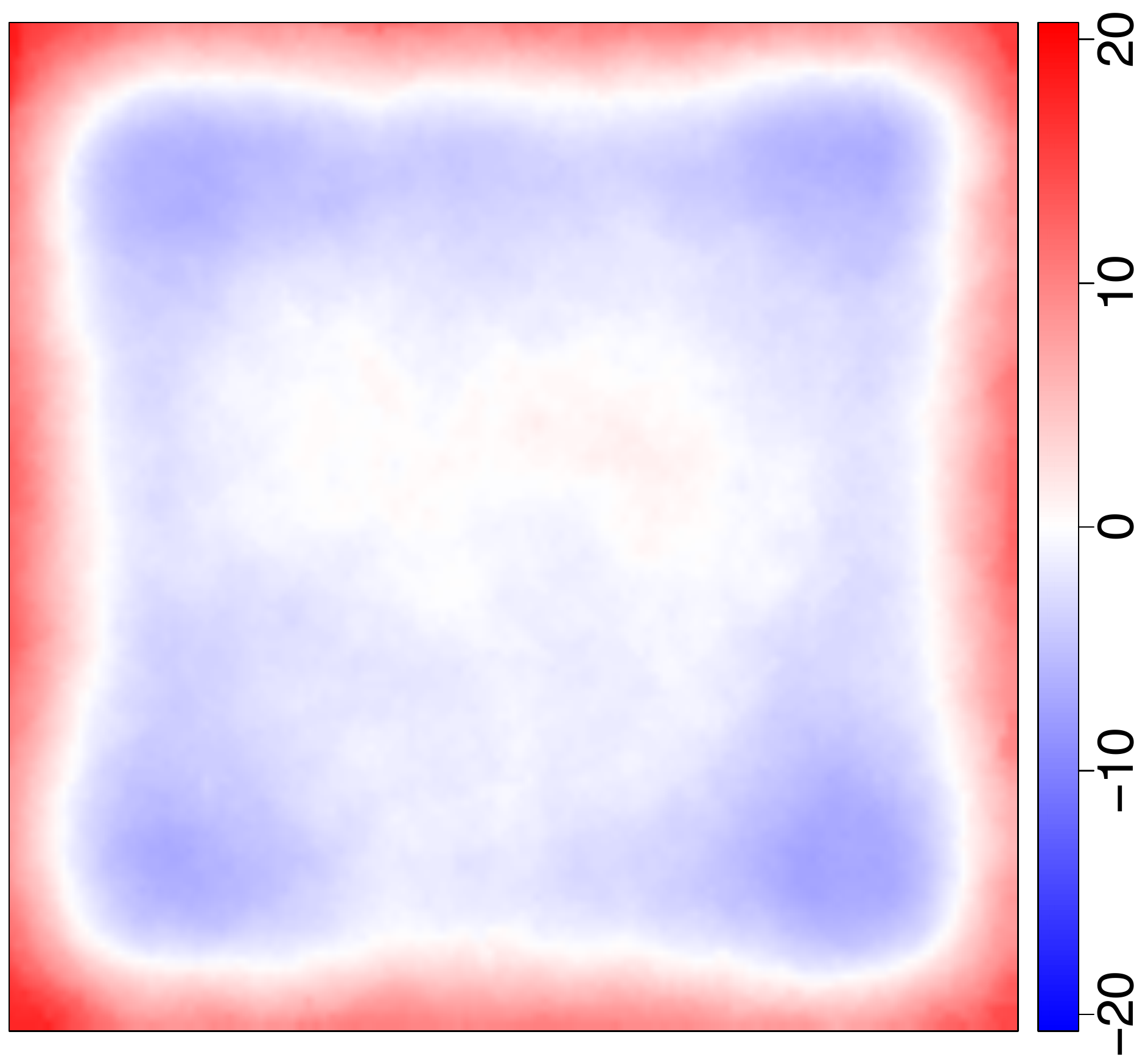}
  \qquad
 \includegraphics[width=0.35\textwidth]{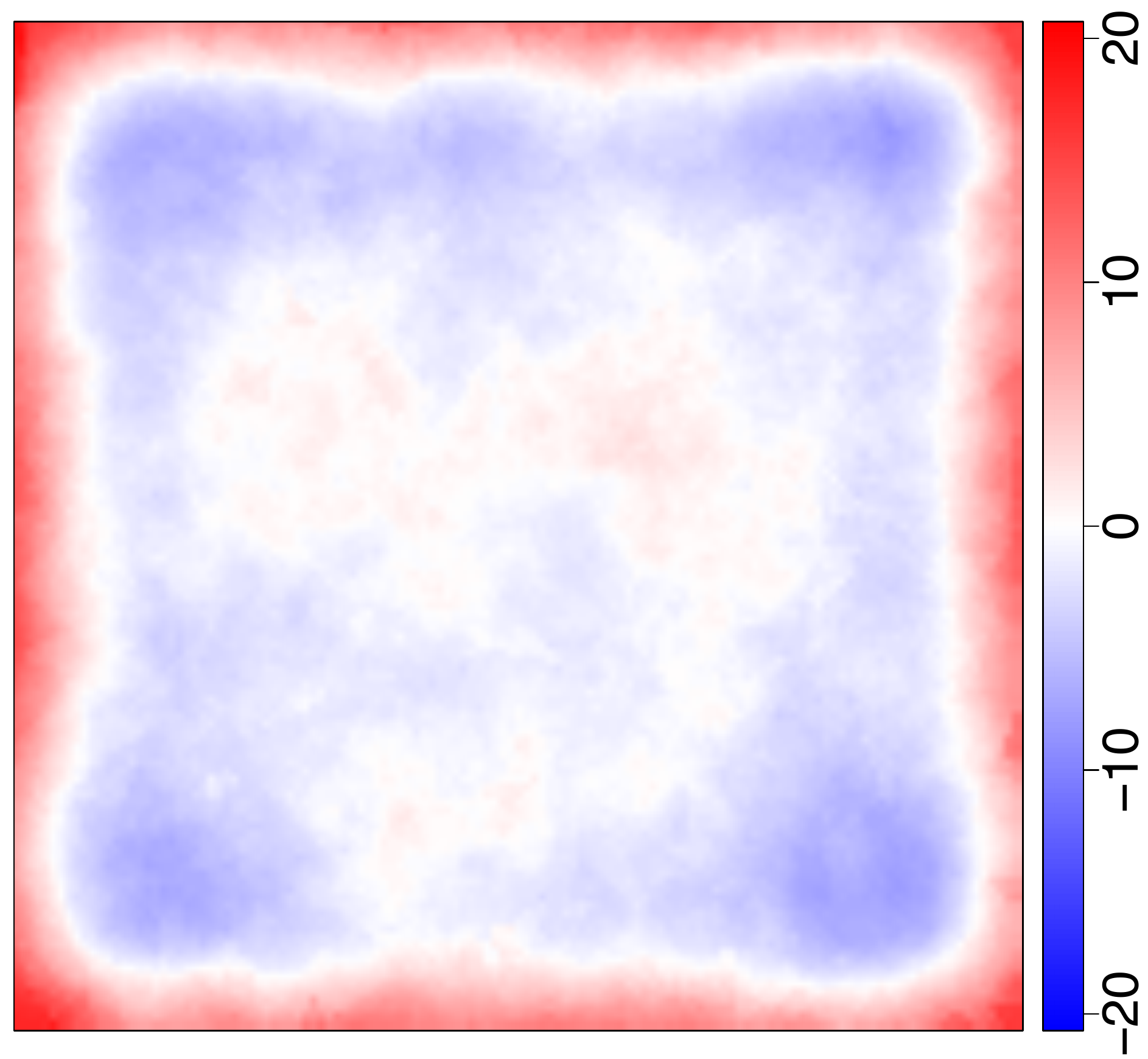}
 \qquad
  \includegraphics[width=0.35\textwidth]{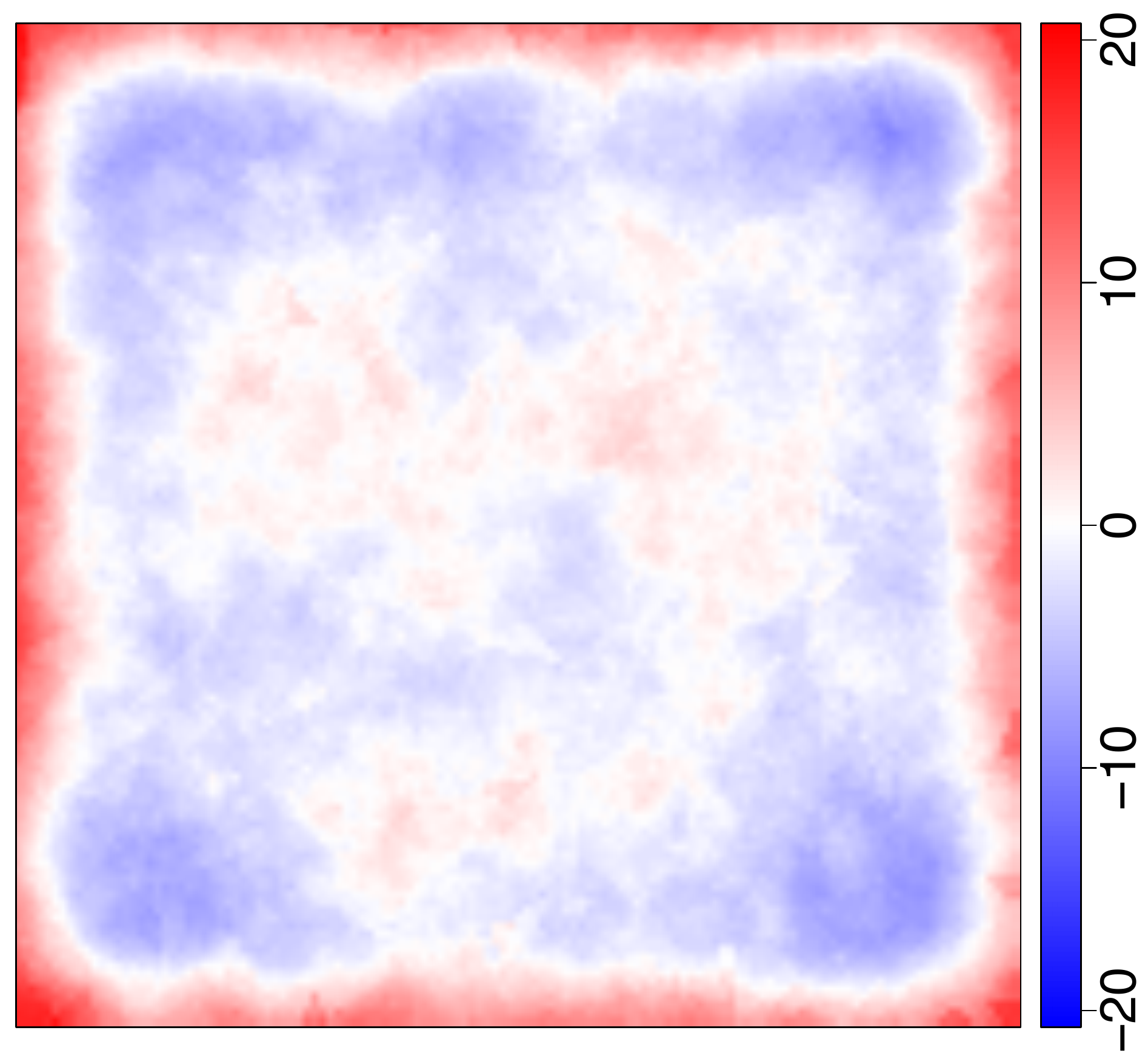}
  \qquad
  \includegraphics[width=0.35\textwidth]{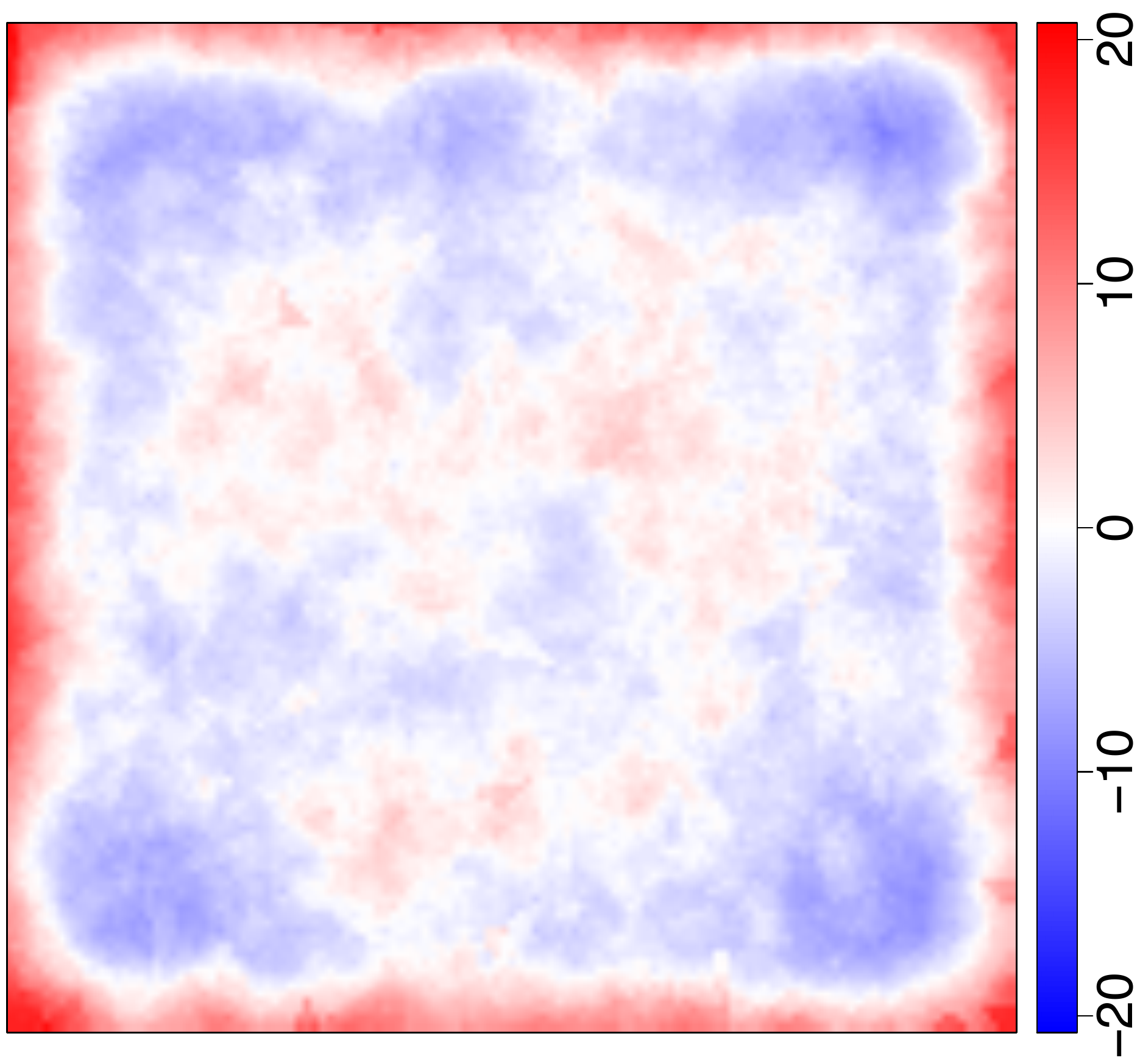}
  \qquad
  \includegraphics[width=0.35\textwidth]{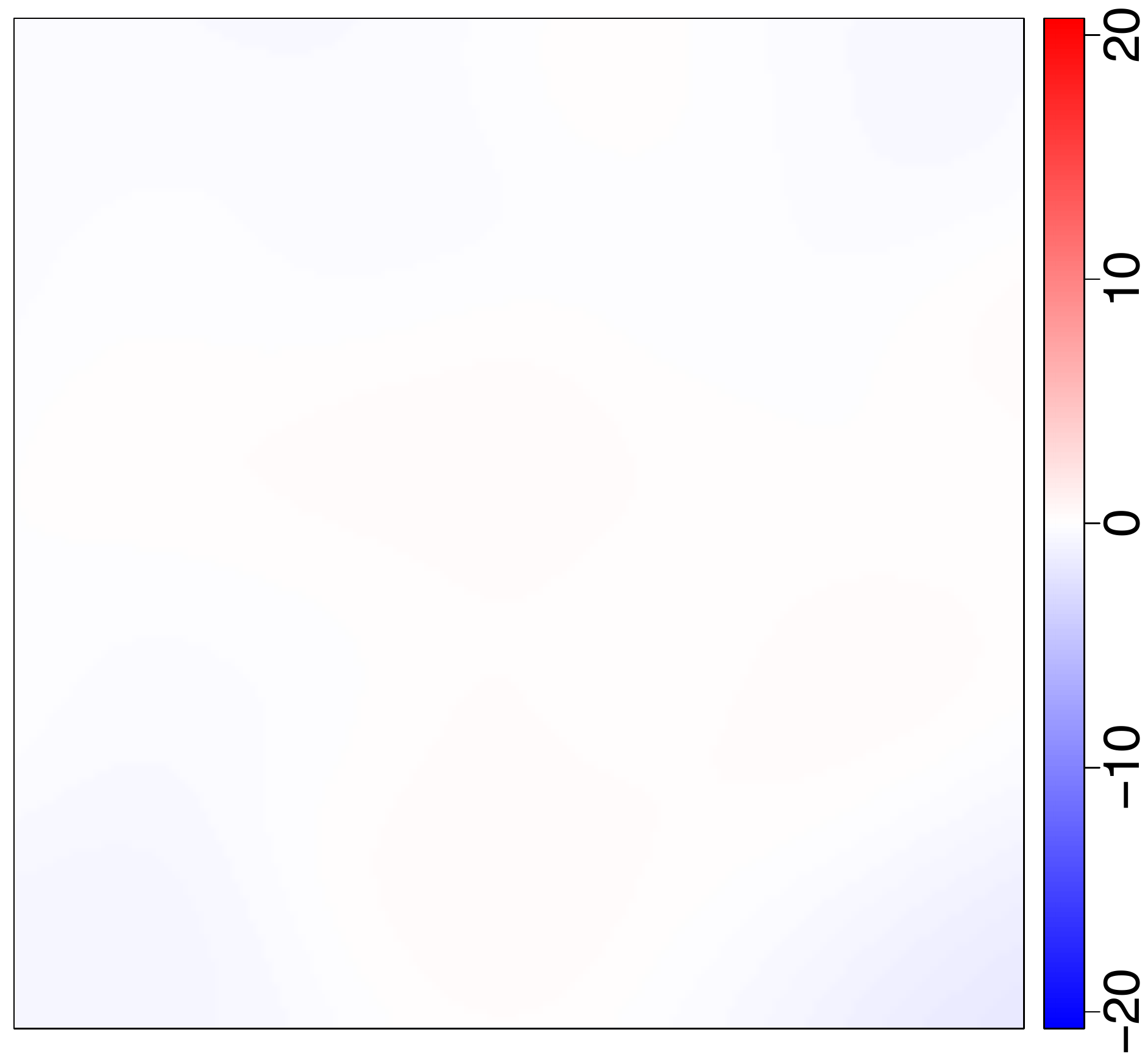}
  \qquad
  \includegraphics[width=0.35\textwidth]{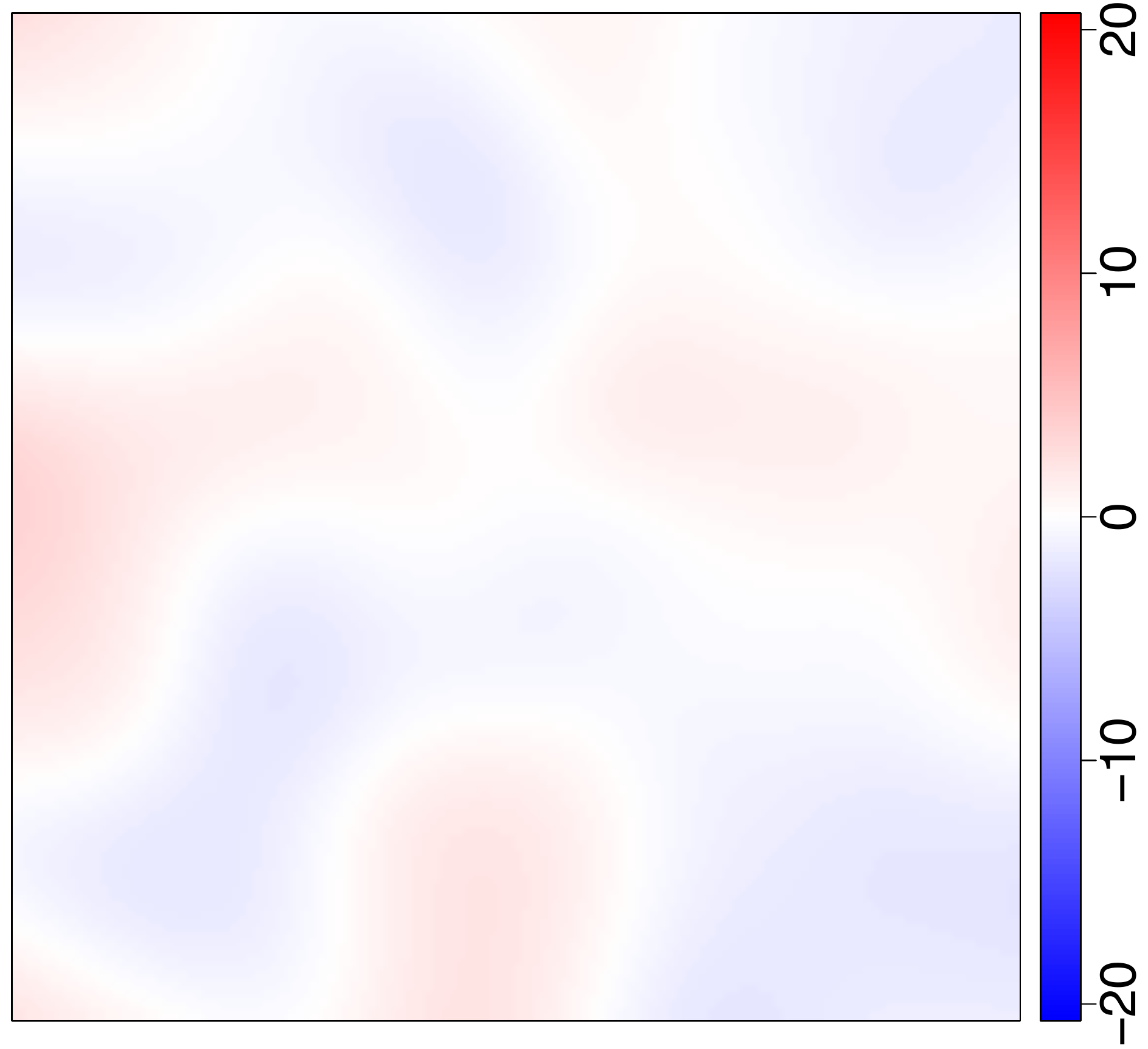}
\caption{Estimated bias for $\widehat{\rho}_{p,m}^{V}(u)$, $u\in W=[0,1]^2$, $m=200$, and kernel estimators, based on $500$ realisations of a homogeneous Poisson process $X\subseteq W=[0,1]^2$ with intensity $\rho=60$. From top-left to bottom-right: $\widehat{\rho}_{p,m}^{V}(u)$ with $p=0.1,0.3,0.5,0.7,0.9,1$; kernel estimators with bandwidths selected using Poisson likelihood cross-validation \citep{BRT15,Load99} (left) and the method of \citet{cronie2018bandwidth} (right) are on the last row.}
\label{f:BiasHomPoiR2}
\end{figure*}

\begin{figure*}[!h]
\centering
\includegraphics[width=0.35\textwidth]{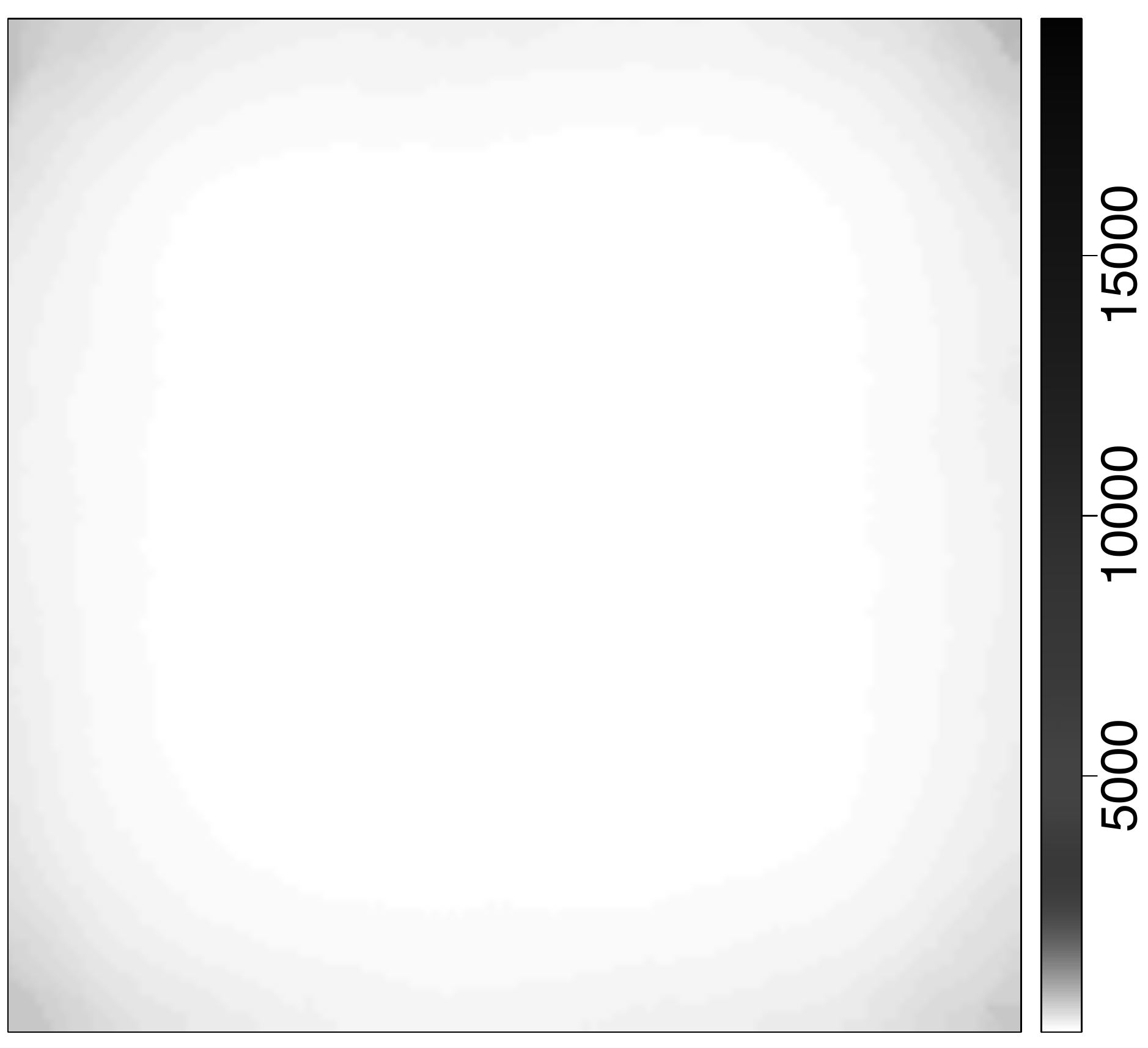}
  \qquad
 \includegraphics[width=0.35\textwidth]{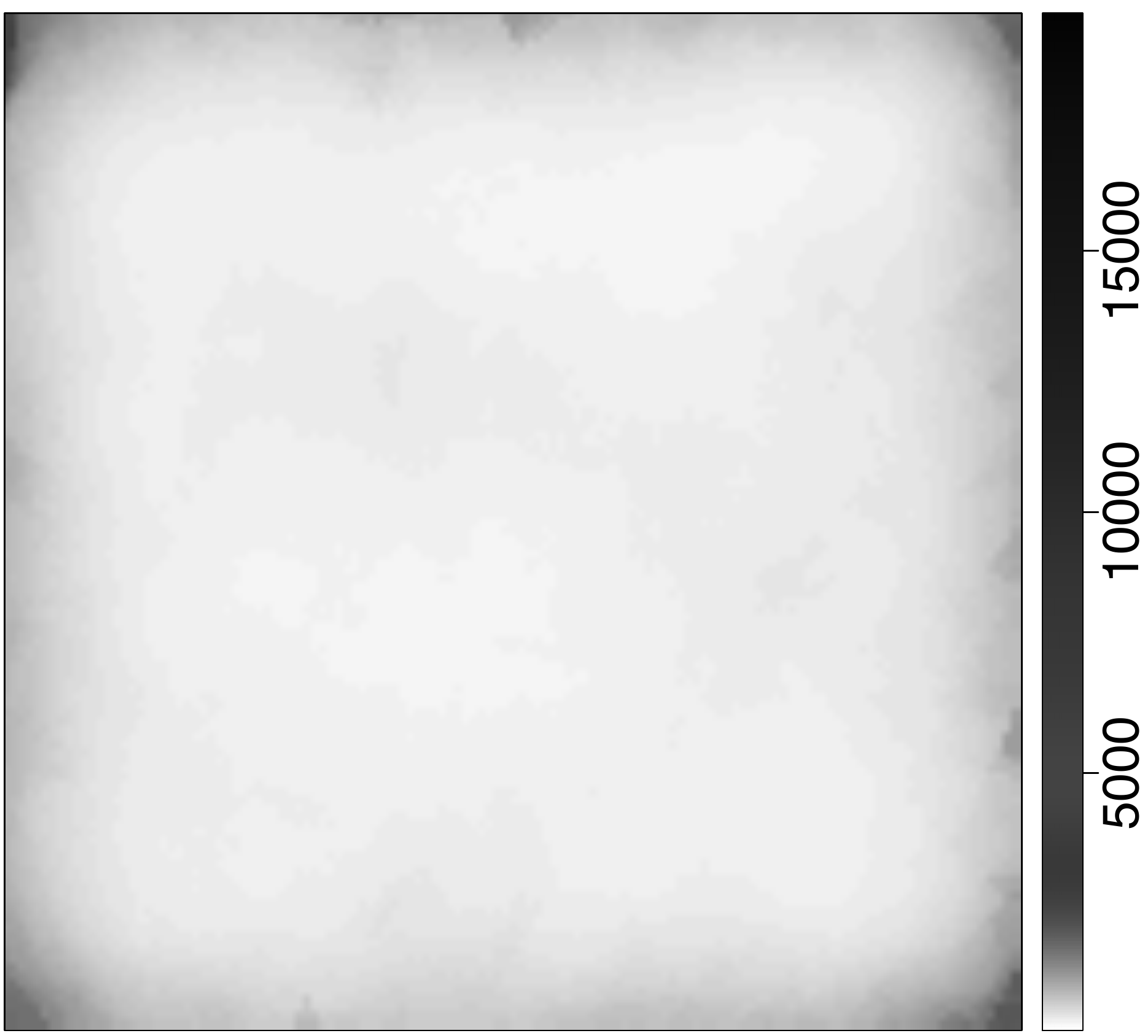}
 \qquad
  \includegraphics[width=0.35\textwidth]{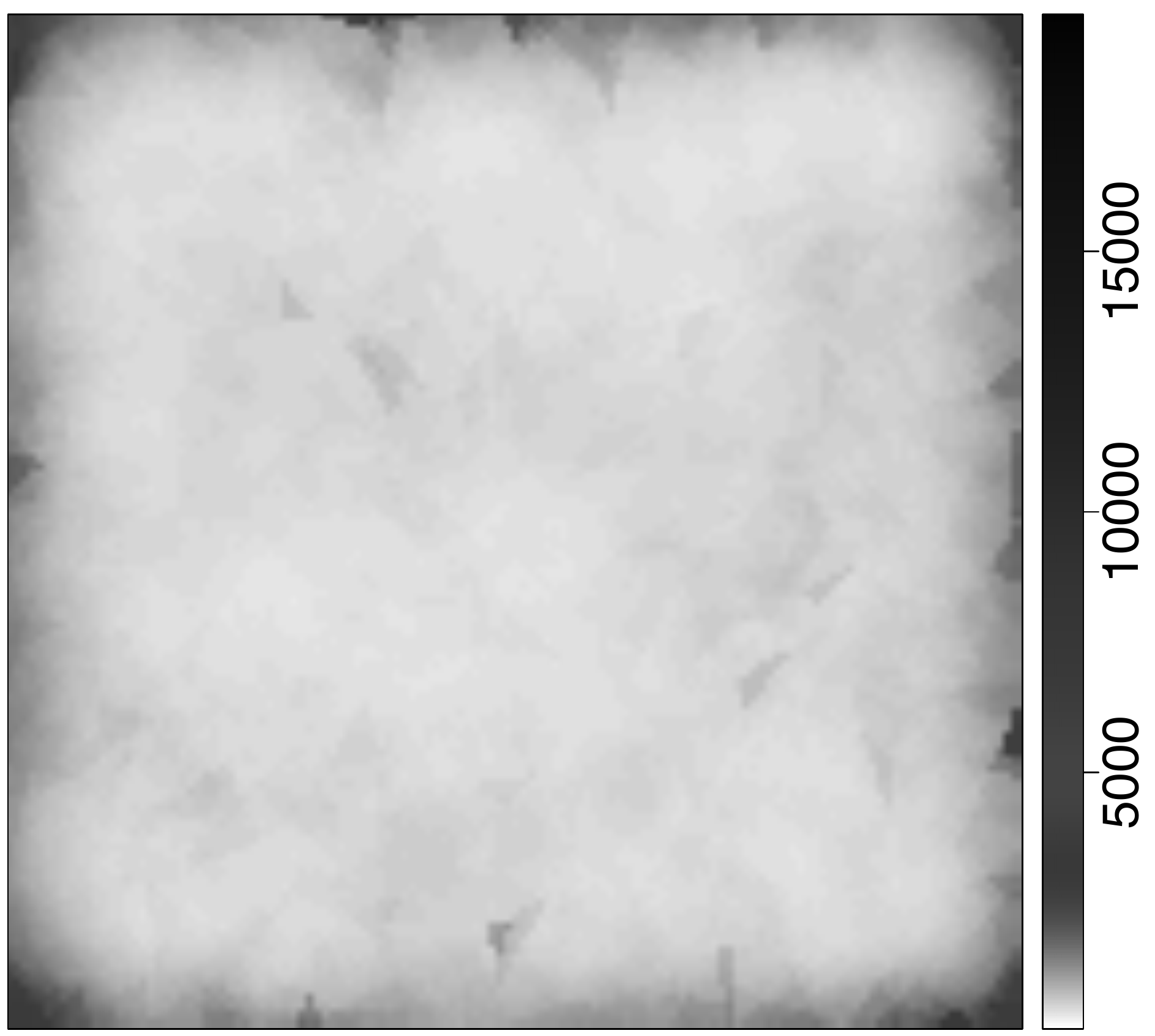}
  \qquad
 \includegraphics[width=0.35\textwidth]{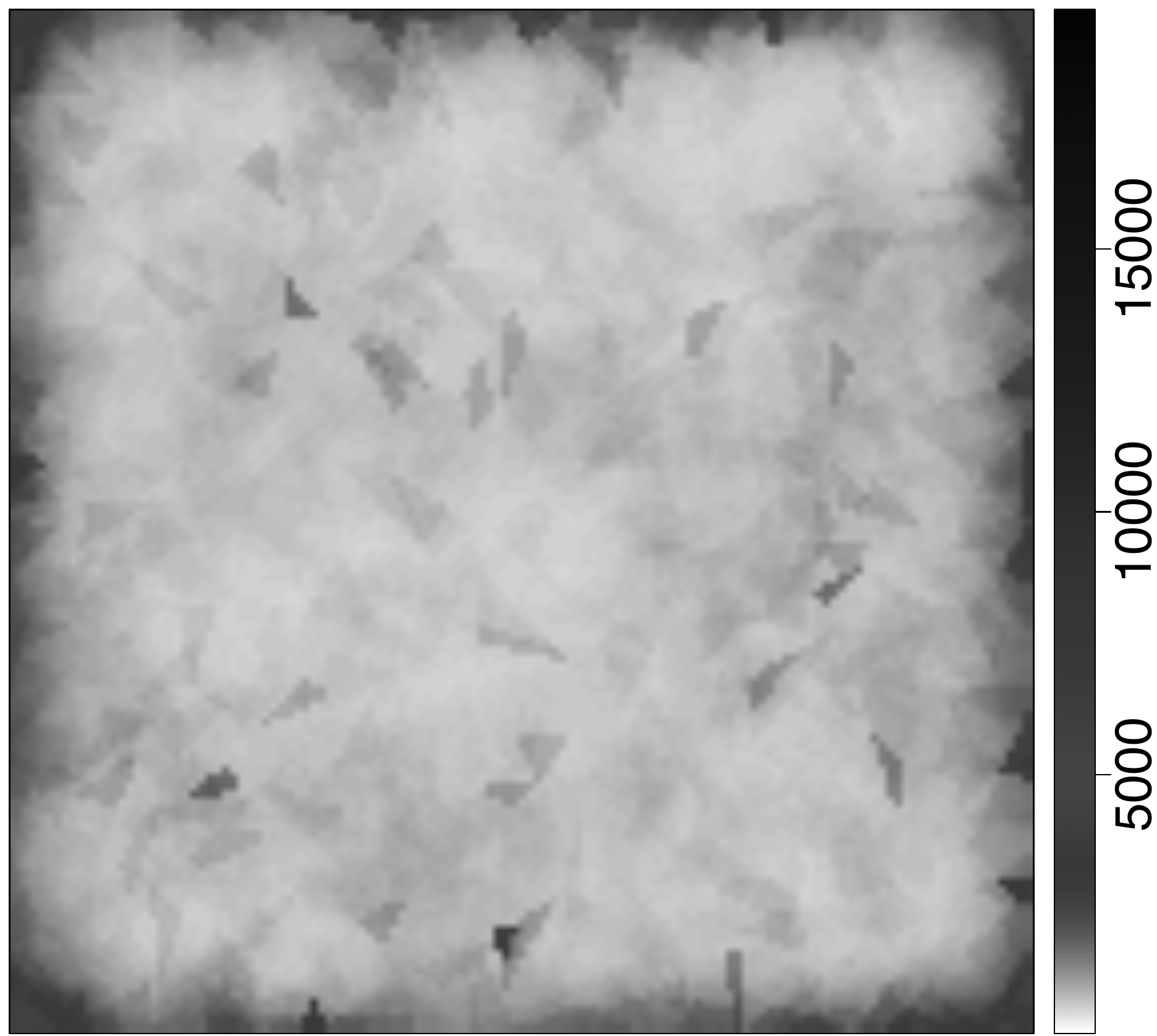}
 \qquad
  \includegraphics[width=0.35\textwidth]{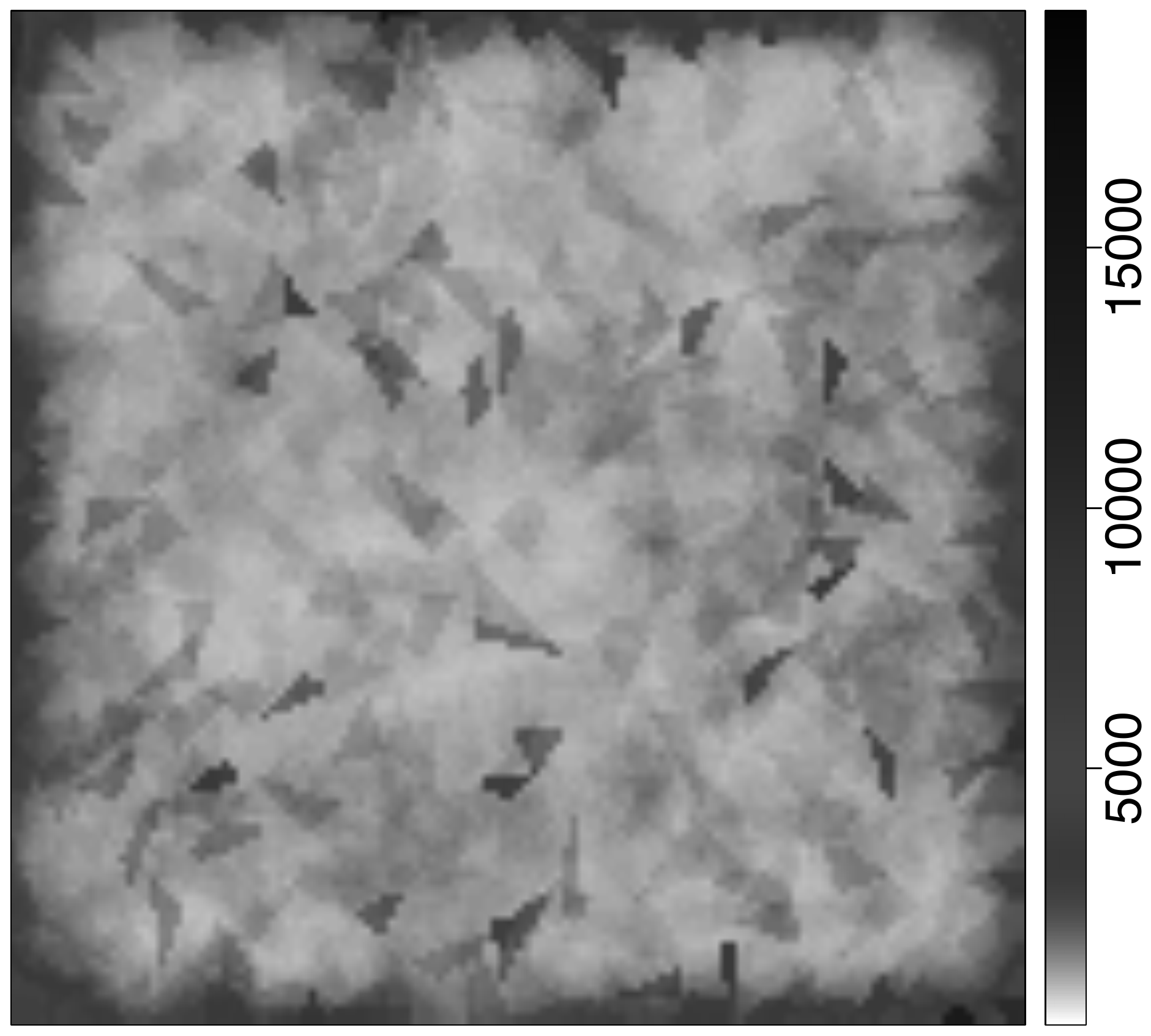}
  \qquad
  \includegraphics[width=0.35\textwidth]{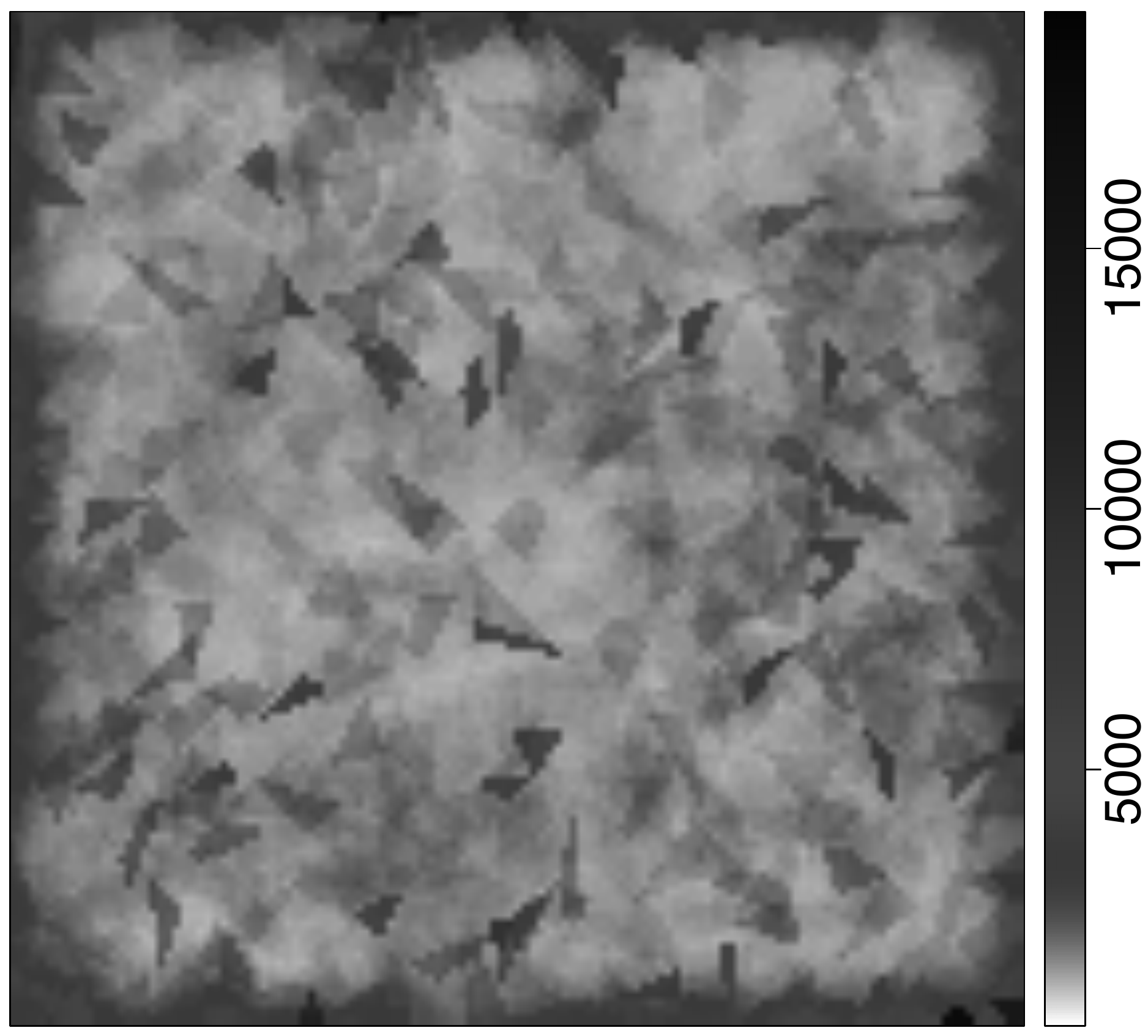}
  \qquad
  \includegraphics[width=0.35\textwidth]{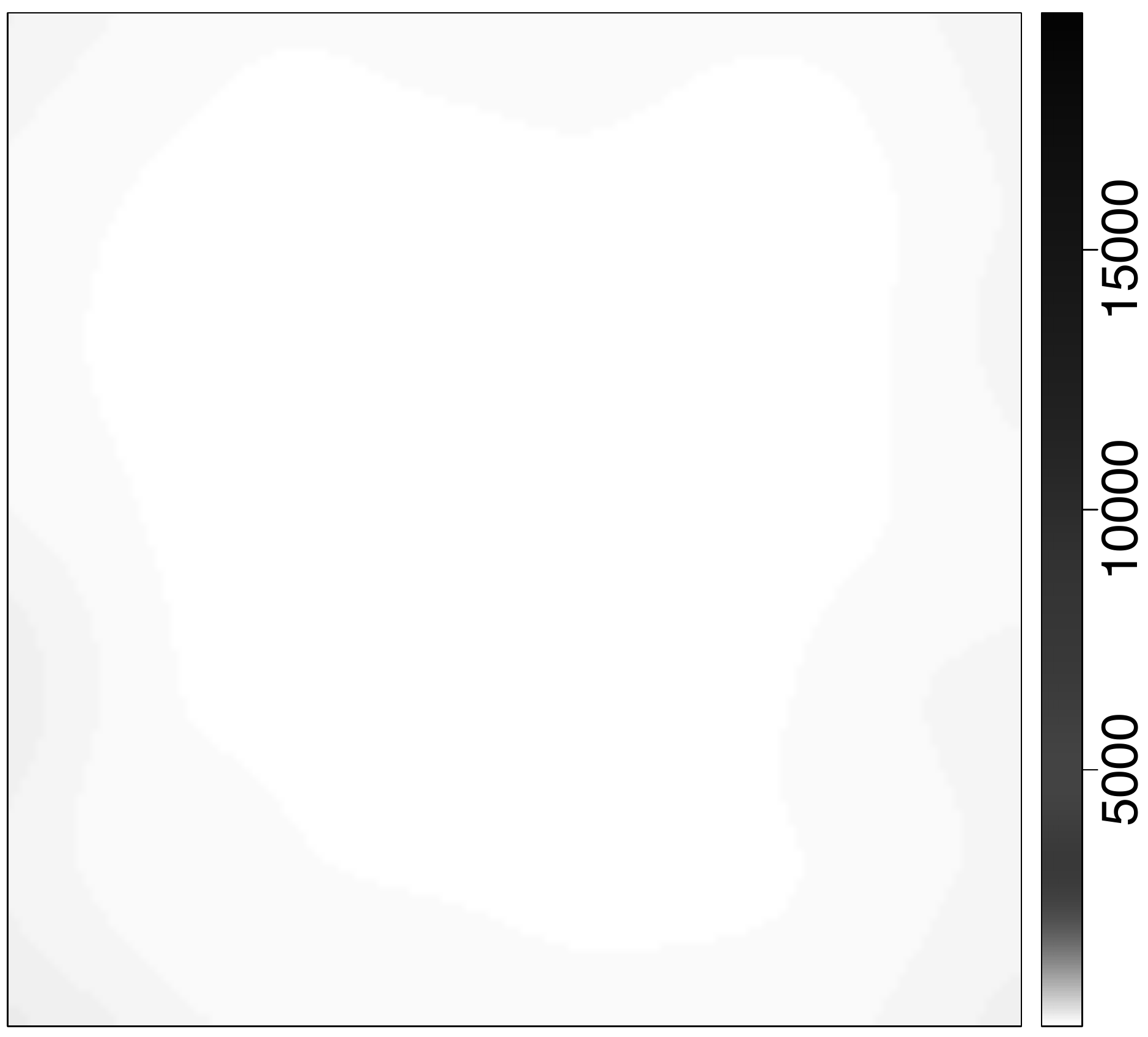}
  \qquad
  \includegraphics[width=0.35\textwidth]{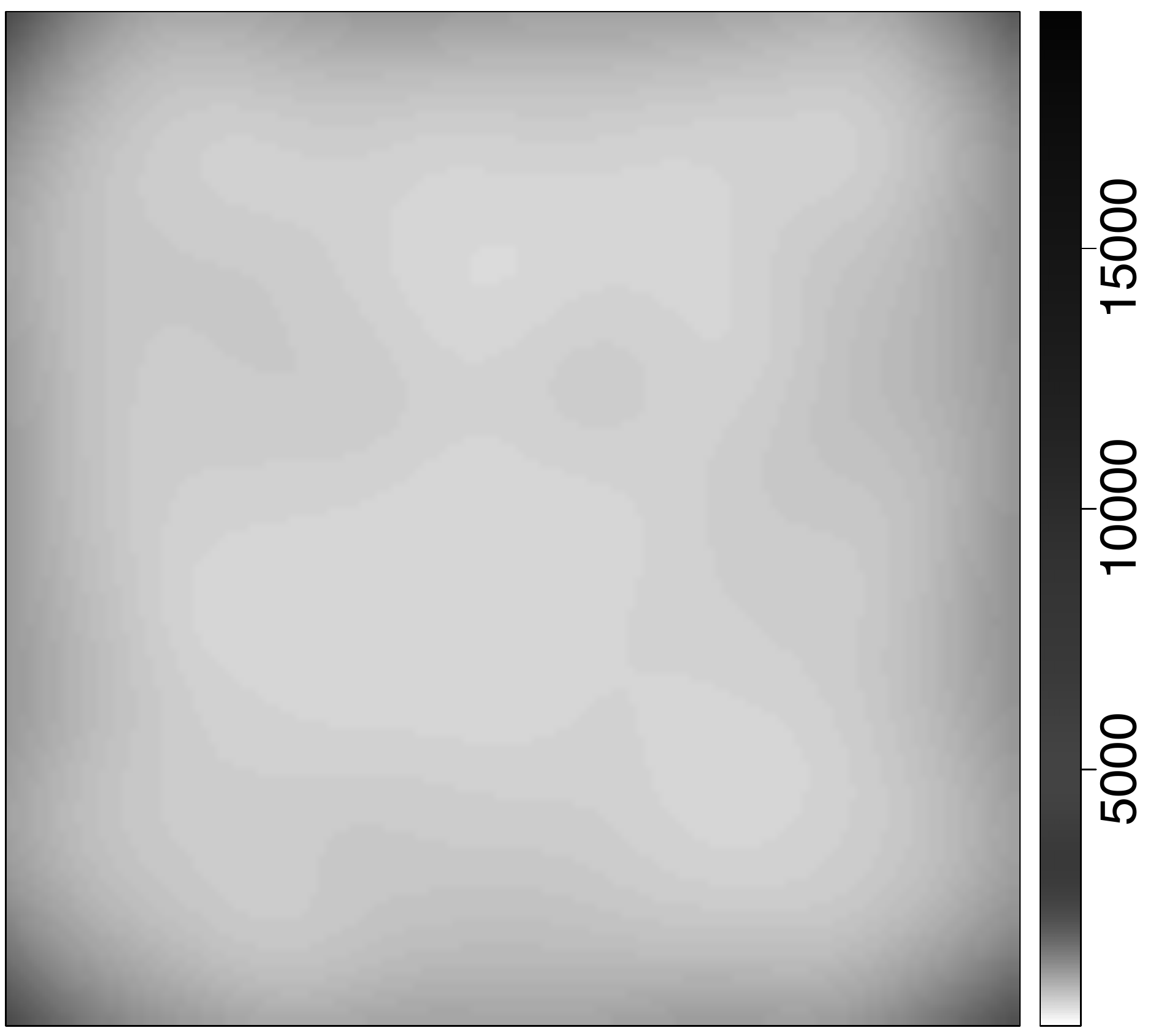}
\caption{
Estimated variance for $\widehat{\rho}_{p,m}^{V}(u)$, $u\in W=[0,1]^2$, $m=200$, and kernel estimators, based on $500$ realisations of a homogeneous Poisson process $X\subseteq W=[0,1]^2$ with intensity $\rho=60$. From top-left to bottom-right: $\widehat{\rho}_{p,m}^{V}(u)$ with $p=0.1,0.3,0.5,0.7,0.9,1$; kernel estimators with bandwidths selected using Poisson likelihood cross-validation \citep{BRT15,Load99} (left) and the method of \citet{cronie2018bandwidth} (right) are on the last row.
}
\label{f:VarHomPoiR2}
\end{figure*}

\begin{figure*}[!h]
\centering
  \includegraphics[width=0.35\textwidth]{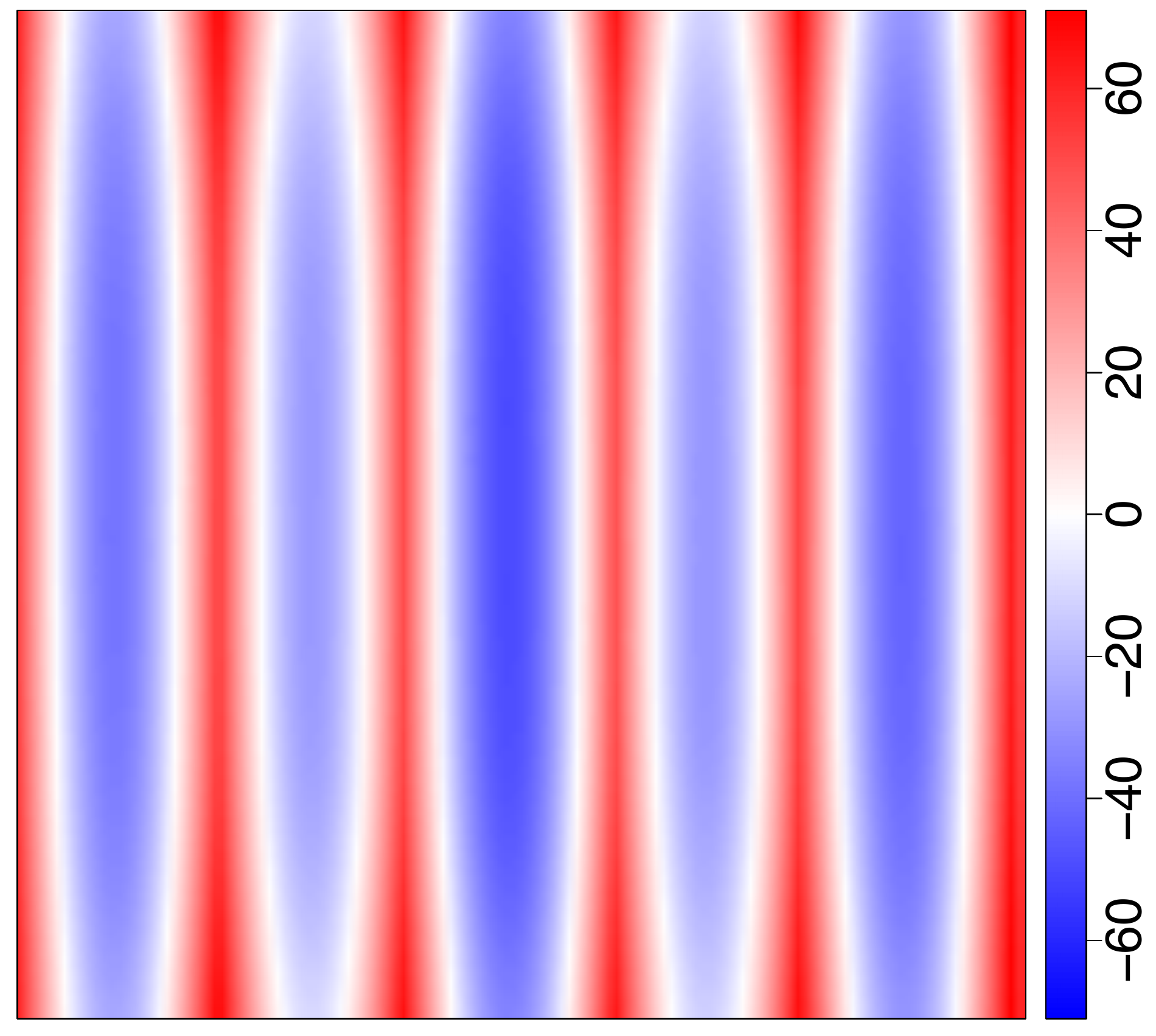}
  \qquad
 \includegraphics[width=0.35\textwidth]{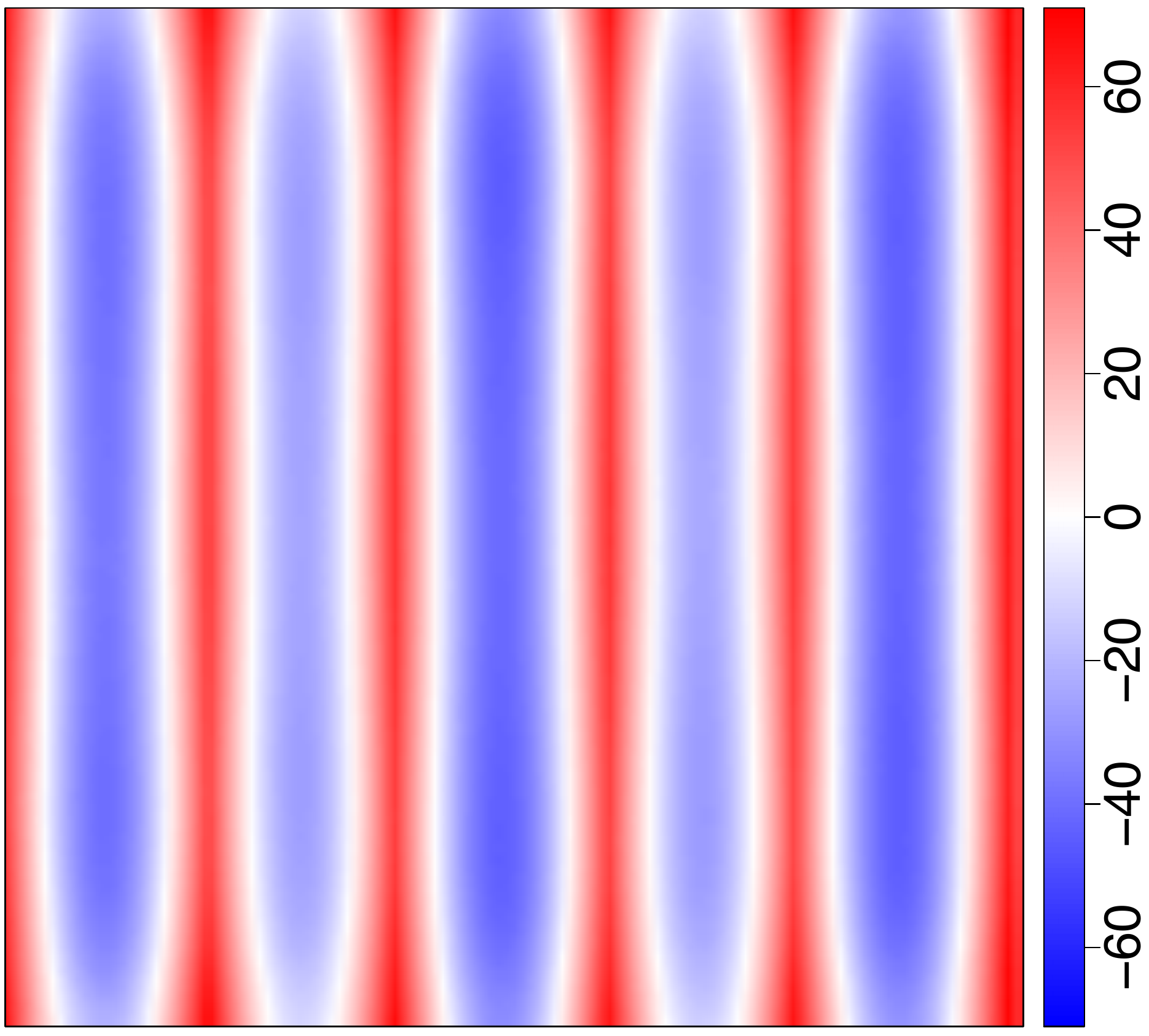}
 \qquad
  \includegraphics[width=0.35\textwidth]{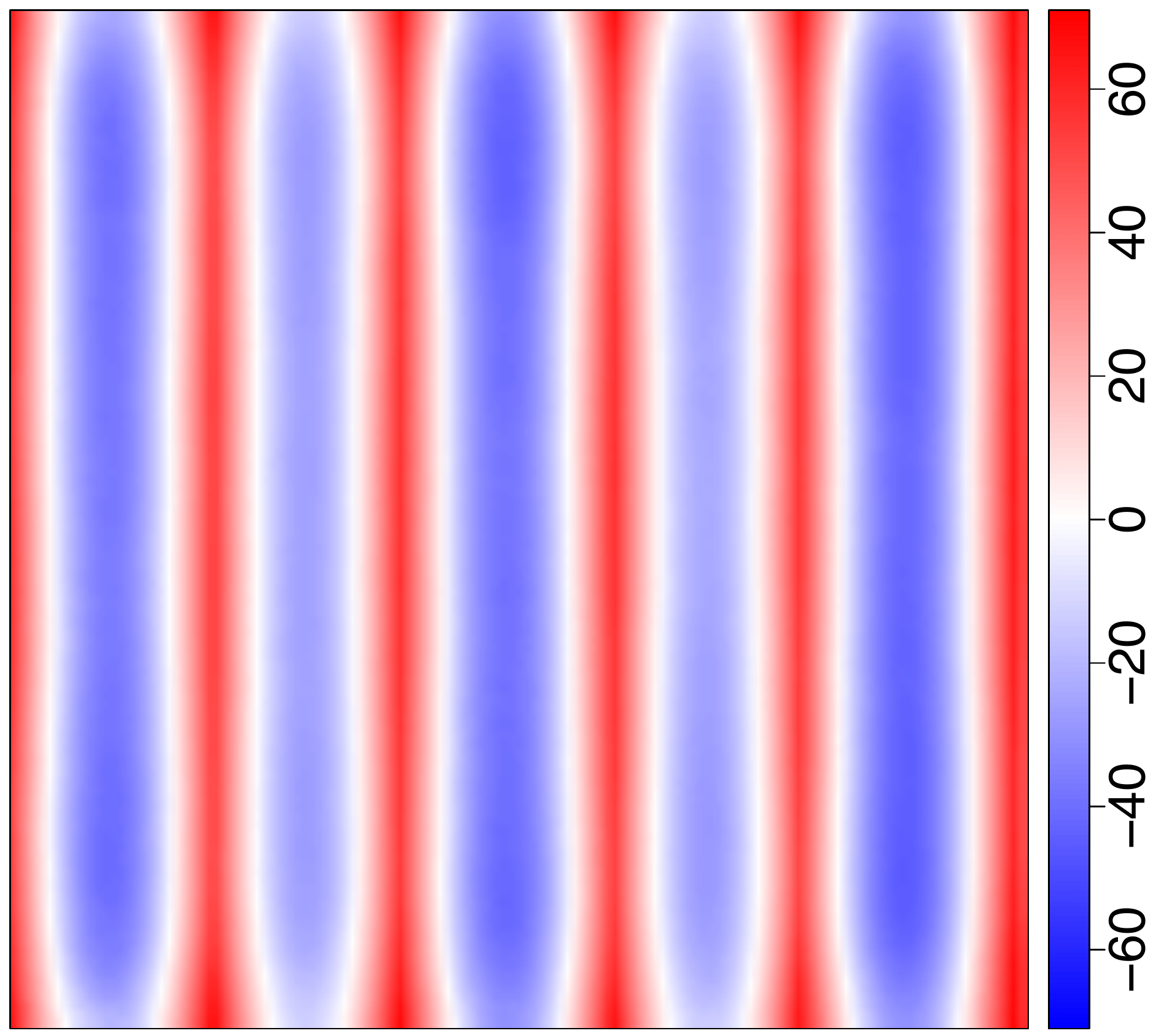}
  \qquad
 \includegraphics[width=0.35\textwidth]{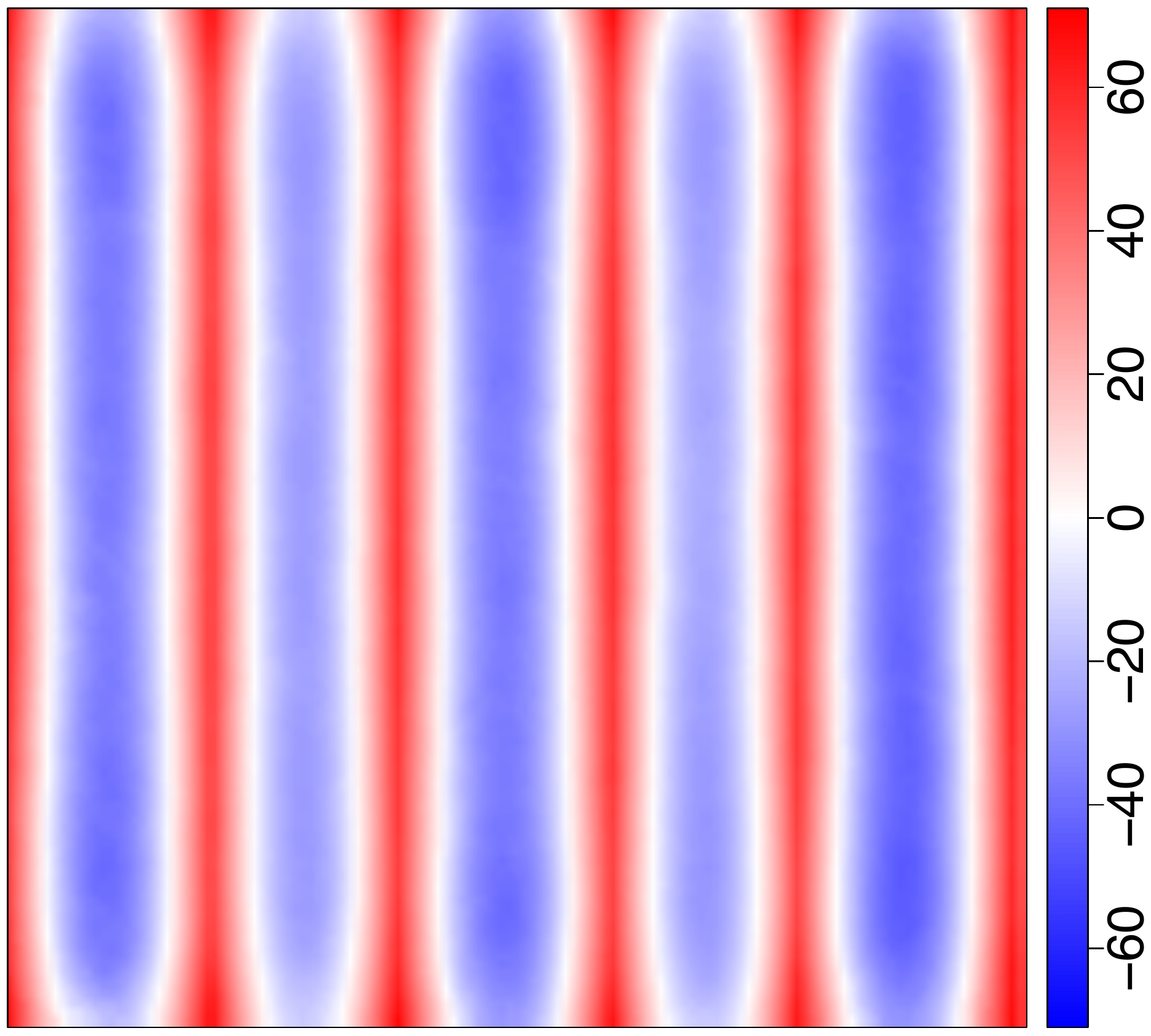}
 \qquad
  \includegraphics[width=0.35\textwidth]{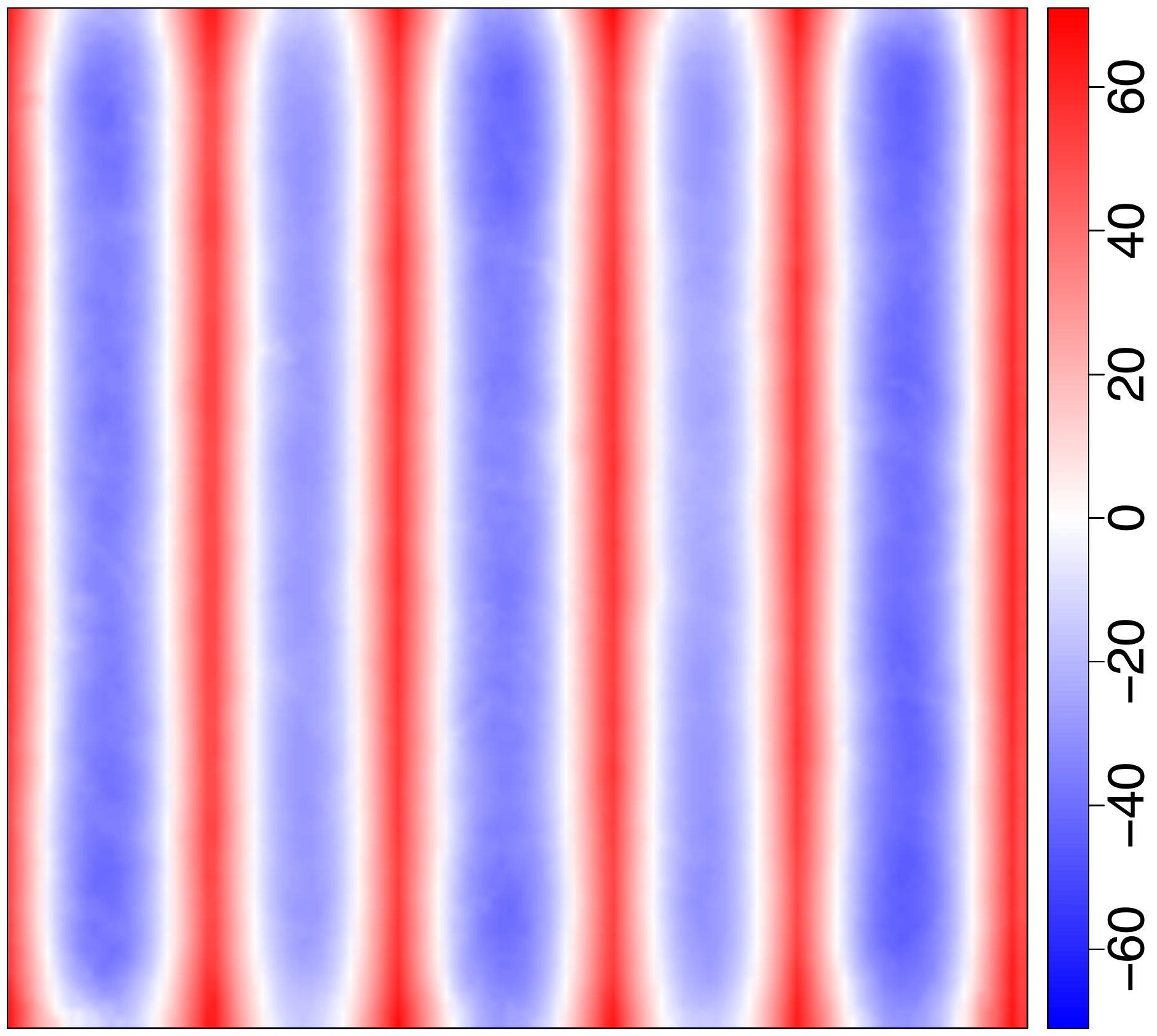}
  \qquad
  \includegraphics[width=0.35\textwidth]{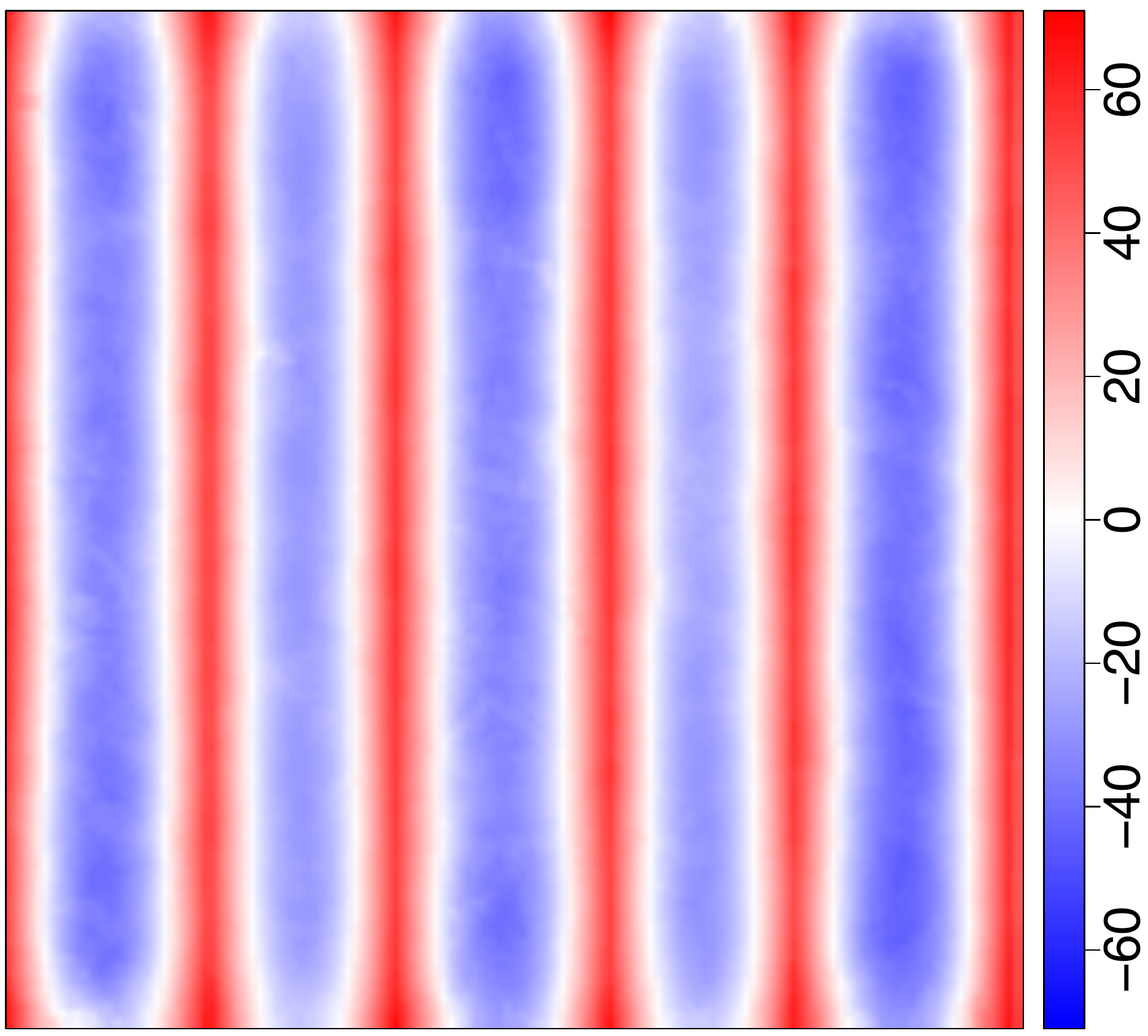}
  \qquad
  \includegraphics[width=0.35\textwidth]{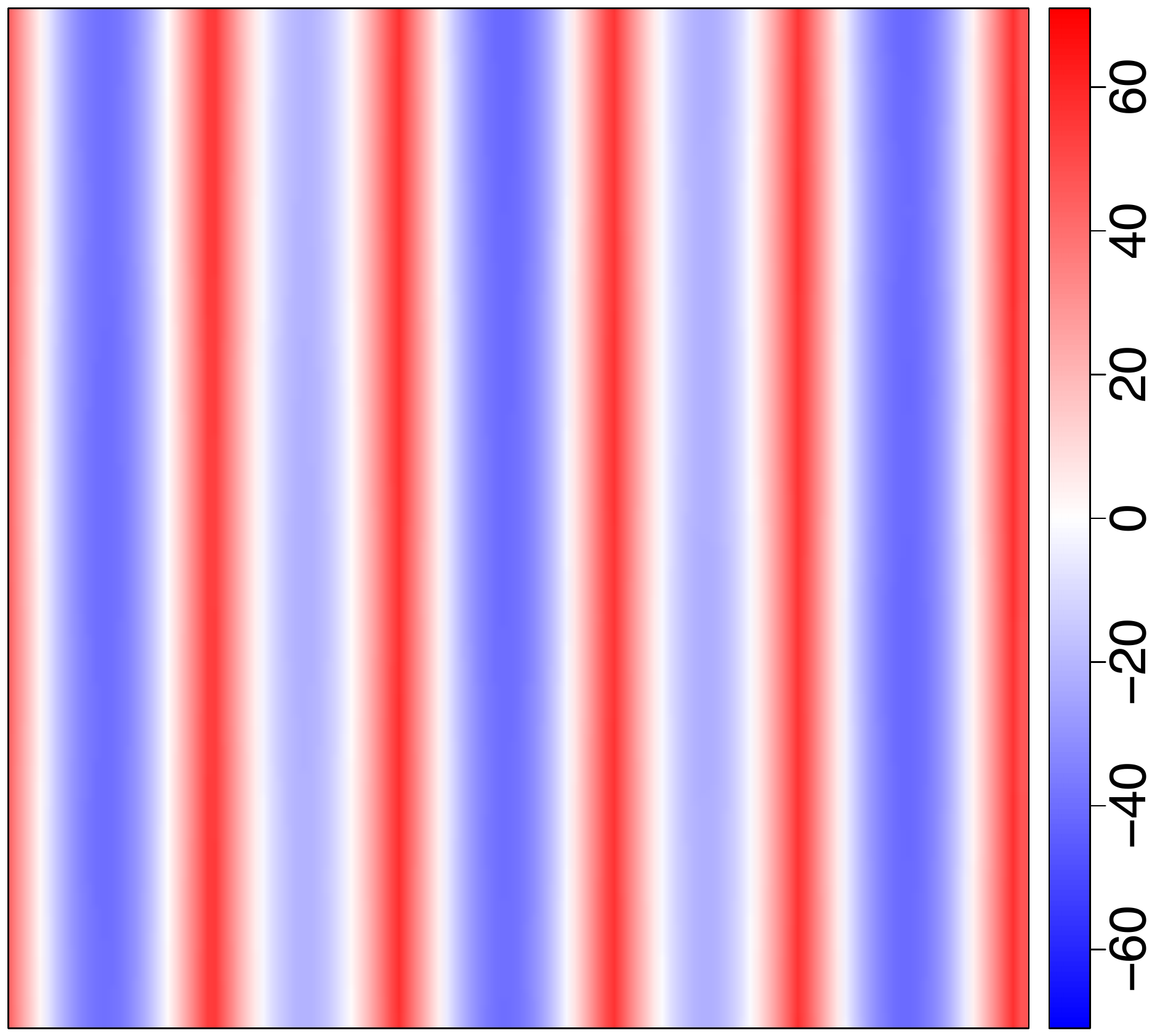}
  \qquad
  \includegraphics[width=0.35\textwidth]{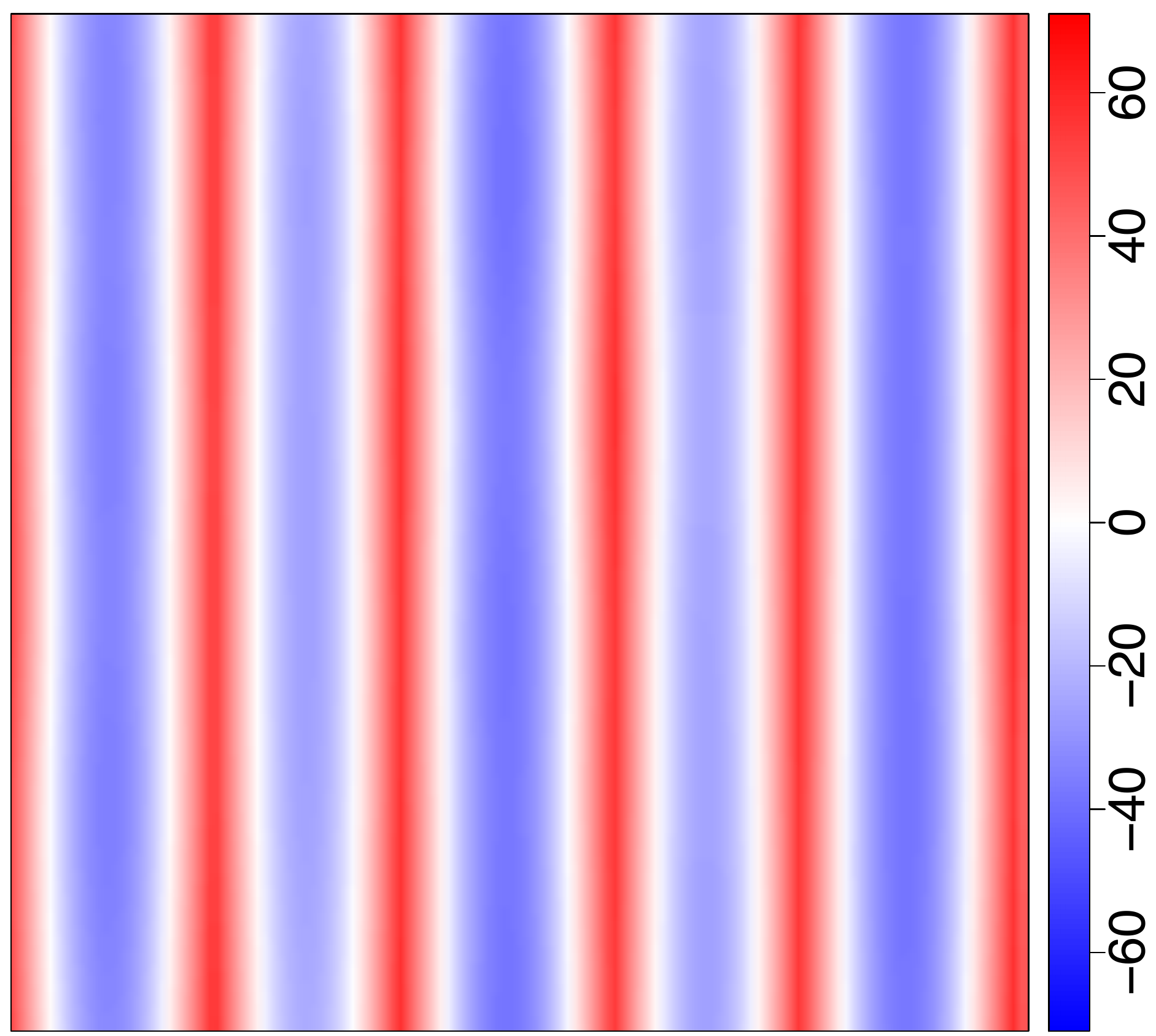}

\caption{
Estimated bias for $\widehat{\rho}_{p,m}^{V}(u)$, $u\in W=[0,1]^2$, $m=200$, and kernel estimators, based on $500$ realisations of an inhomogeneous Poisson process $X\subseteq W=[0,1]^2$ with intensity $\rho(x,y)=|10+90\sin(16x)|$. From top-left to bottom-right: $\widehat{\rho}_{p,m}^{V}(u)$ with $p=0.1,0.3,0.5,0.7,0.9,1$; kernel estimators with bandwidths selected using Poisson likelihood cross-validation \citep{BRT15,Load99} (left) and the method of \citet{cronie2018bandwidth} (right) are on the last row.
}
\label{f:BiasInhomPoiR2}
\end{figure*}

\begin{figure*}[!h]
\centering
  \includegraphics[width=0.35\textwidth]{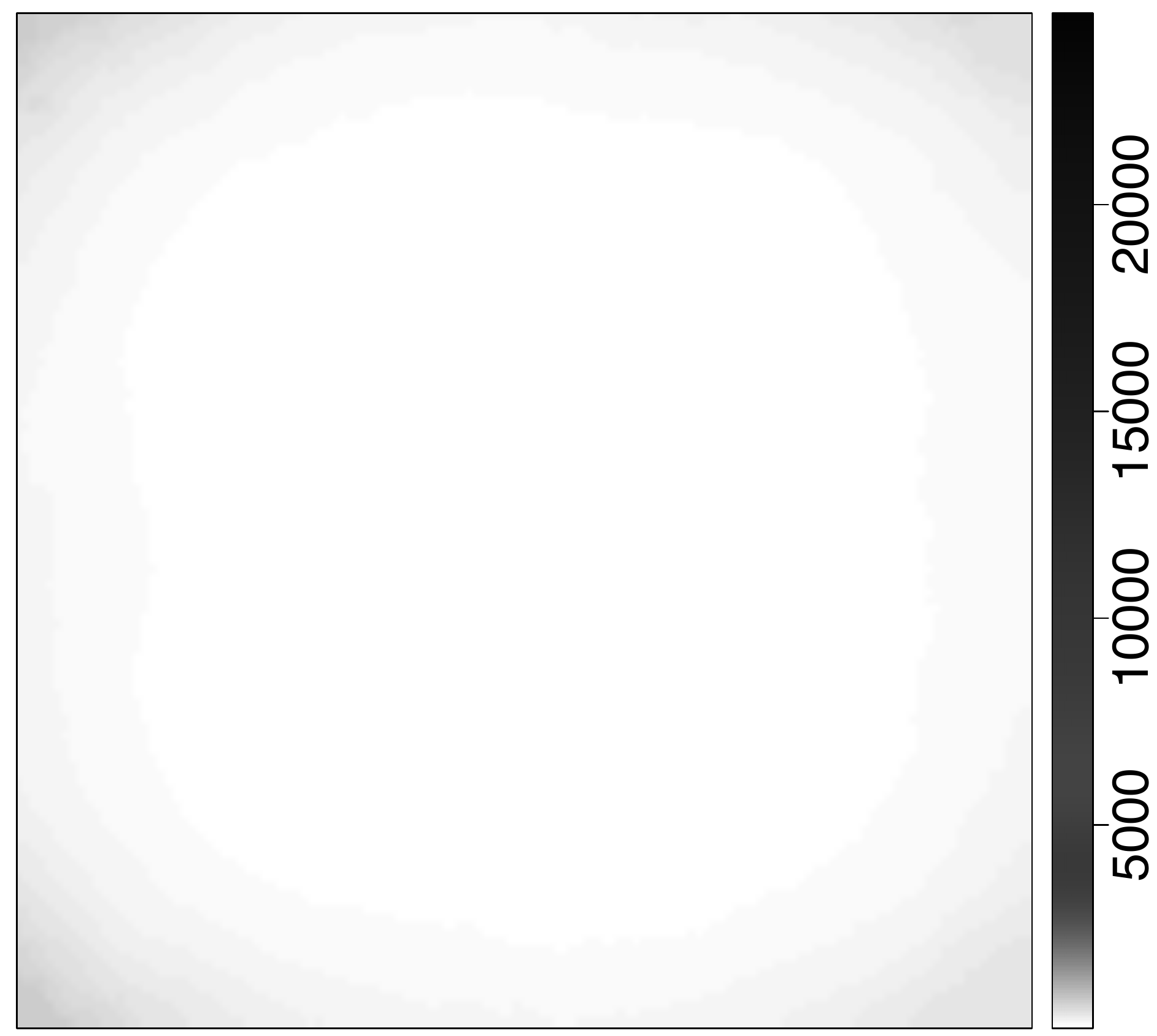}
  \qquad
 \includegraphics[width=0.35\textwidth]{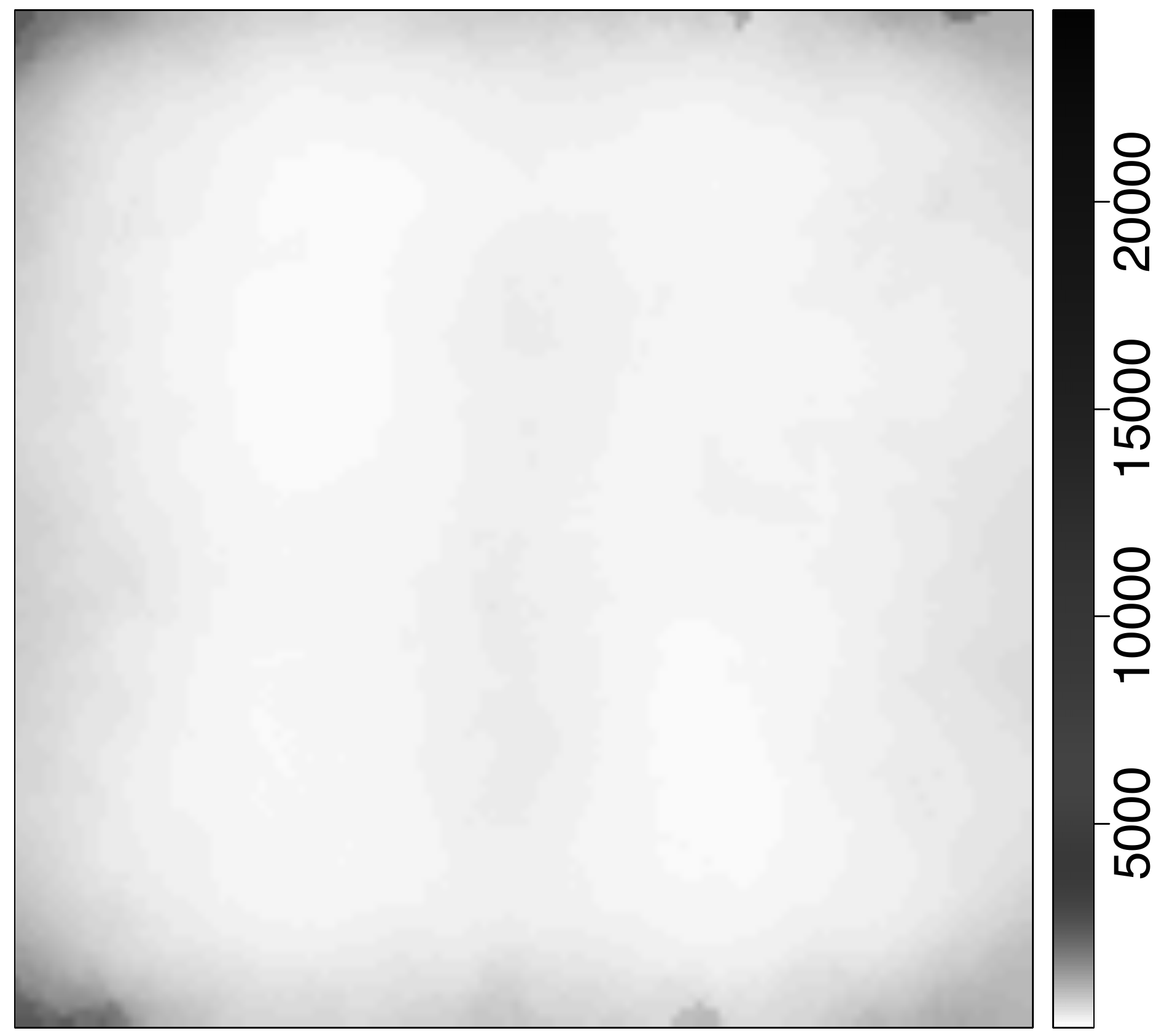}
 \qquad
  \includegraphics[width=0.35\textwidth]{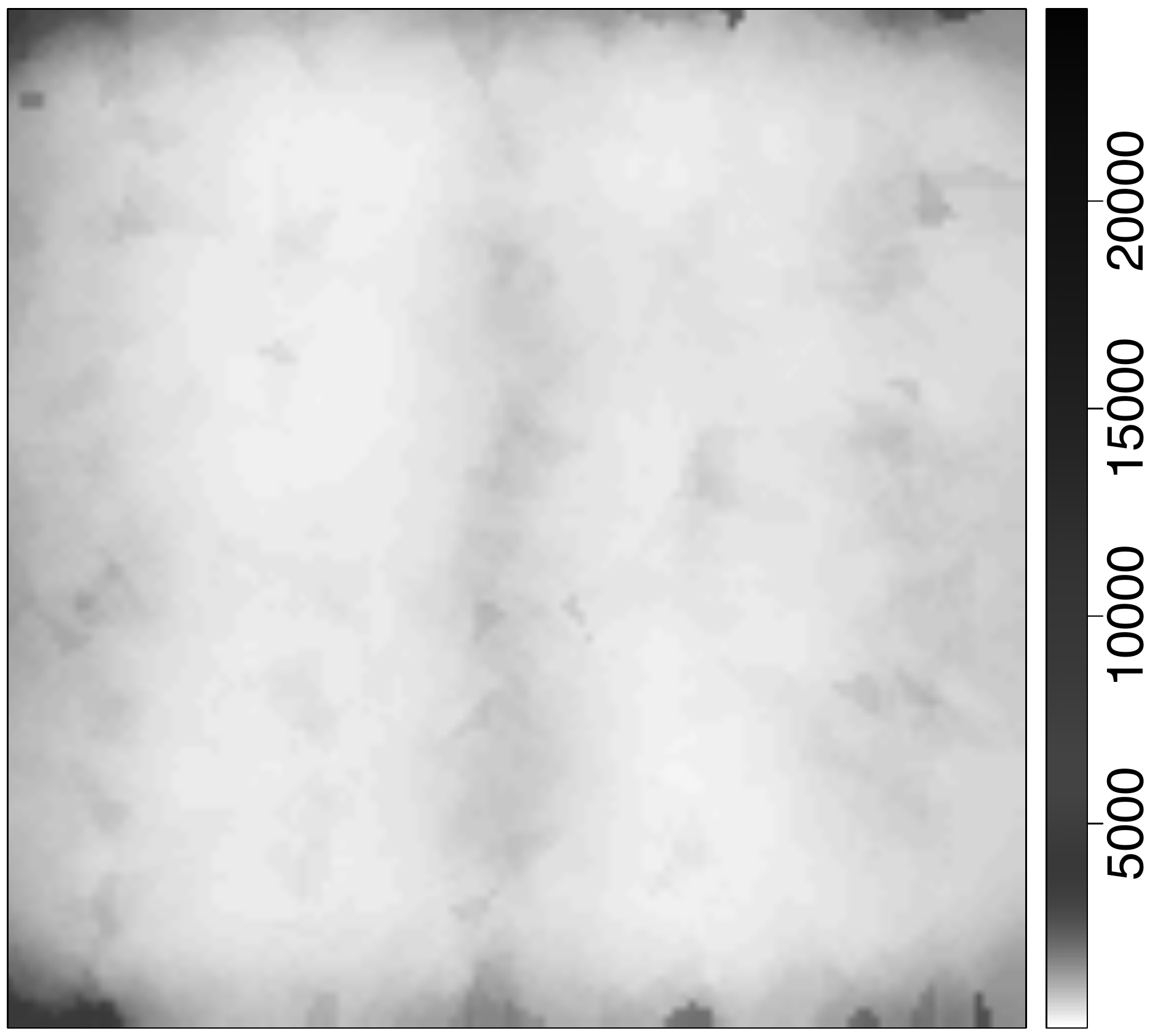}
  \qquad
 \includegraphics[width=0.35\textwidth]{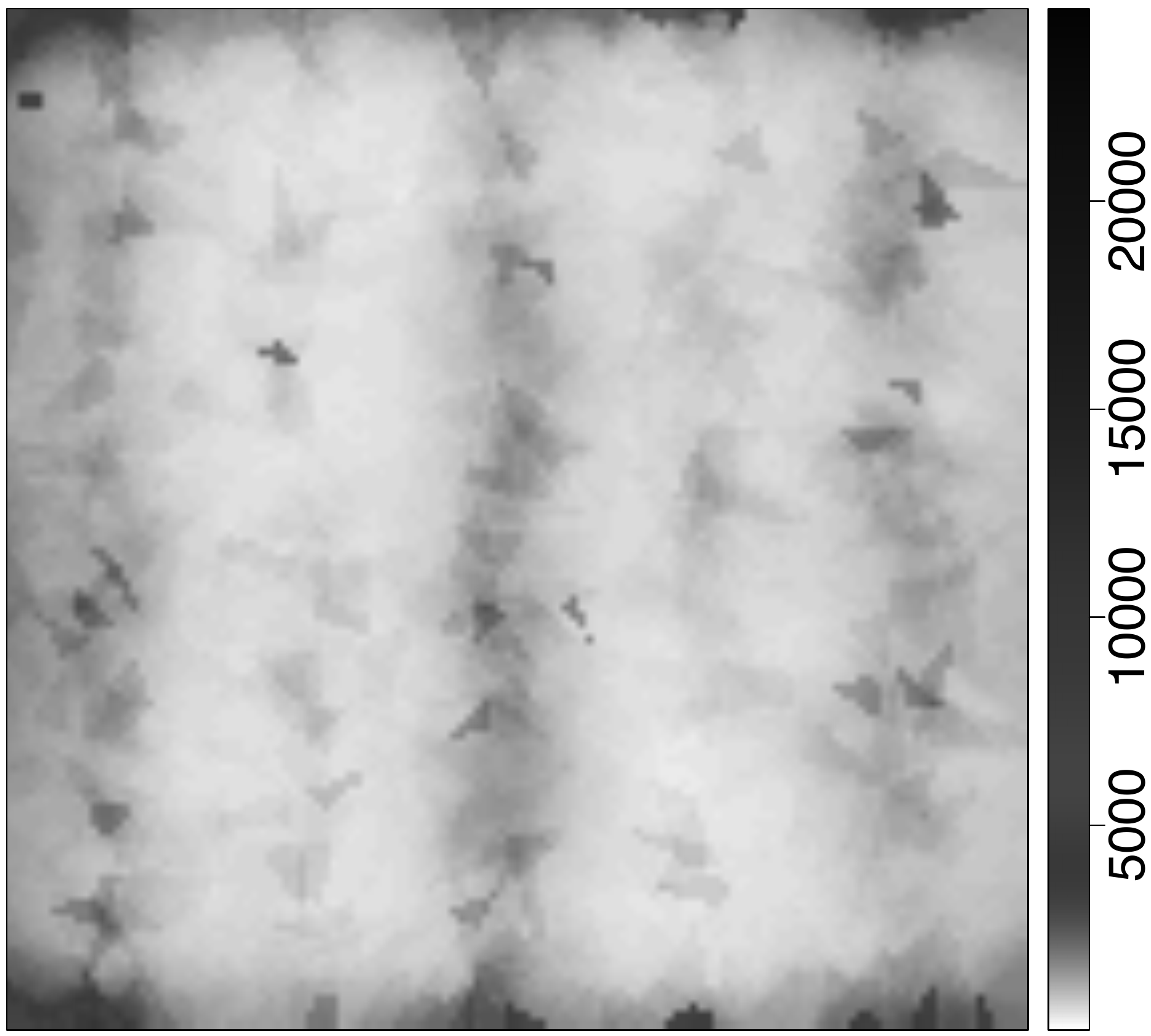}
 \qquad
  \includegraphics[width=0.35\textwidth]{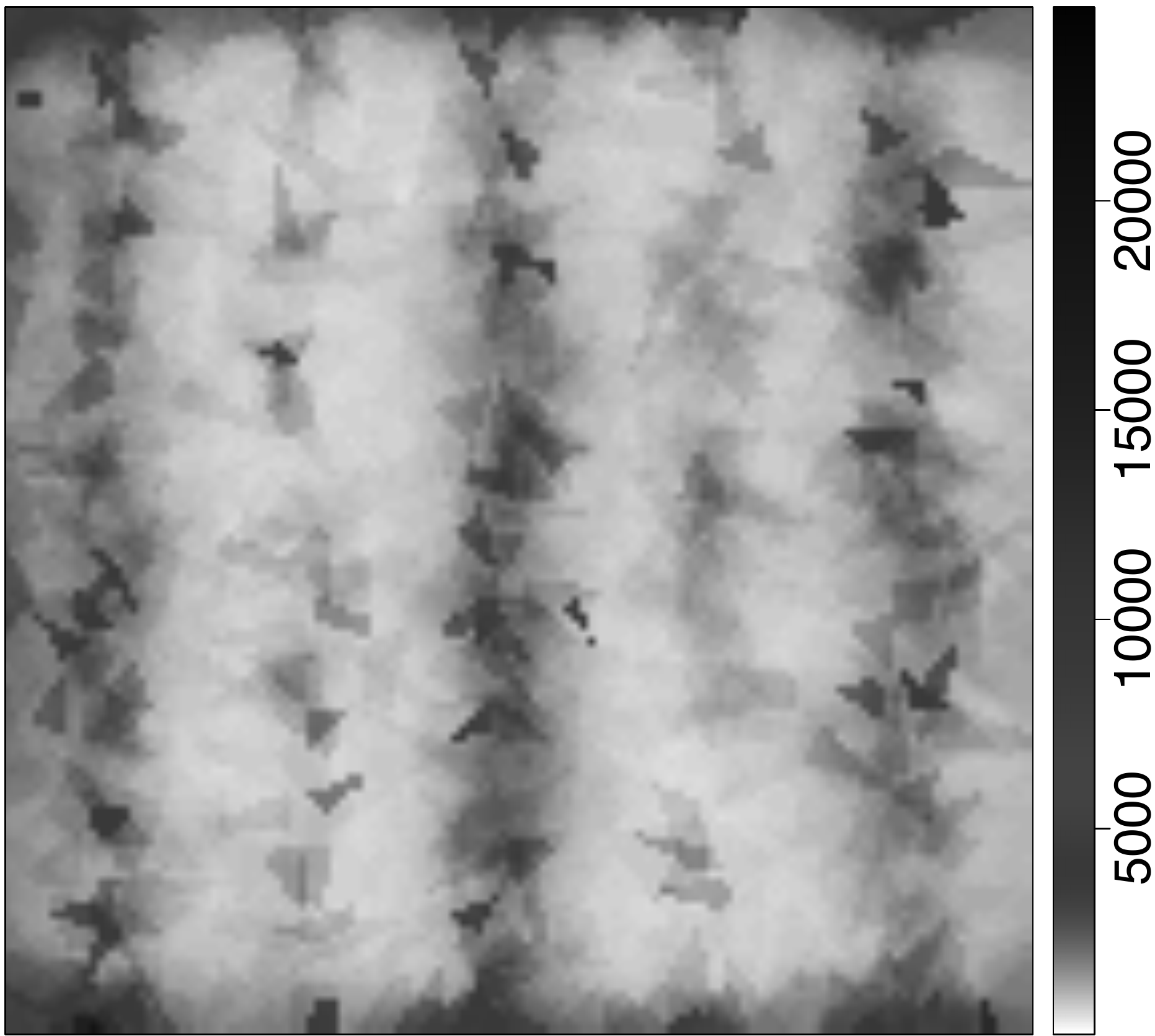}
  \qquad
  \includegraphics[width=0.35\textwidth]{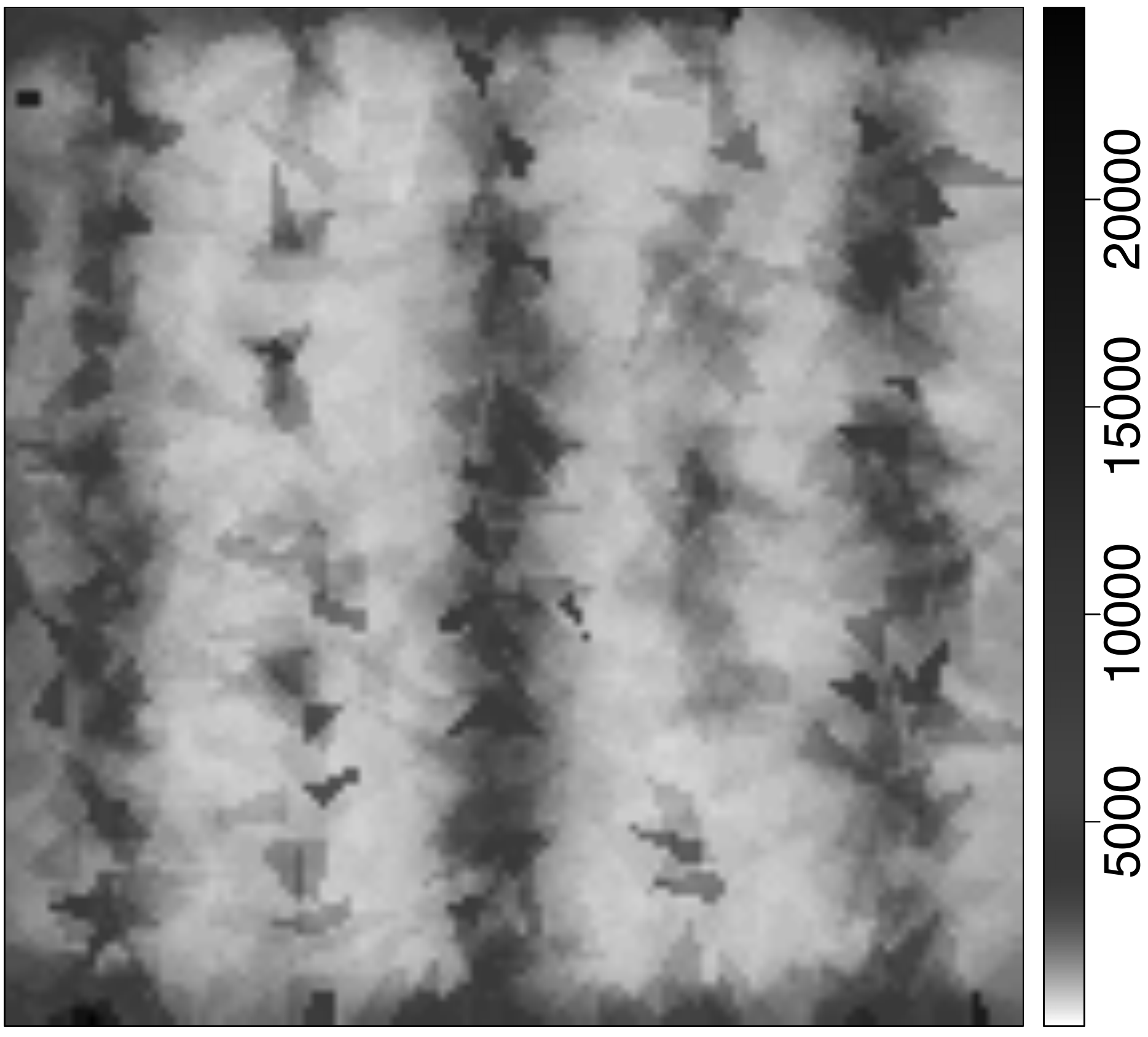}
  \qquad
  \includegraphics[width=0.35\textwidth]{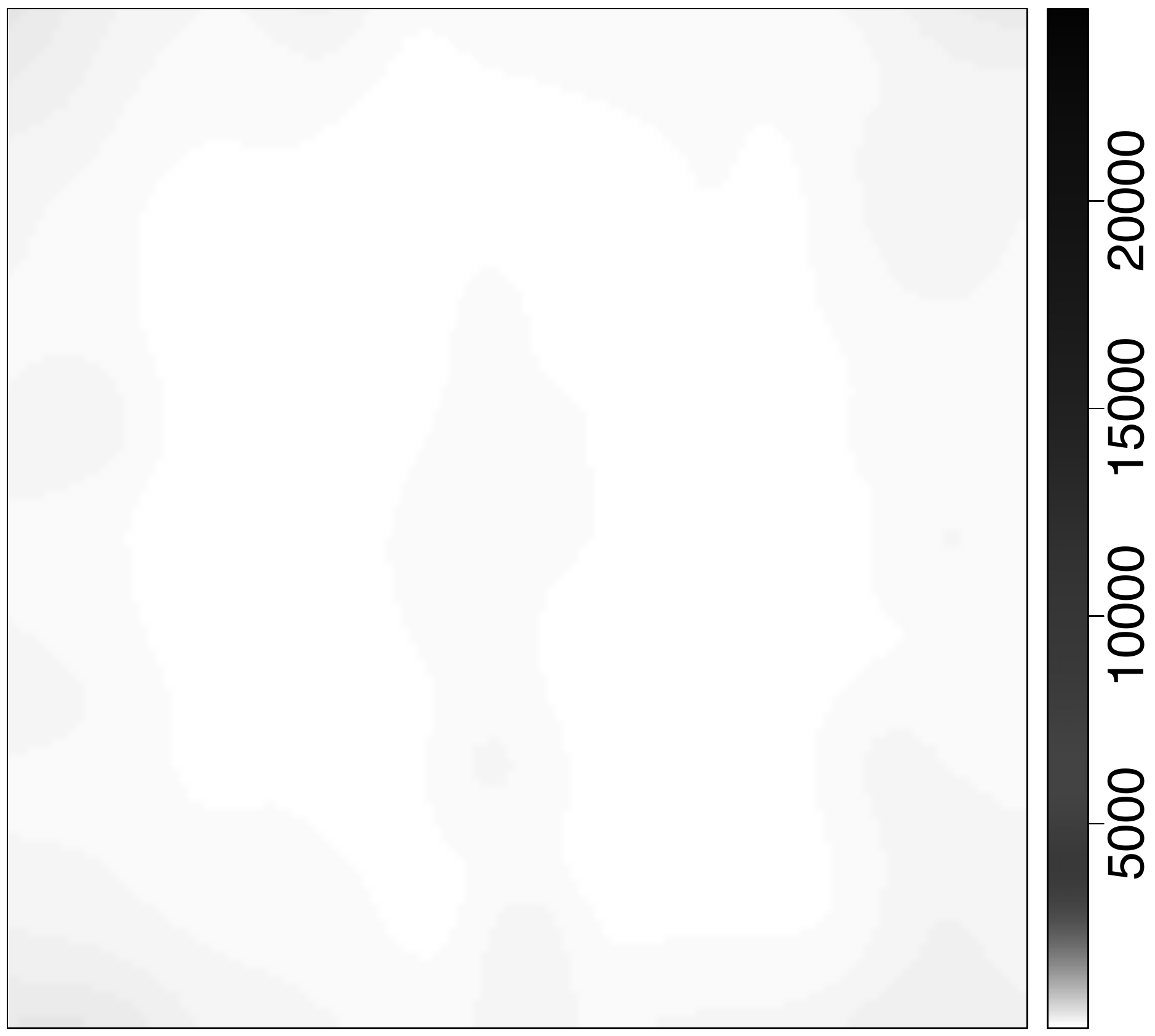}
  \qquad
  \includegraphics[width=0.35\textwidth]{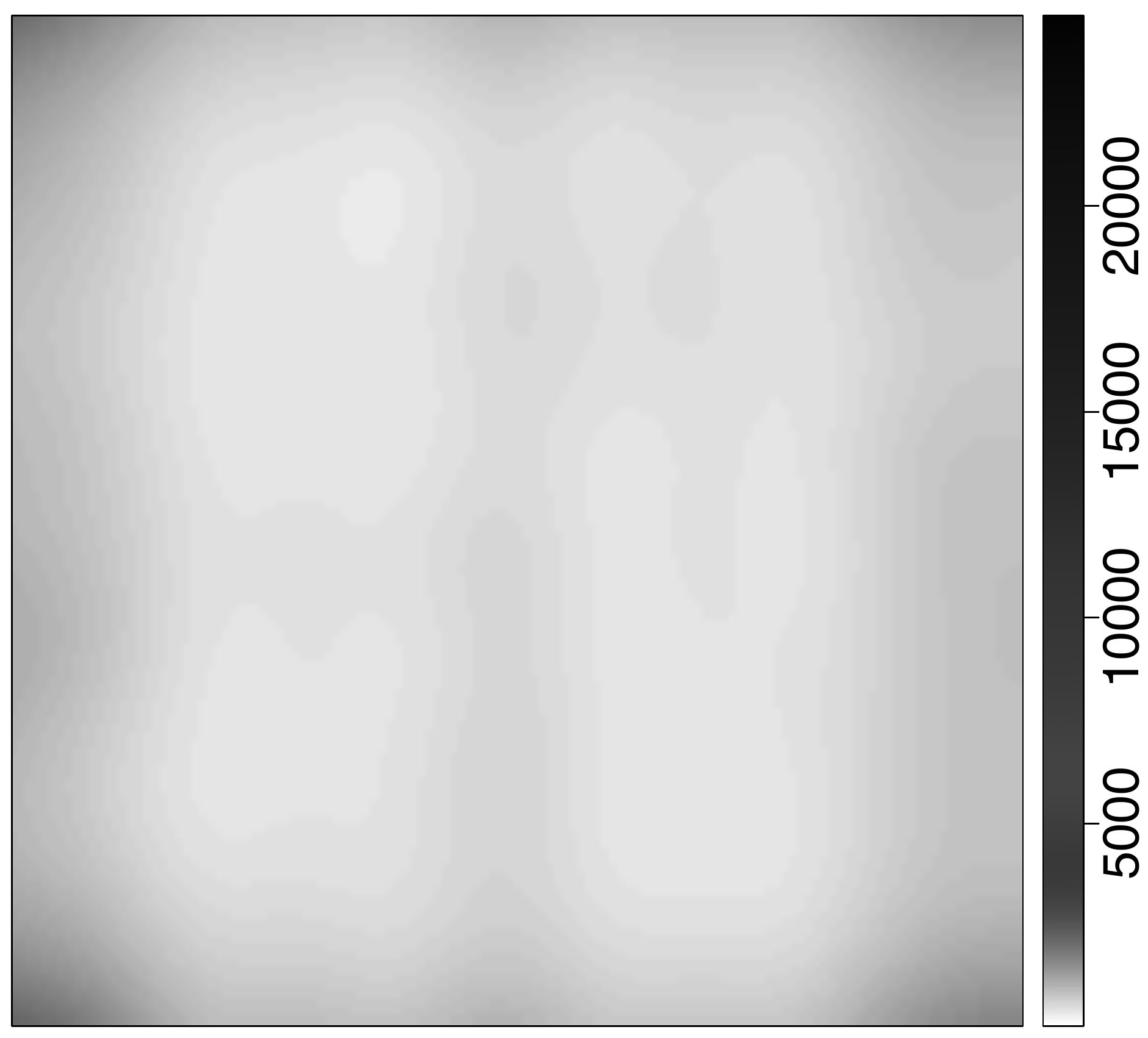}
\caption{
Estimated bias for $\widehat{\rho}_{p,m}^{V}(u)$, $u\in W=[0,1]^2$, $m=200$, and kernel estimators, based on $500$ realisations of an inhomogeneous Poisson process $X\subseteq W=[0,1]^2$ with intensity $\rho(x,y)=|10+90\sin(16x)|$. From top-left to bottom-right: $\widehat{\rho}_{p,m}^{V}(u)$ with $p=0.1,0.3,0.5,0.7,0.9,1$; kernel estimators with bandwidths selected using Poisson likelihood cross-validation \citep{BRT15,Load99} (left) and the method of \citet{cronie2018bandwidth} (right) are on the last row.
}
\label{f:VarInhomPoiR2}
\end{figure*}

\begin{figure*}[!h]
\centering
 \includegraphics[width=0.35\textwidth]{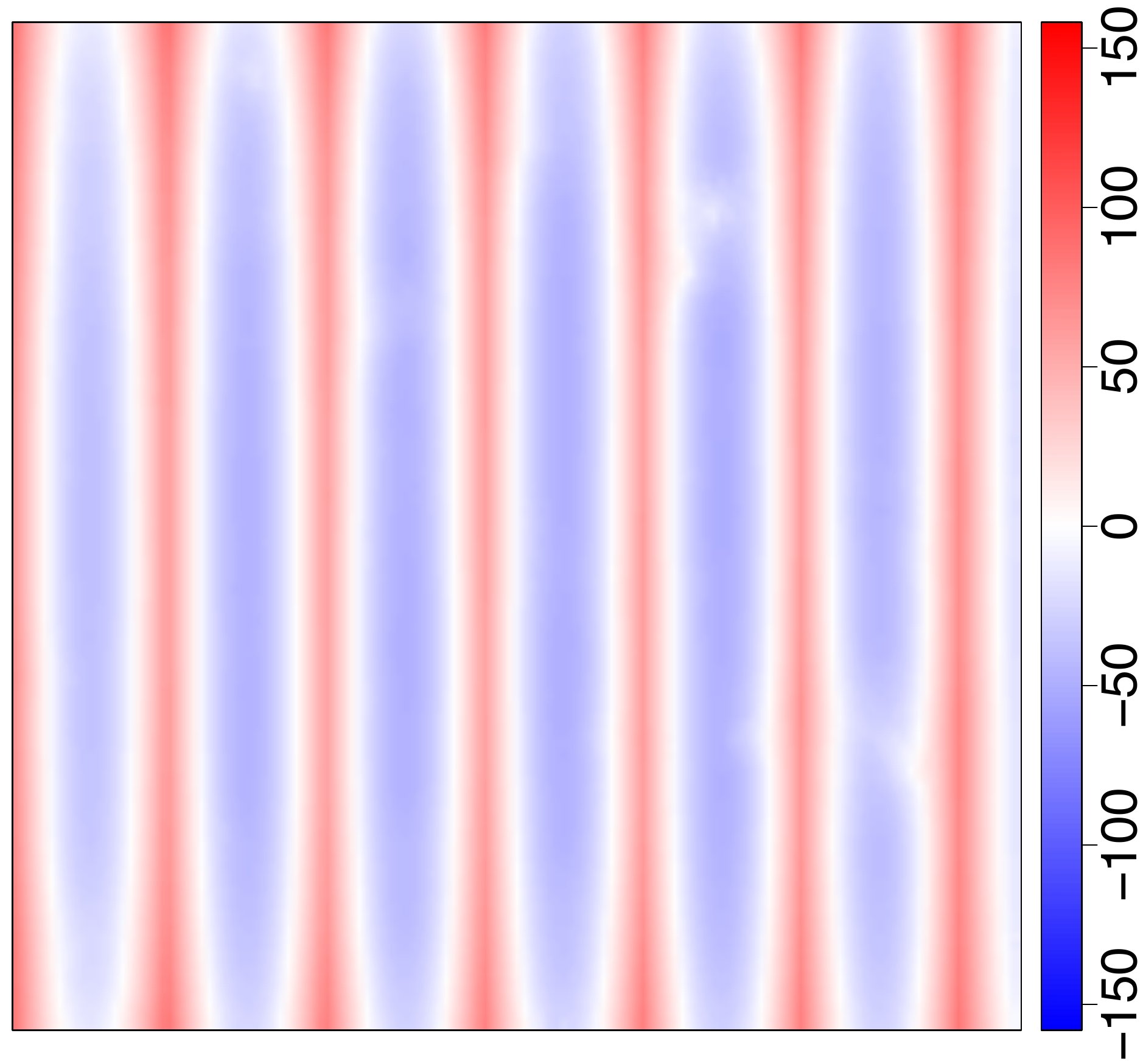}
  \qquad
 \includegraphics[width=0.35\textwidth]{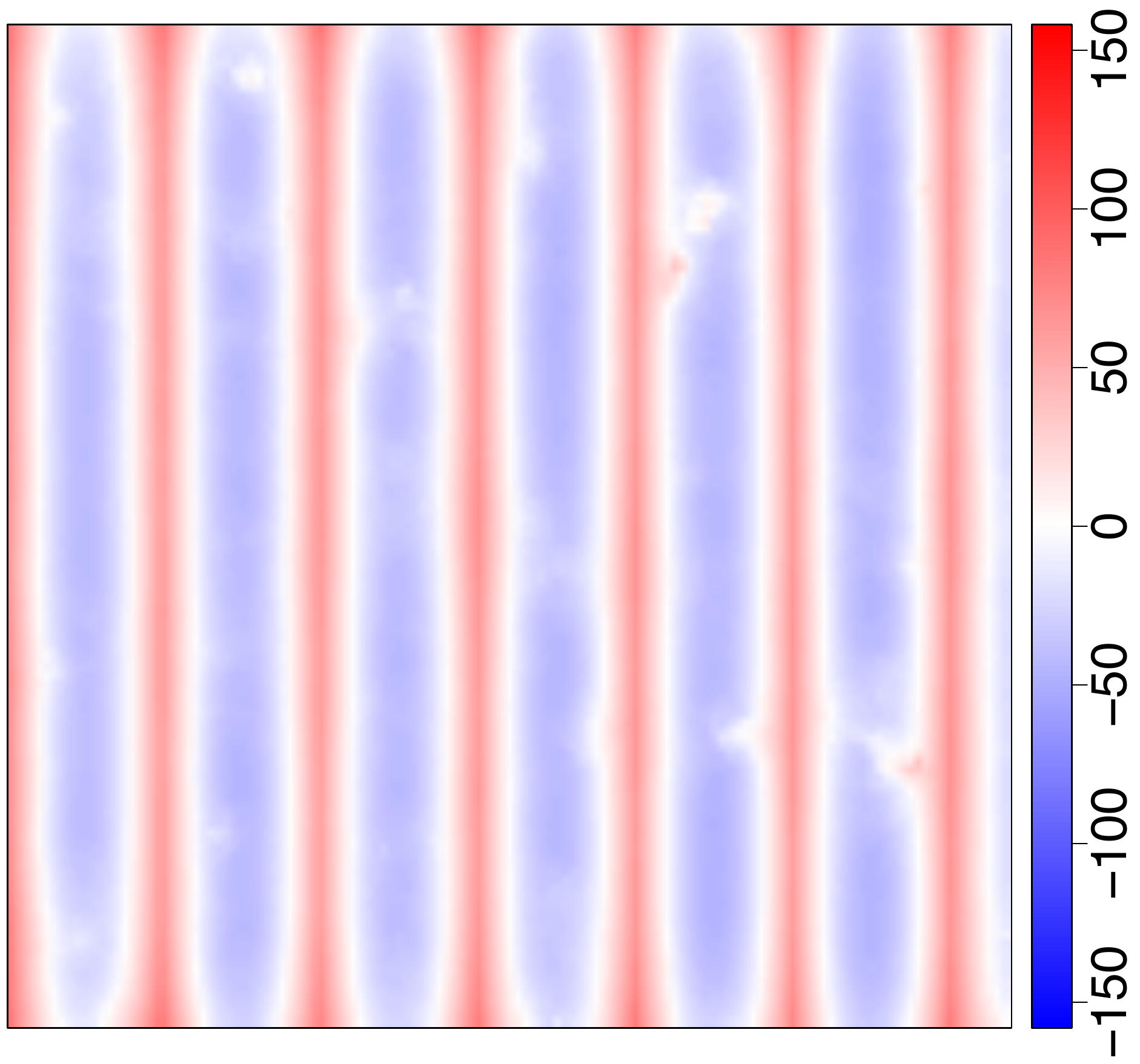}
 \qquad
  \includegraphics[width=0.35\textwidth]{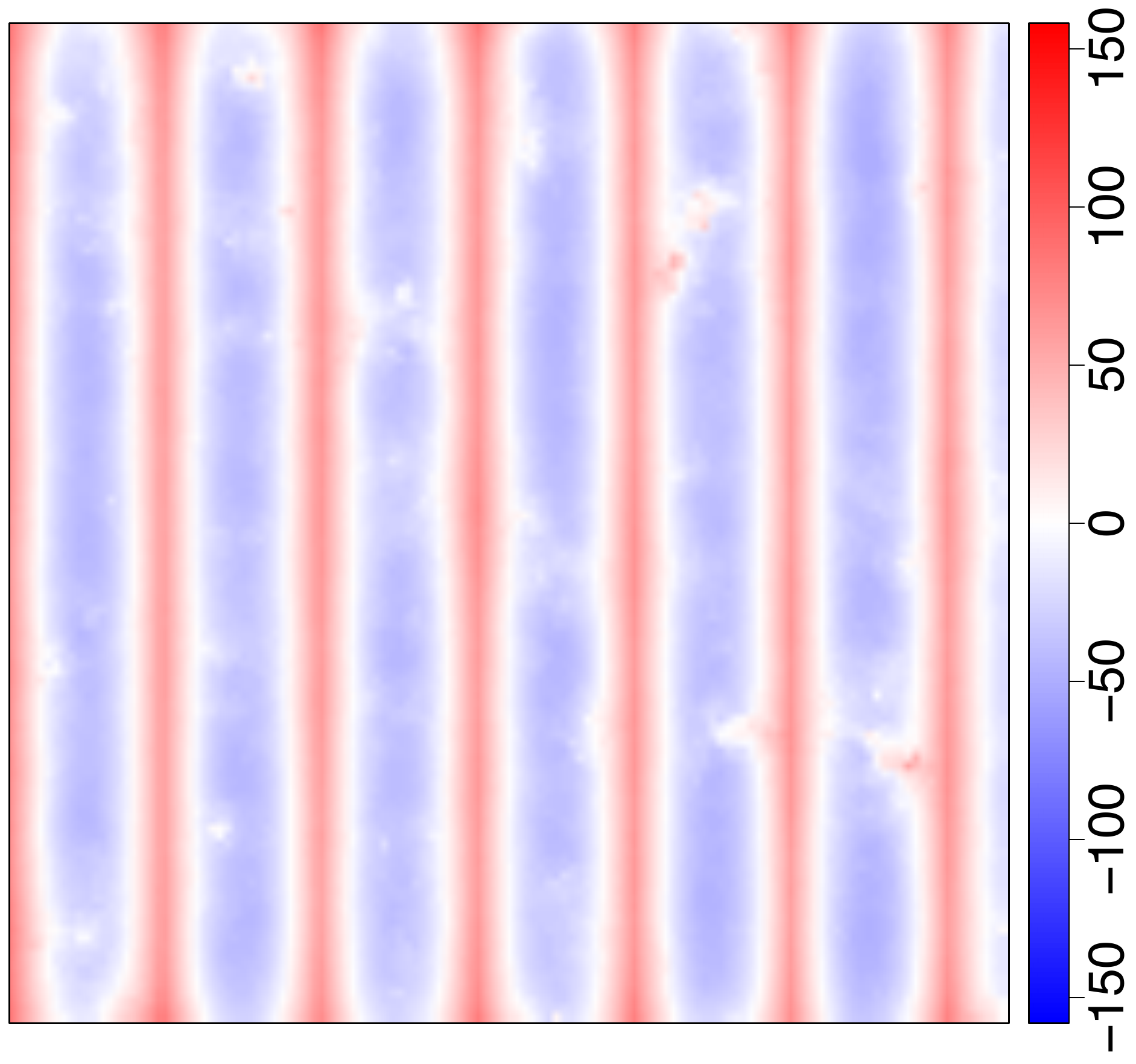}
  \qquad
 \includegraphics[width=0.35\textwidth]{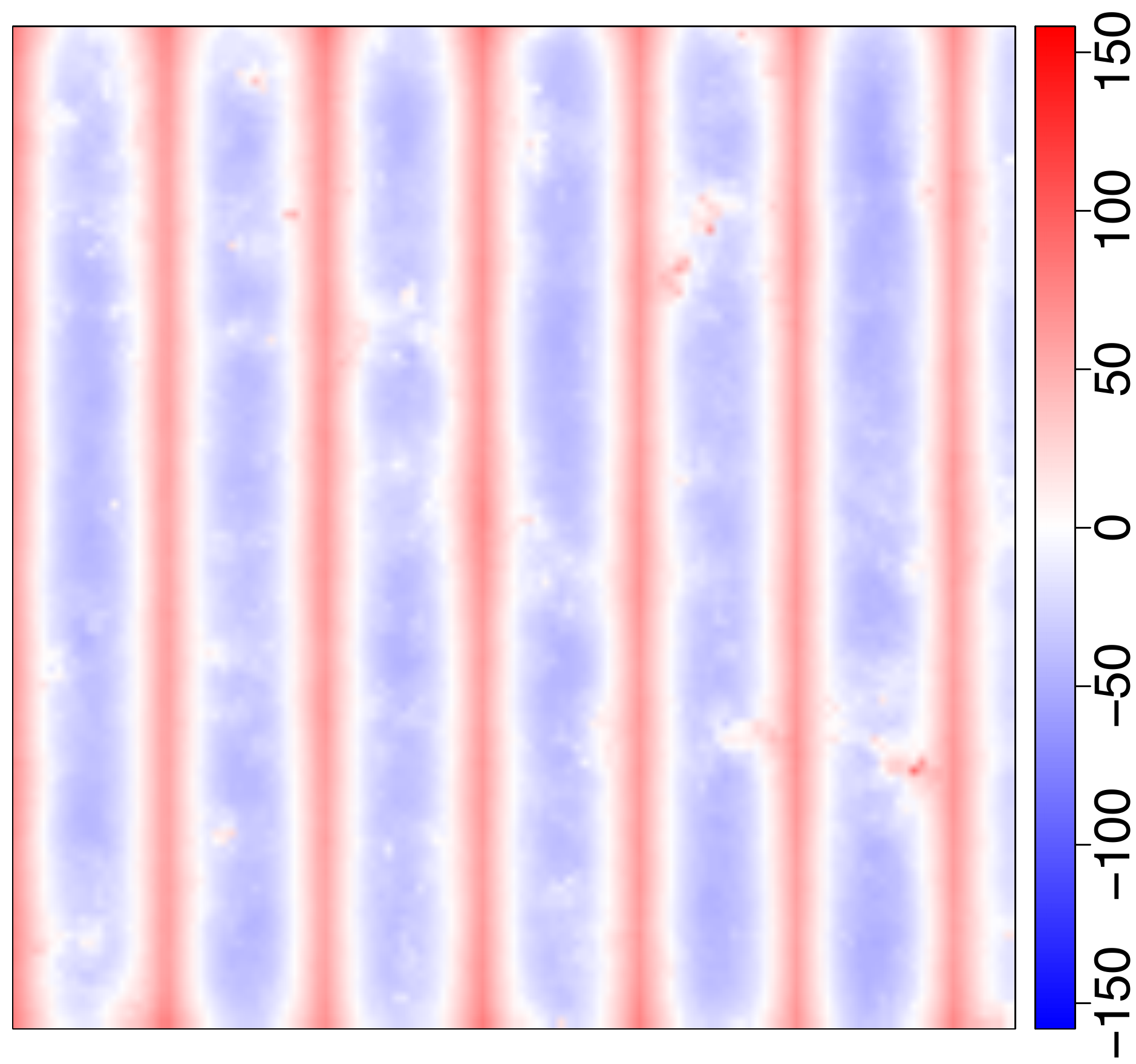}
 \qquad
  \includegraphics[width=0.35\textwidth]{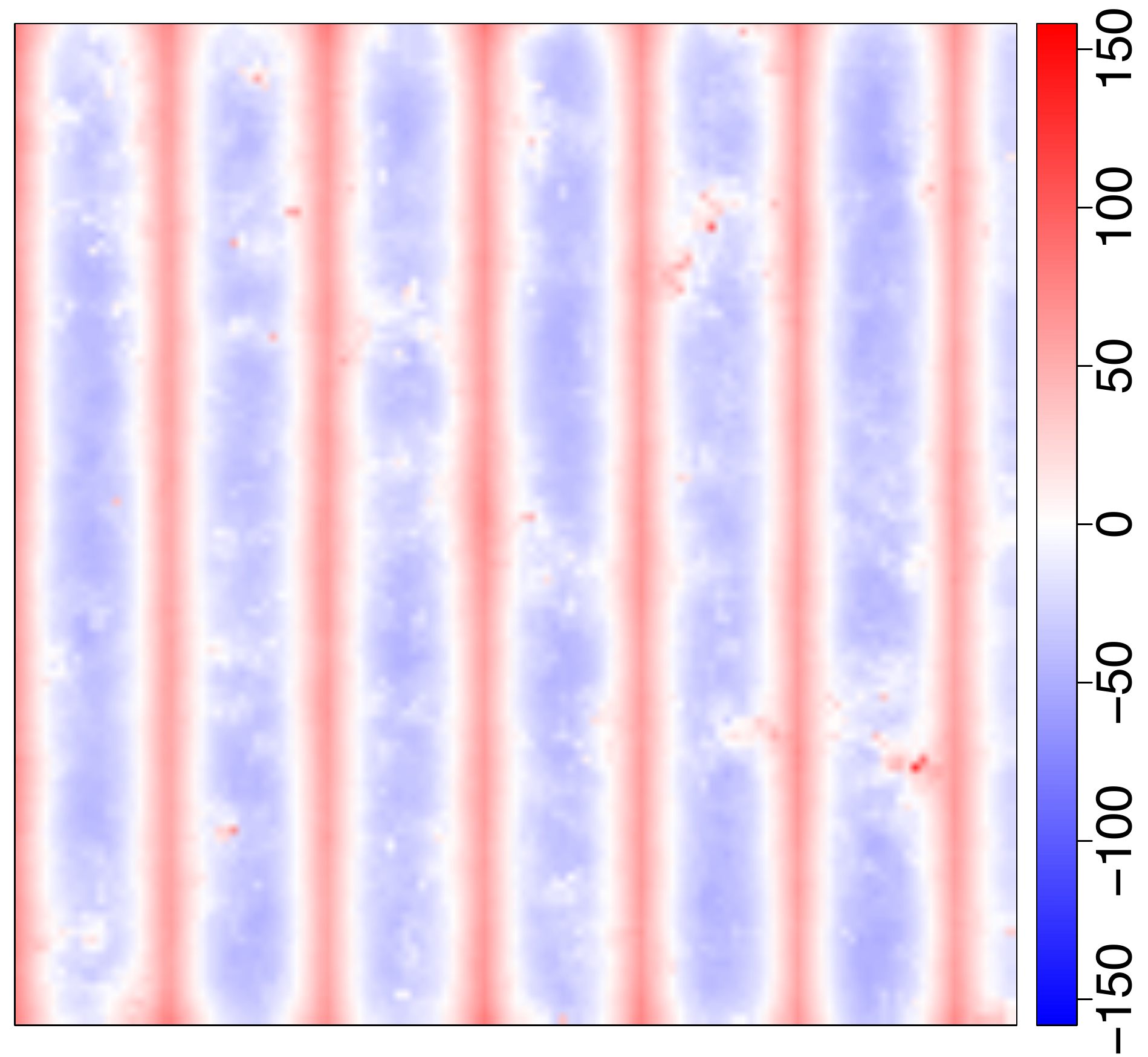}
  \qquad
  \includegraphics[width=0.35\textwidth]{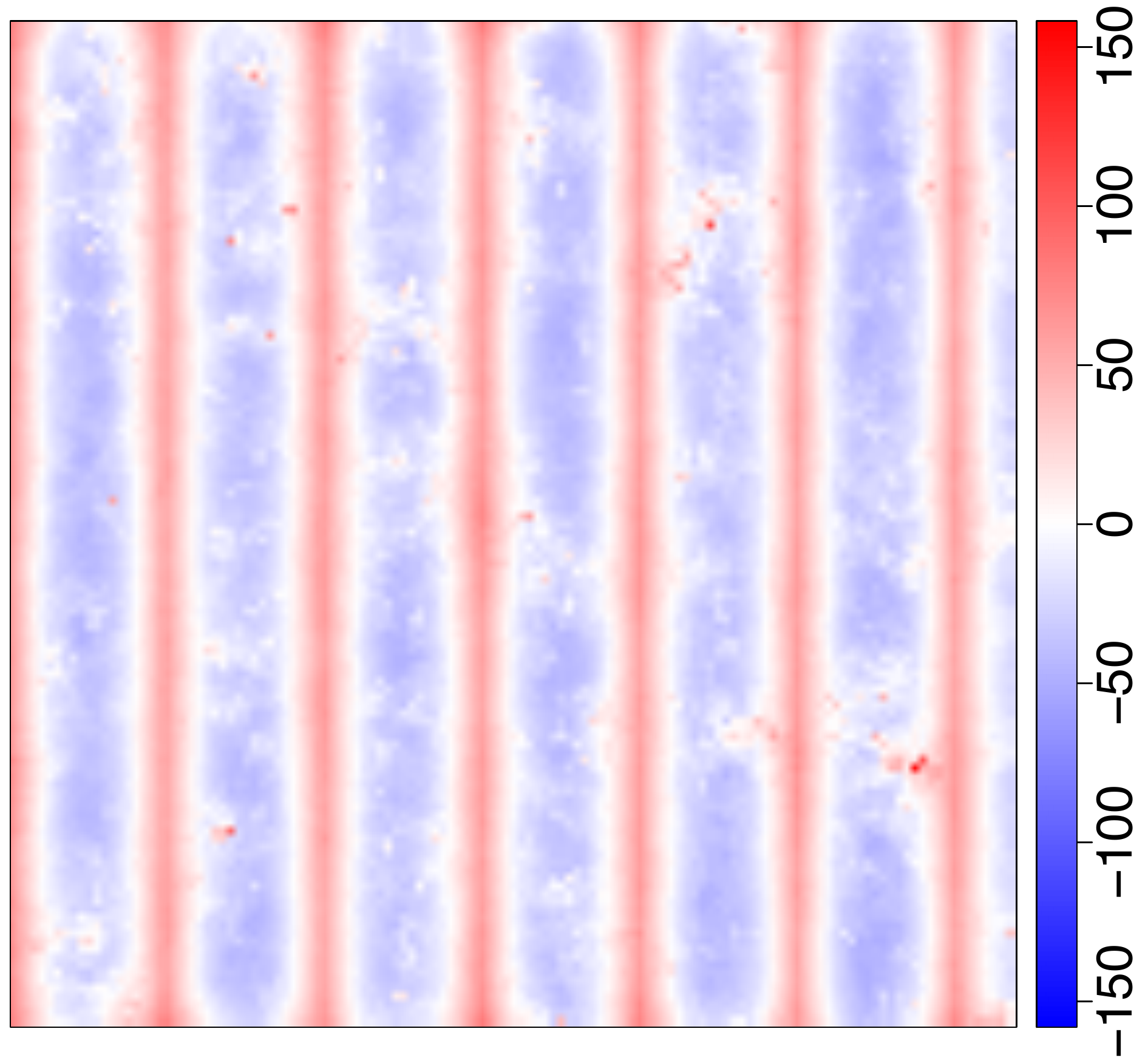}
  \qquad
  \includegraphics[width=0.35\textwidth]{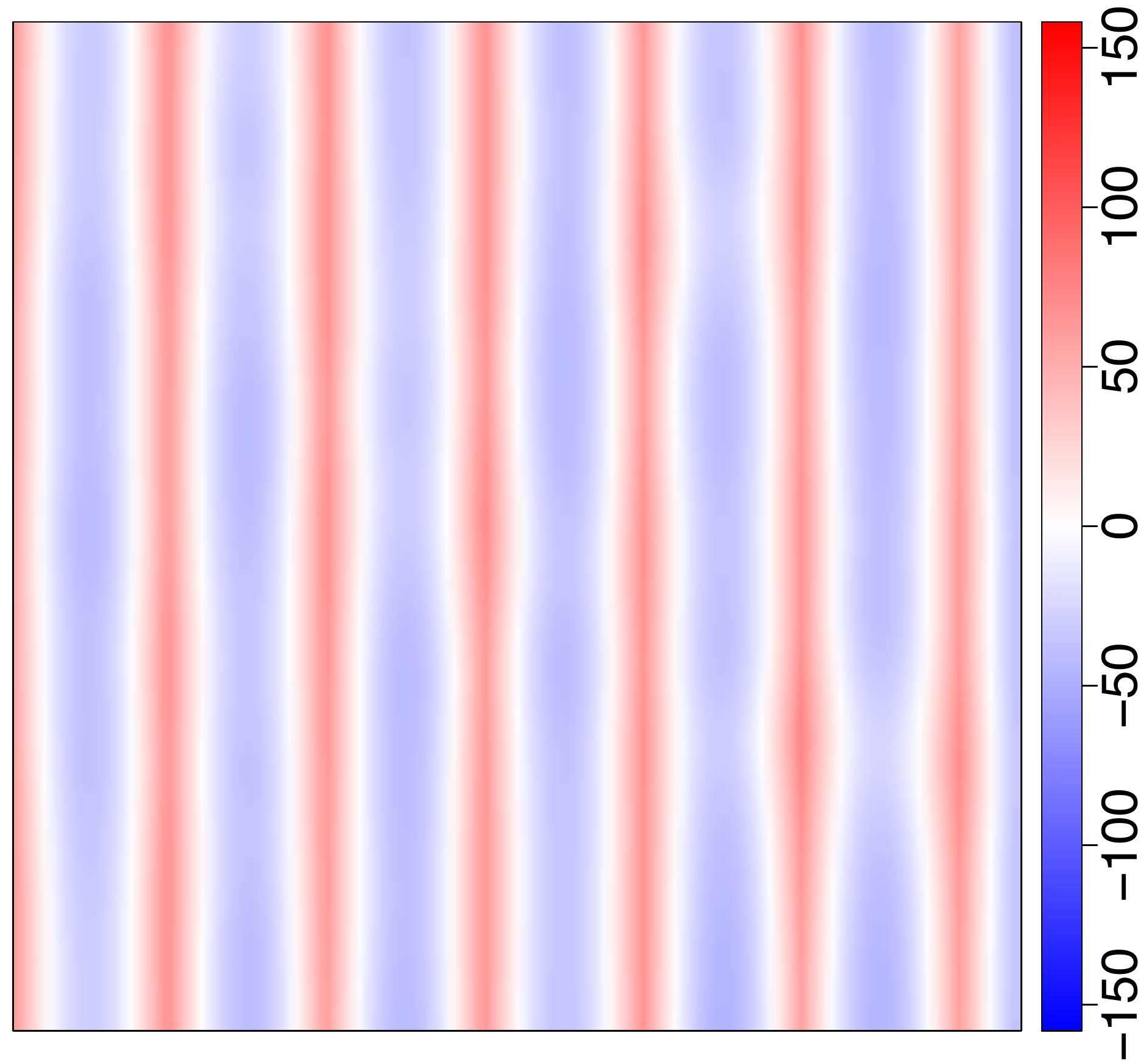}
  \qquad
  \includegraphics[width=0.35\textwidth]{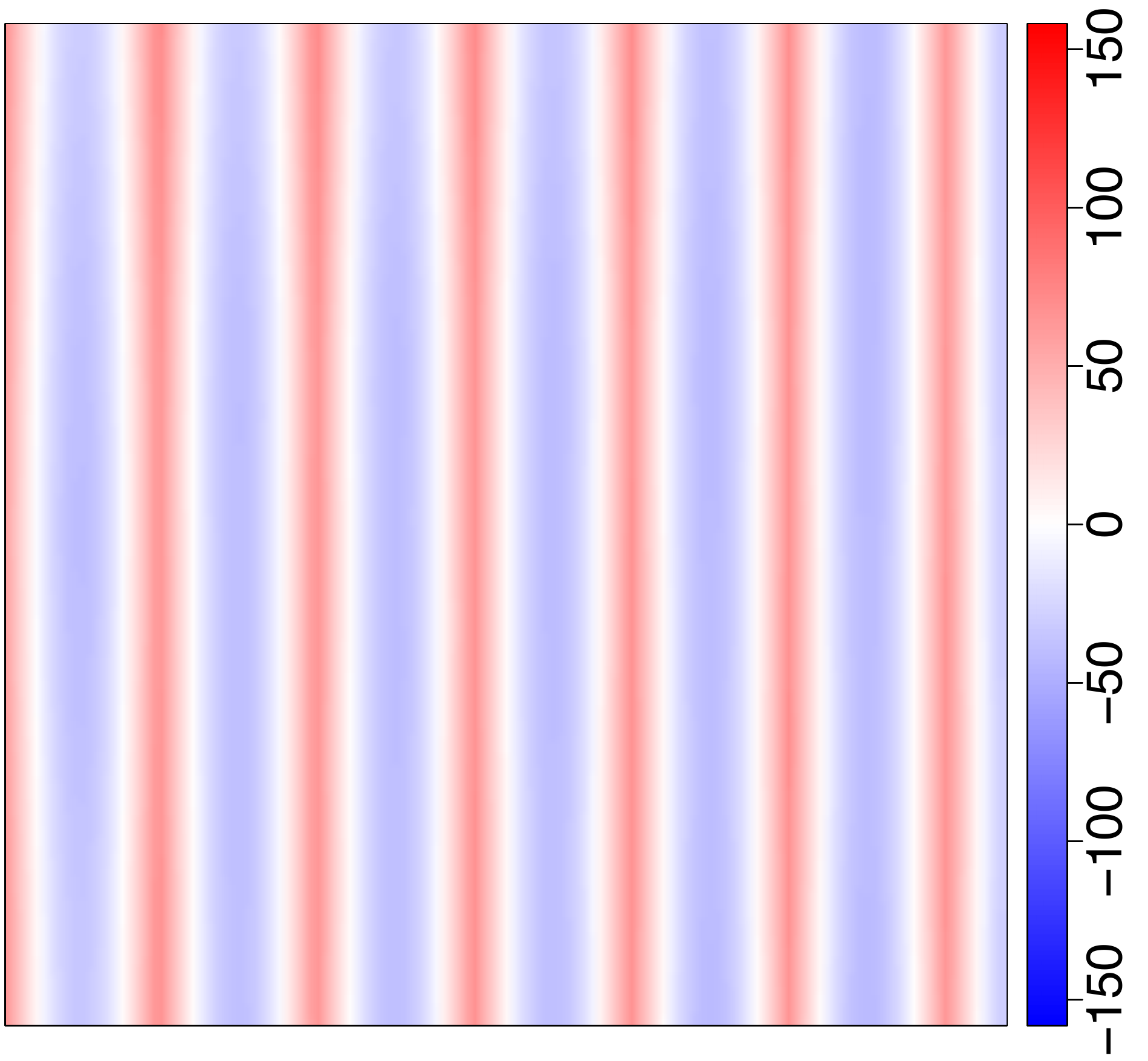}
\caption{
Estimated bias for $\widehat{\rho}_{p,m}^{V}(u)$, $u\in W=[0,1]^2$, $m=200$, and kernel estimators, based on $500$ realisations of a log-Gaussian Cox process $X\subseteq W=[0,1]^2$ where the driving Gaussian random field has mean function $(x,y)\mapsto\log(40|\sin(20x)|)$ and covariance function $((x_1,y_1),(x_2,y_2))\mapsto2\exp\{-\|(x_1,y_1)-(x_2,y_2)\|/0.1\}$. From top-left to bottom-right: $\widehat{\rho}_{p,m}^{V}(u)$ with $p=0.1,0.3,0.5,0.7,0.9,1$; kernel estimators with bandwidths selected using Poisson likelihood cross-validation \citep{BRT15,Load99} (left) and the method of \citet{cronie2018bandwidth} (right) are on the last row.
}
\label{f:BiasLGCPR2}
\end{figure*}

\begin{figure*}[!h]
\centering
 \includegraphics[width=0.35\textwidth]{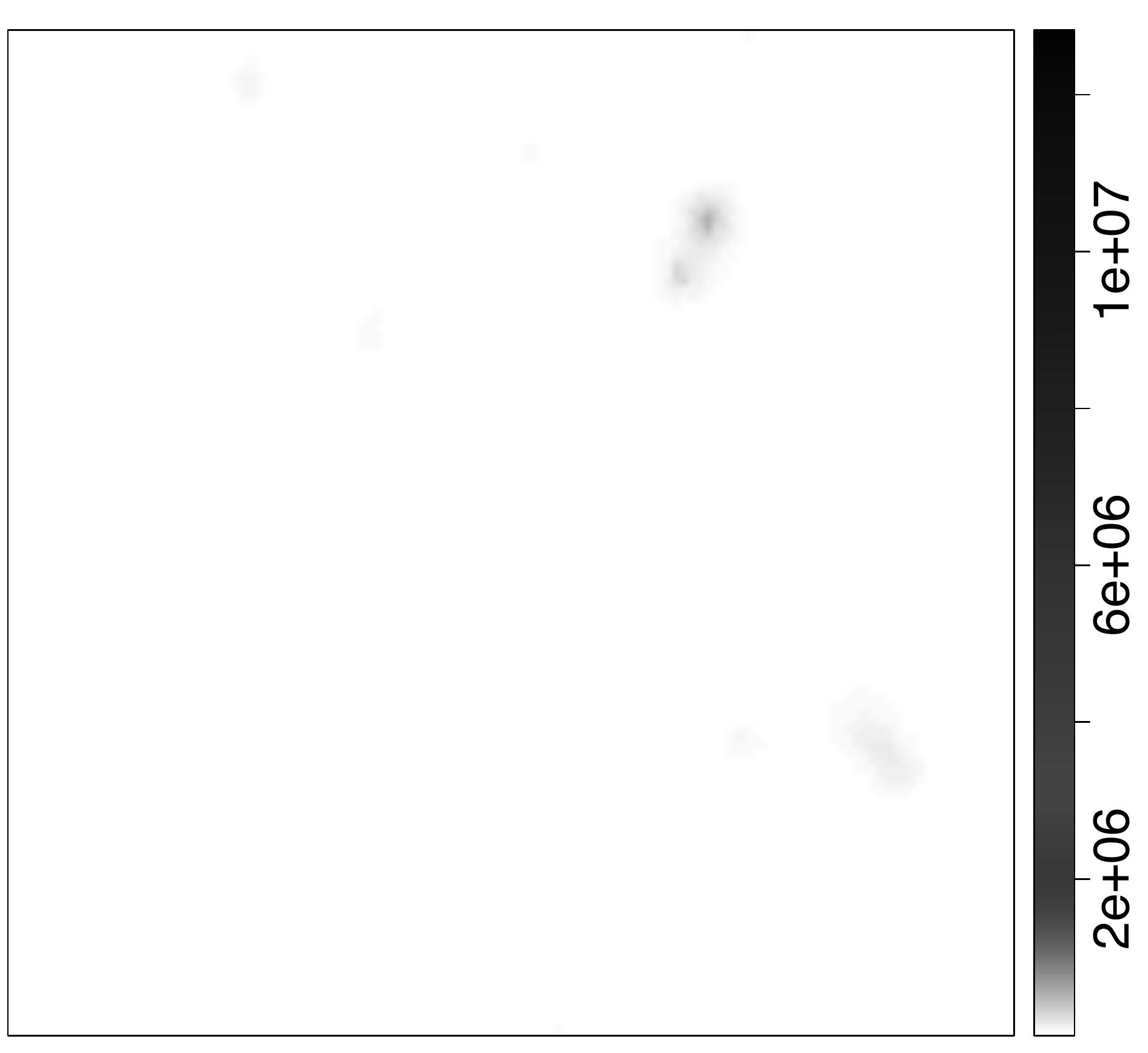}
  \qquad
 \includegraphics[width=0.35\textwidth]{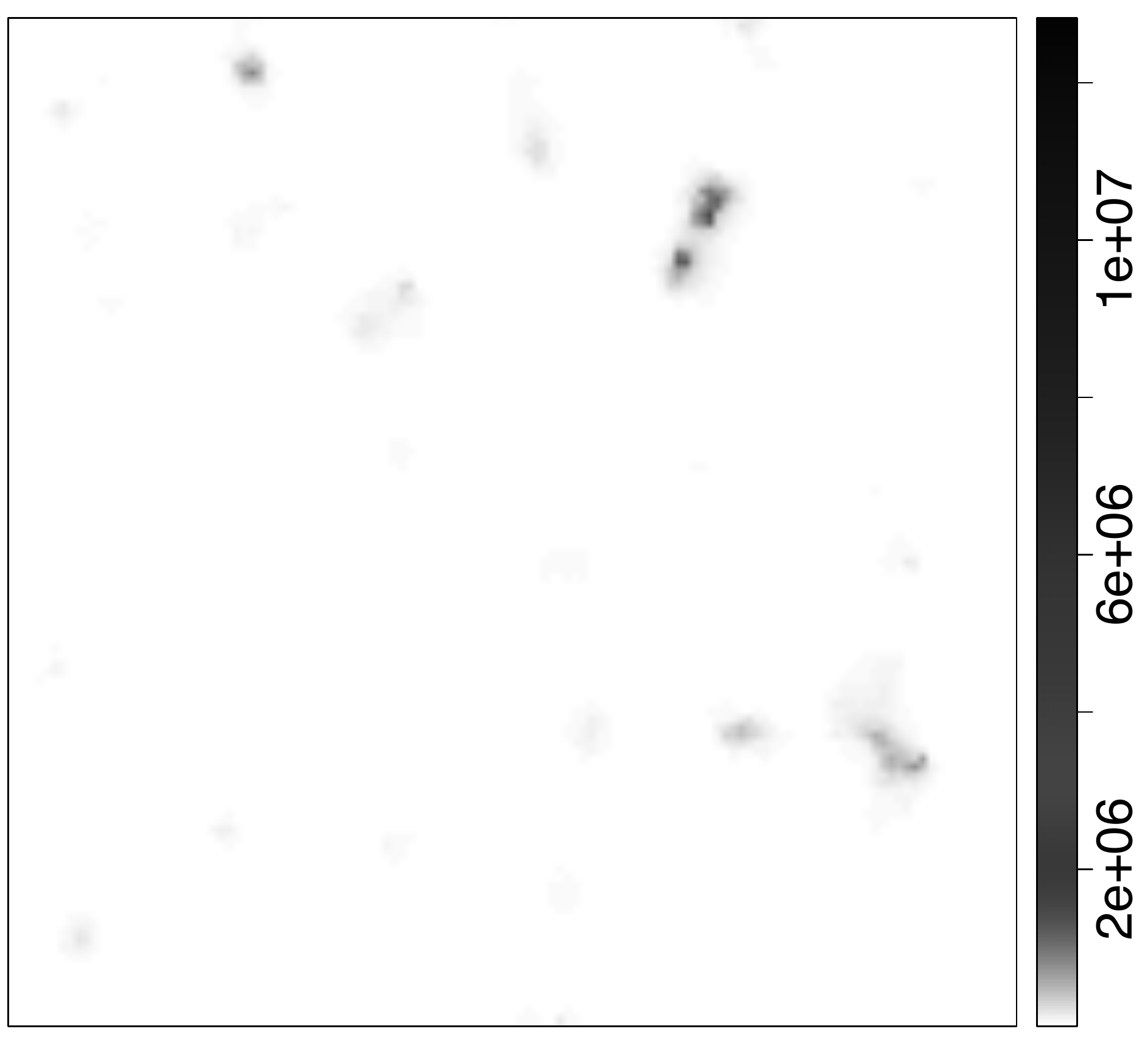}
 \qquad
  \includegraphics[width=0.35\textwidth]{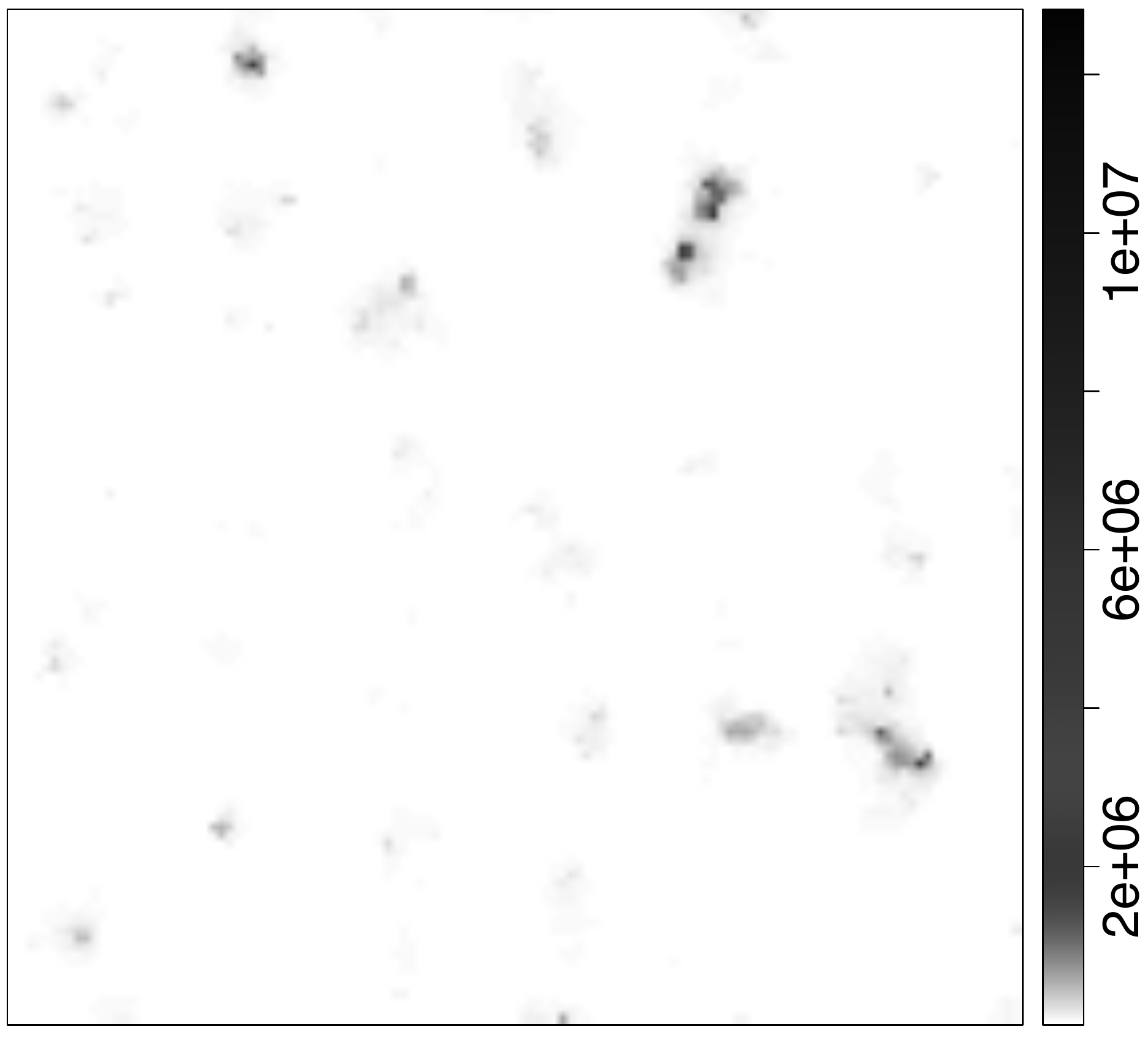}
  \qquad
 \includegraphics[width=0.35\textwidth]{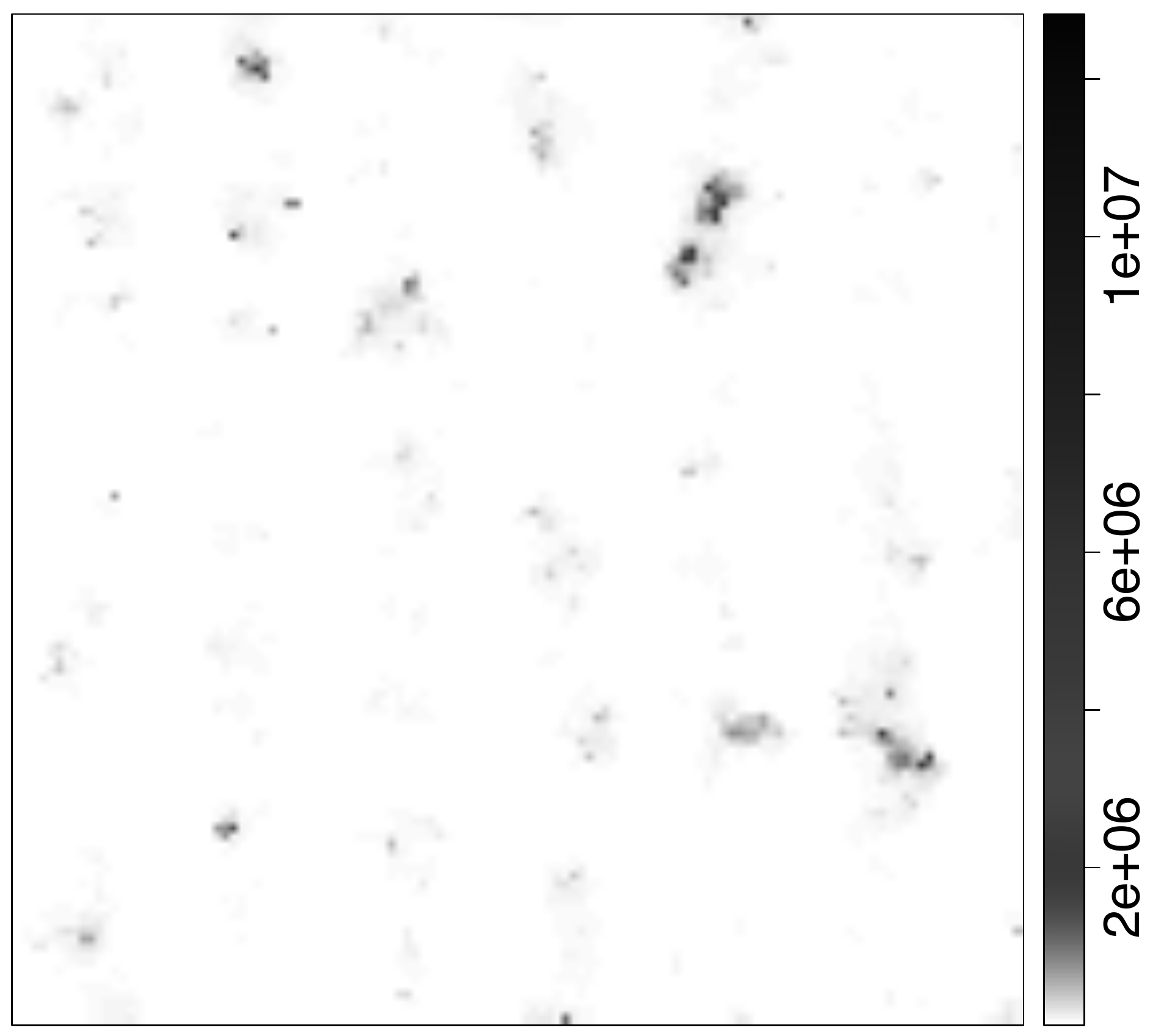}
 \qquad
  \includegraphics[width=0.35\textwidth]{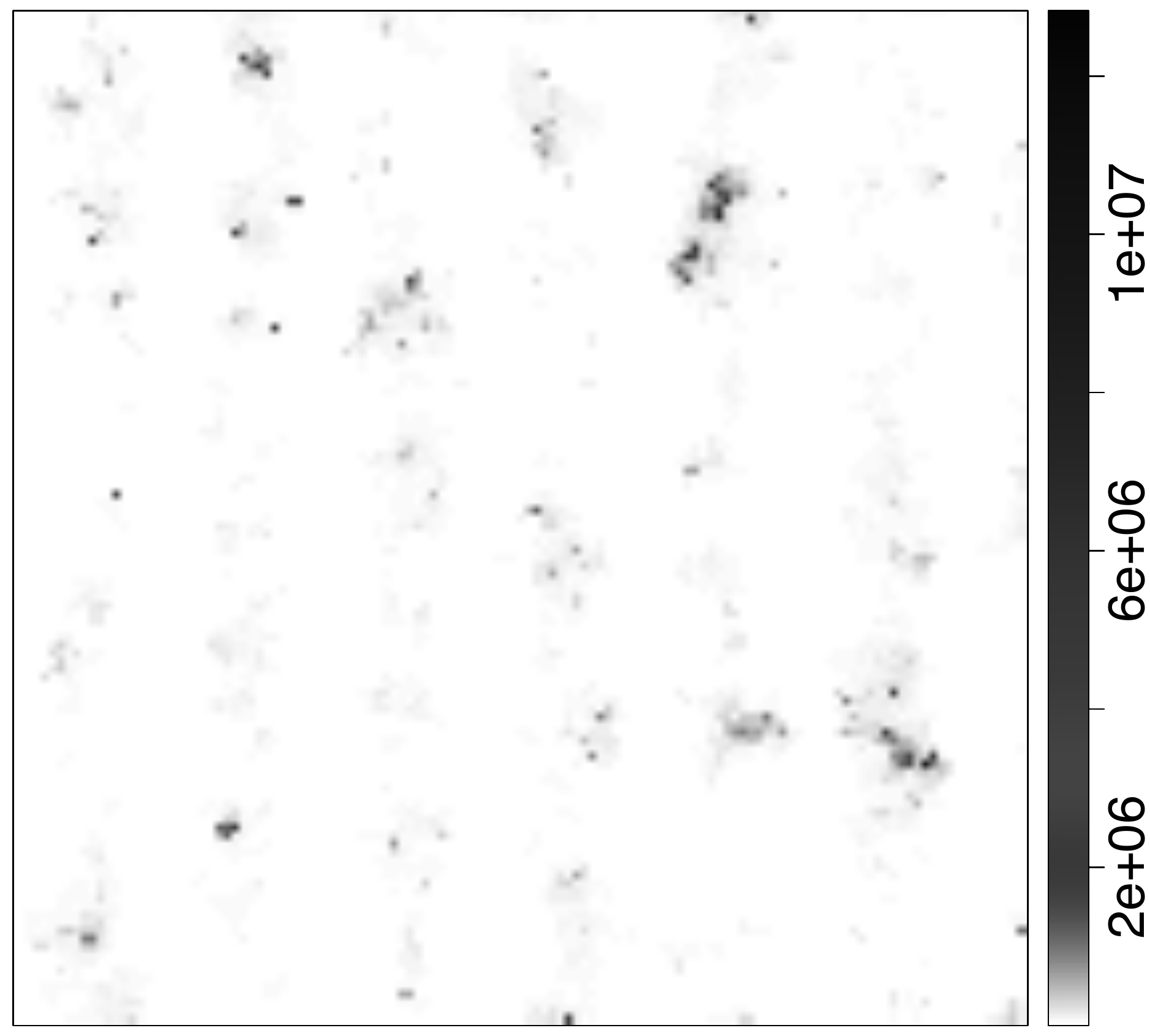}
  \qquad
  \includegraphics[width=0.35\textwidth]{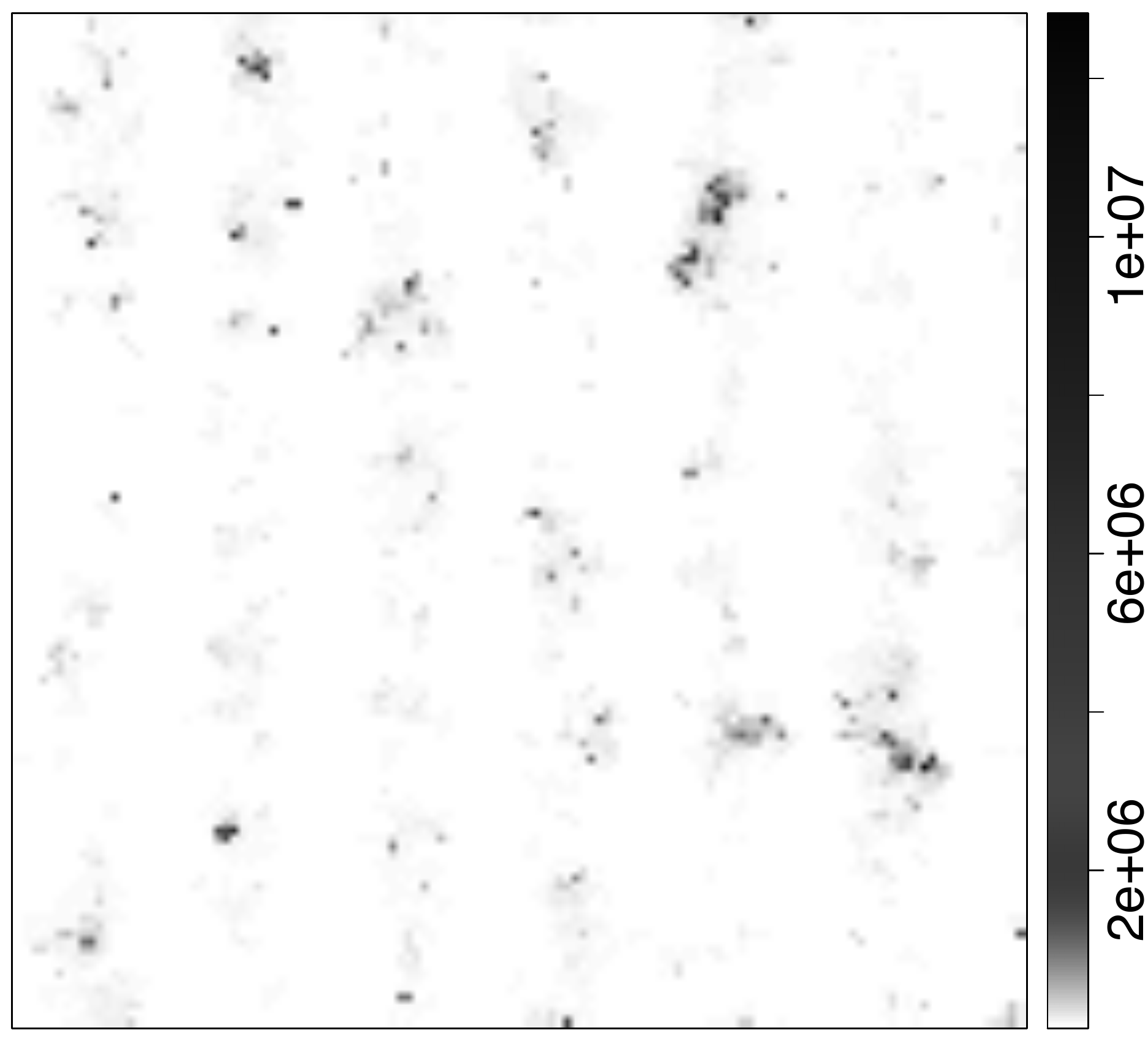}
  \qquad
  \includegraphics[width=0.35\textwidth]{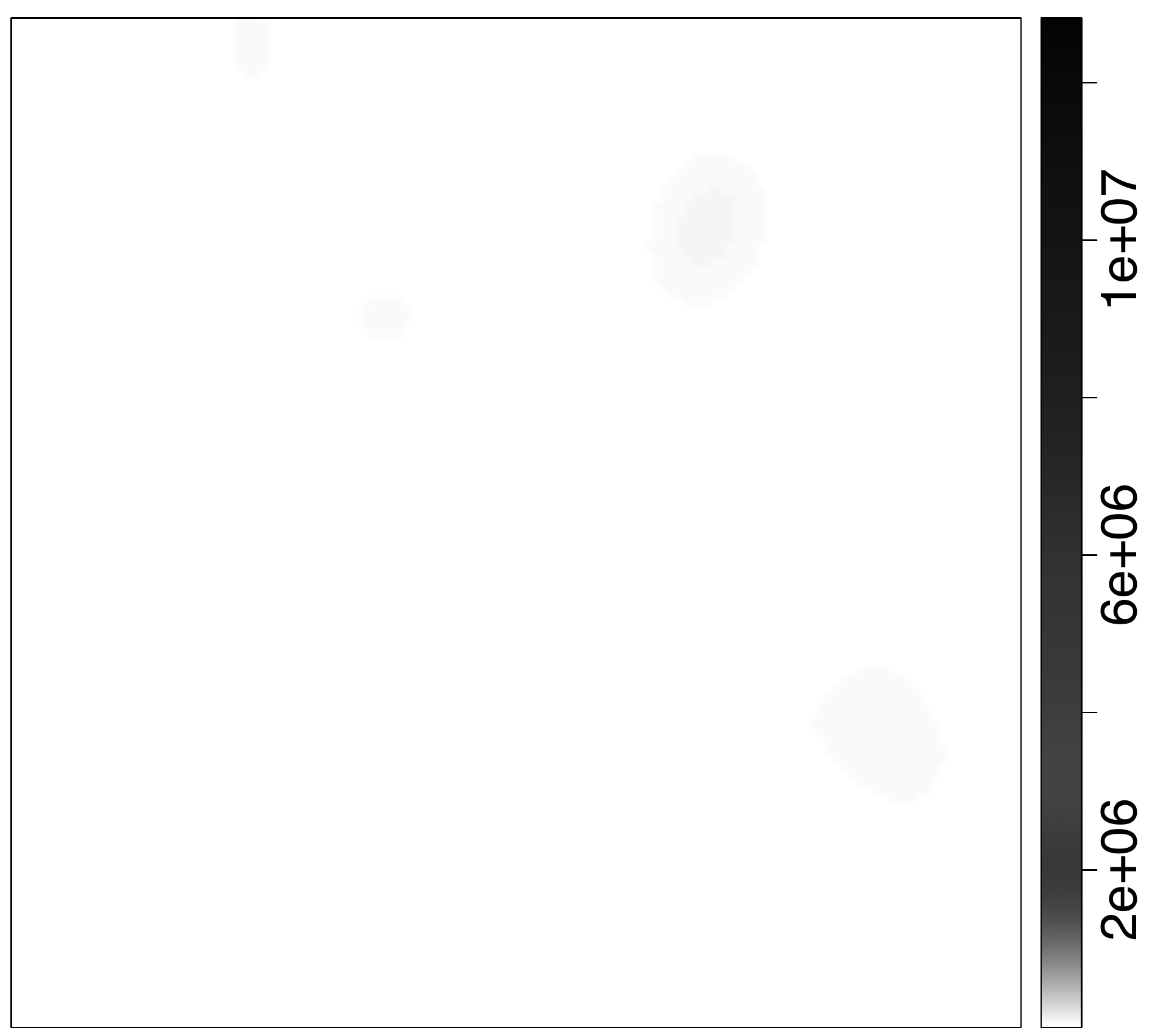}
  \qquad
  \includegraphics[width=0.35\textwidth]{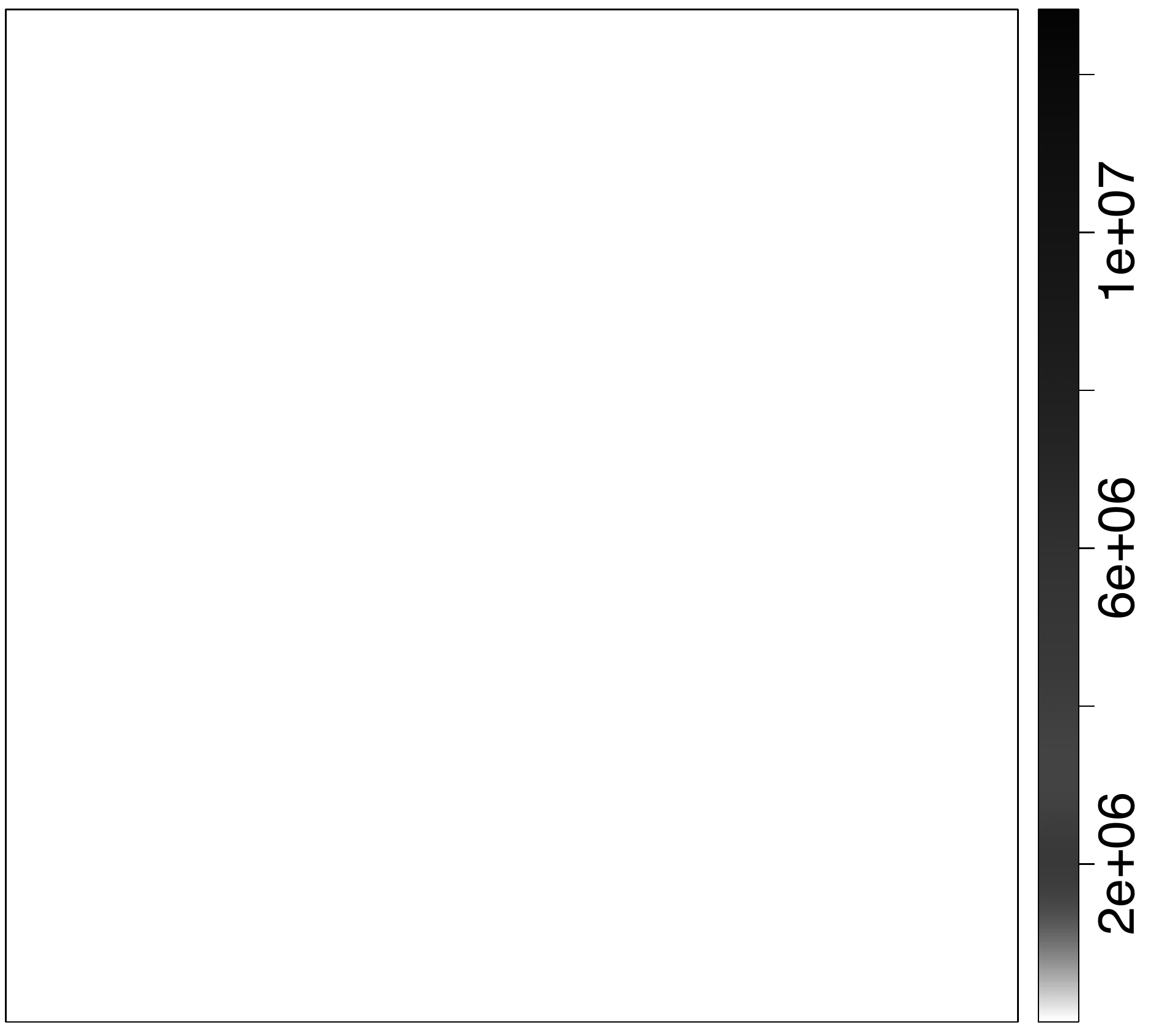}
\caption{
Estimated variance for $\widehat{\rho}_{p,m}^{V}(u)$, $u\in W=[0,1]^2$, $m=200$, and kernel estimators, based on $500$ realisations of a log-Gaussian Cox process $X\subseteq W=[0,1]^2$ where the driving Gaussian random field has mean function $(x,y)\mapsto\log(40|\sin(20x)|)$ and covariance function $((x_1,y_1),(x_2,y_2))\mapsto2\exp\{-\|(x_1,y_1)-(x_2,y_2)\|/0.1\}$. From top-left to bottom-right: $\widehat{\rho}_{p,m}^{V}(u)$ with $p=0.1,0.3,0.5,0.7,0.9,1$; kernel estimators with bandwidths selected using Poisson likelihood cross-validation \citep{BRT15,Load99} (left) and the method of \citet{cronie2018bandwidth} (right) are on the last row.
}
\label{f:VarLGCPR2}
\end{figure*}

\begin{figure*}[!h]
\centering
 \includegraphics[width=0.35\textwidth]{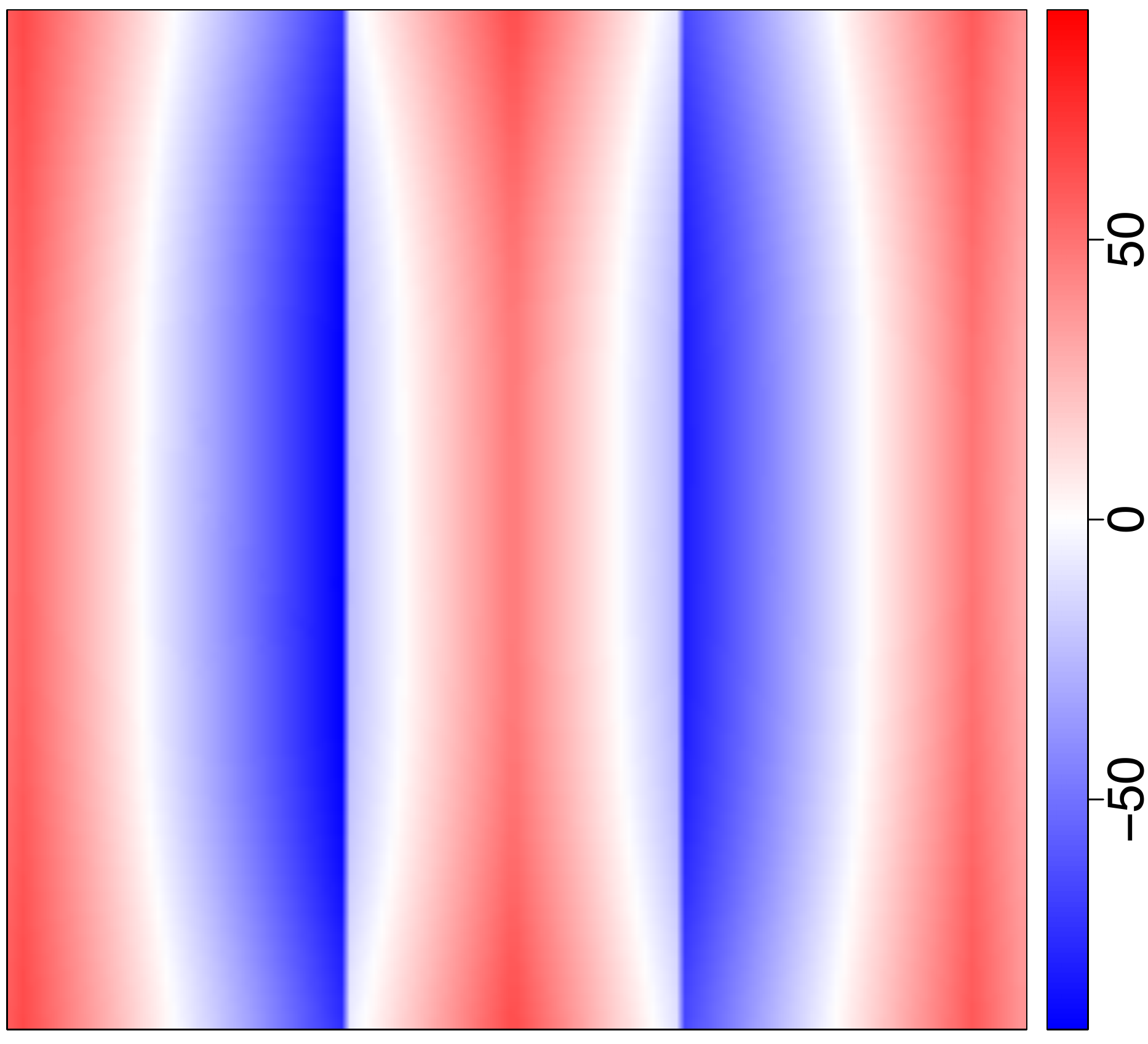}
  \qquad
 \includegraphics[width=0.35\textwidth]{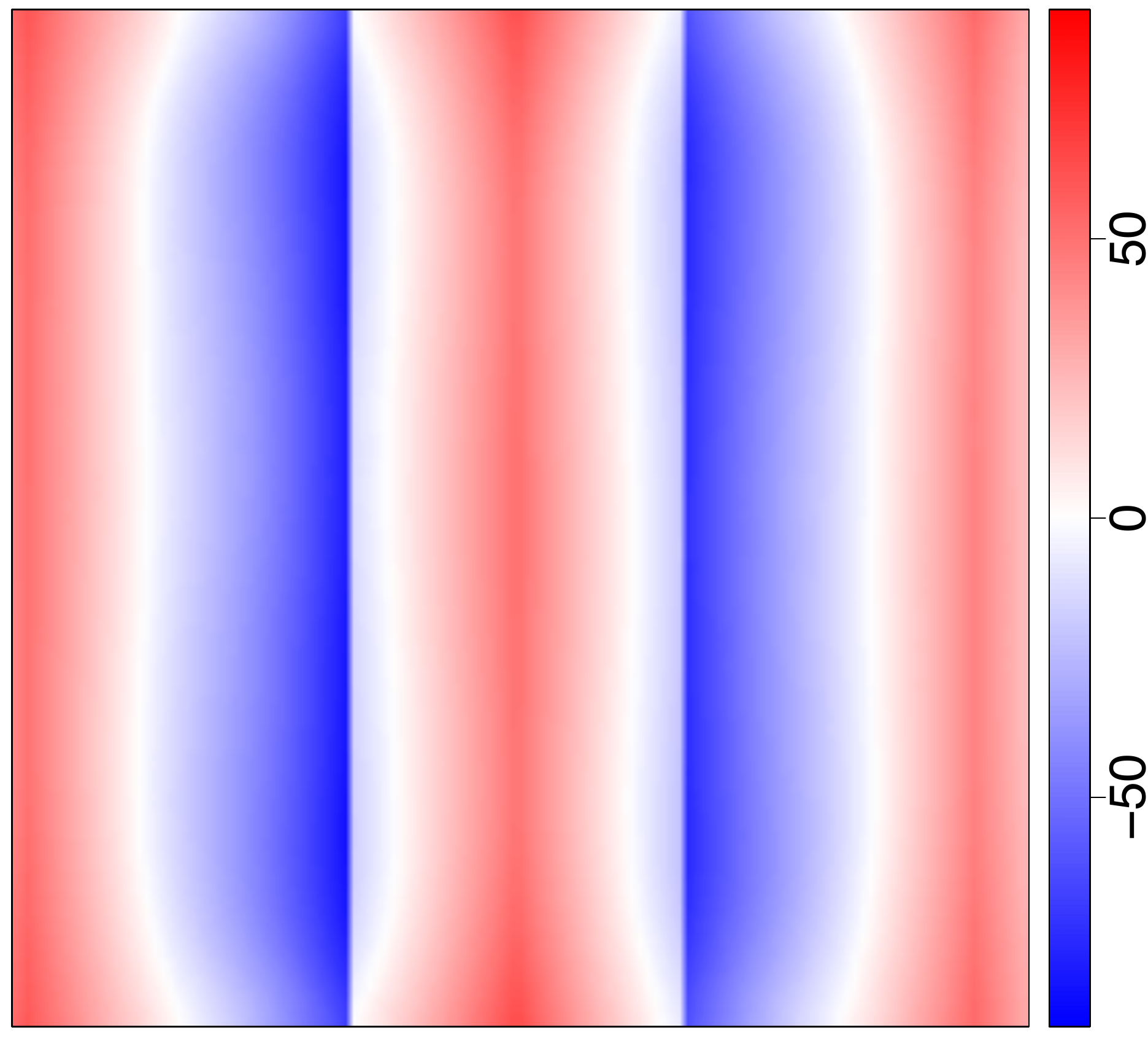}
 \qquad
  \includegraphics[width=0.35\textwidth]{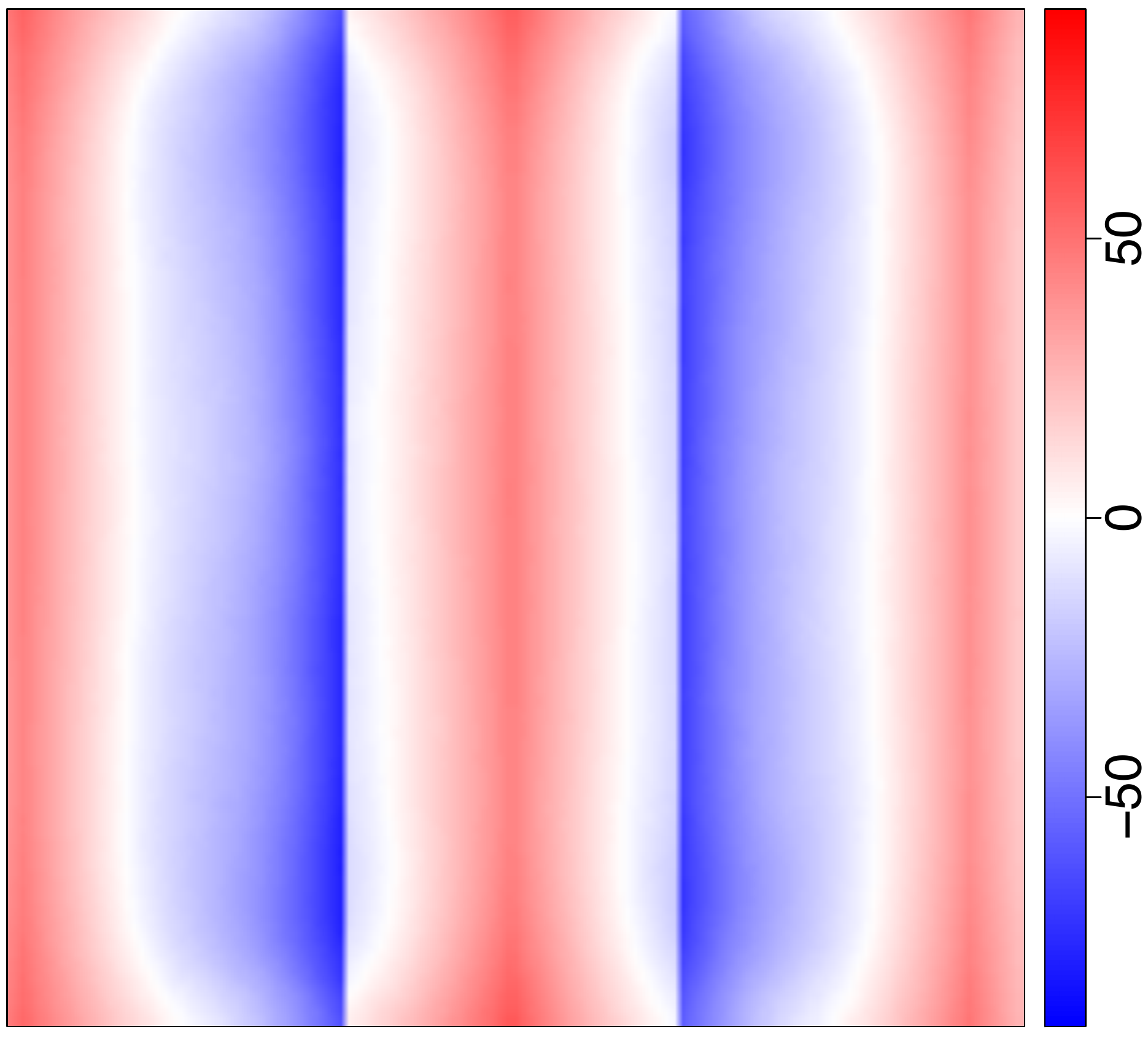}
  \qquad
 \includegraphics[width=0.35\textwidth]{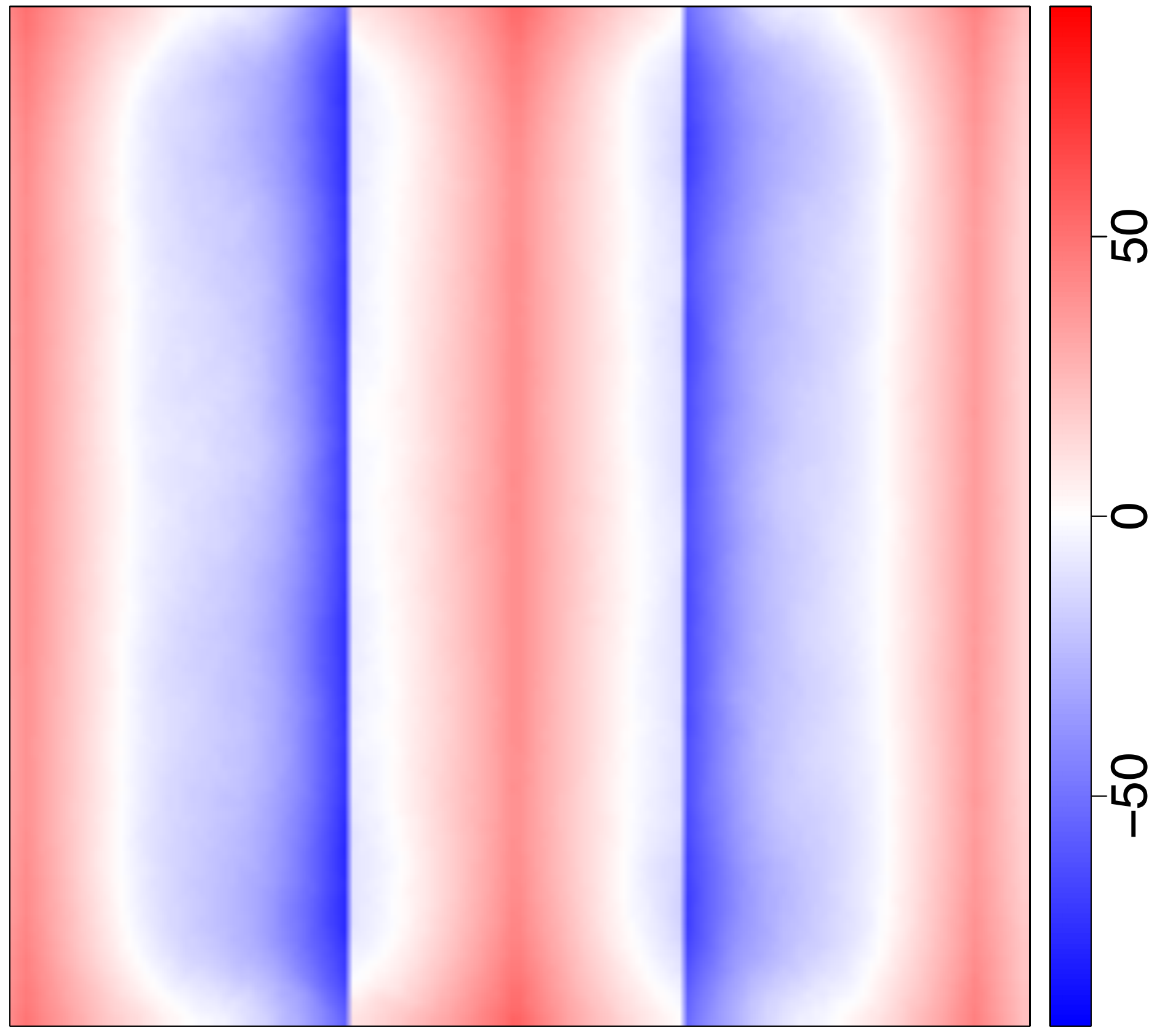}
 \qquad
  \includegraphics[width=0.35\textwidth]{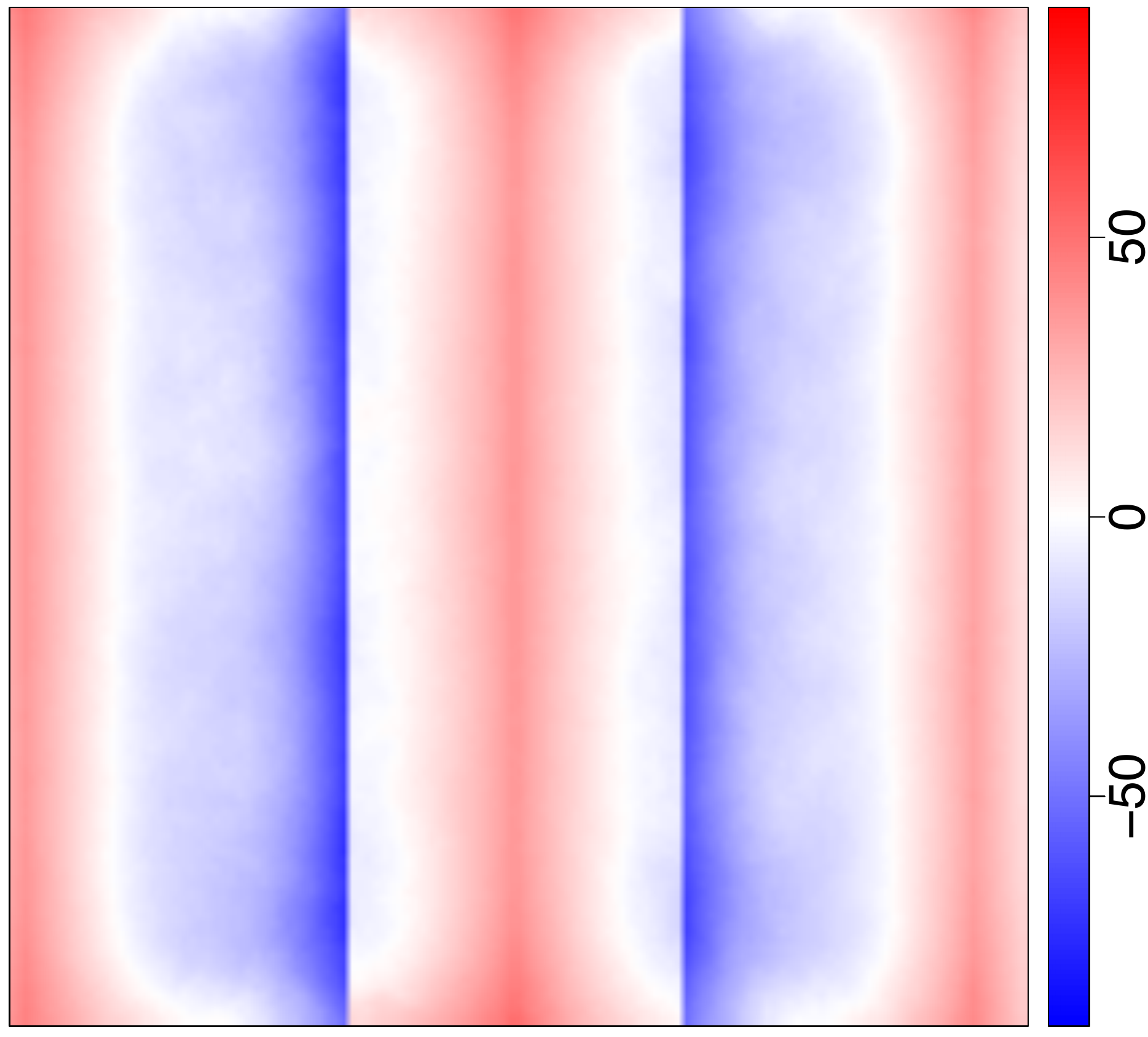}
  \qquad
  \includegraphics[width=0.35\textwidth]{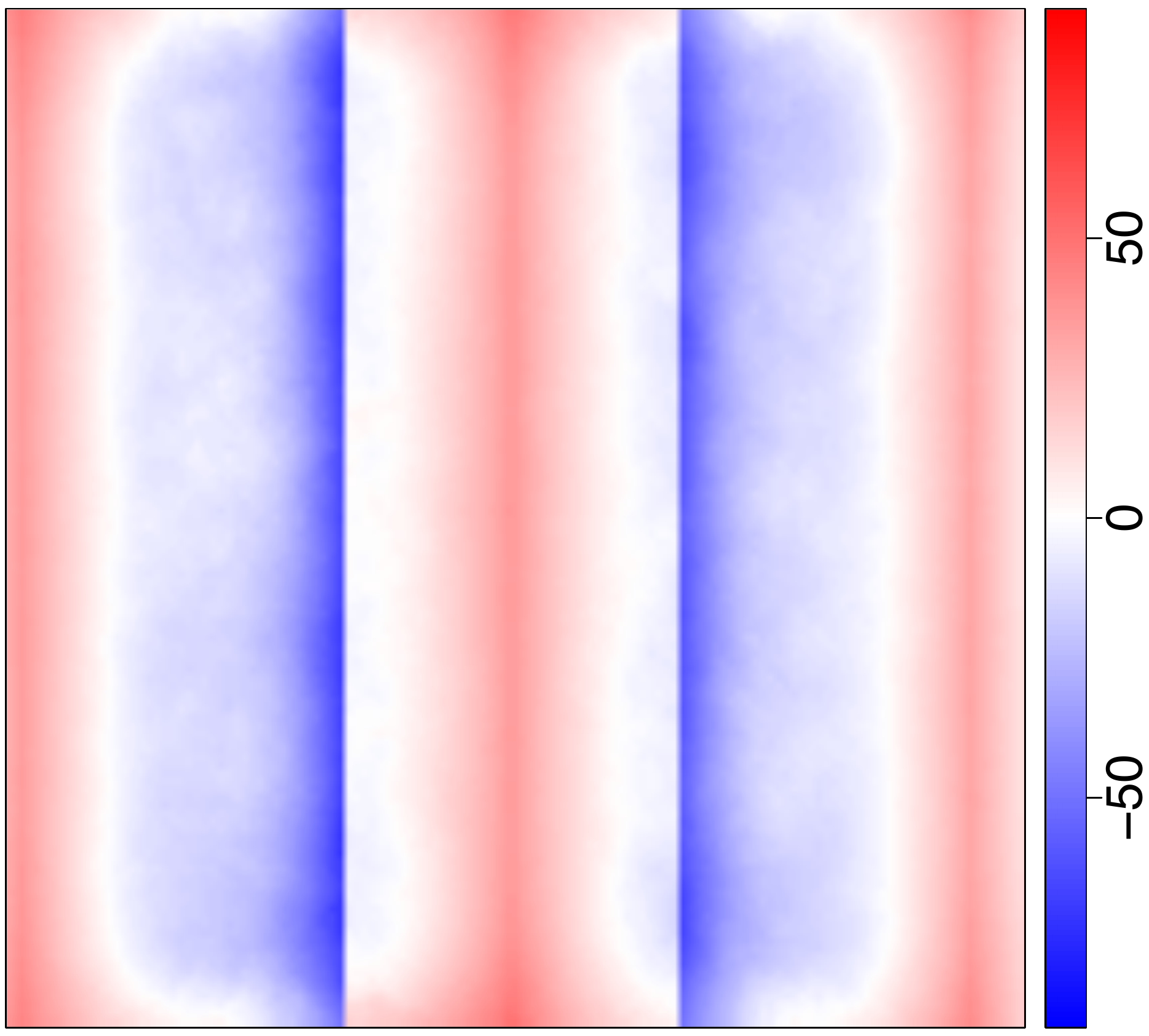}
  \qquad
  \includegraphics[width=0.35\textwidth]{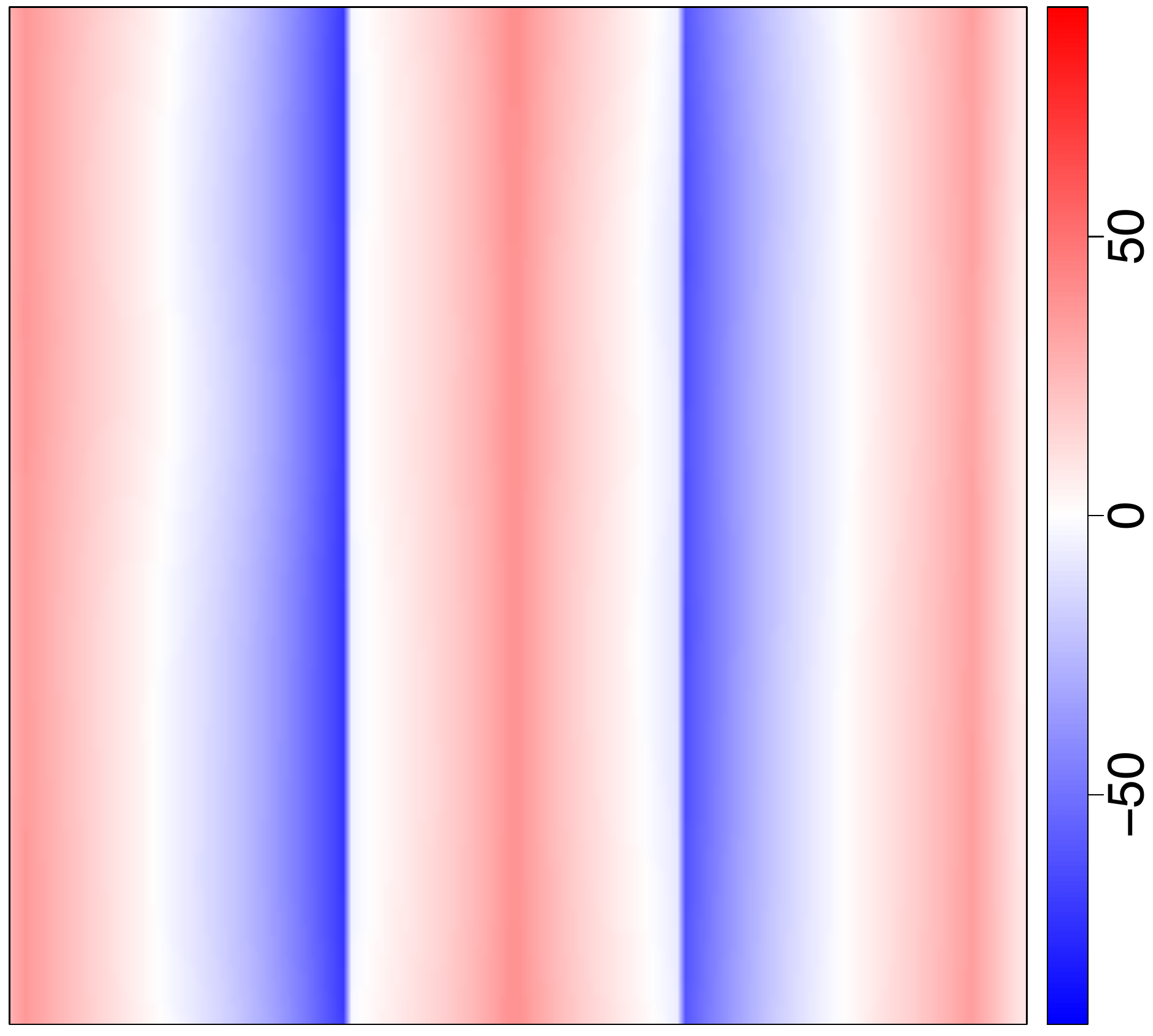}
  \qquad
  \includegraphics[width=0.35\textwidth]{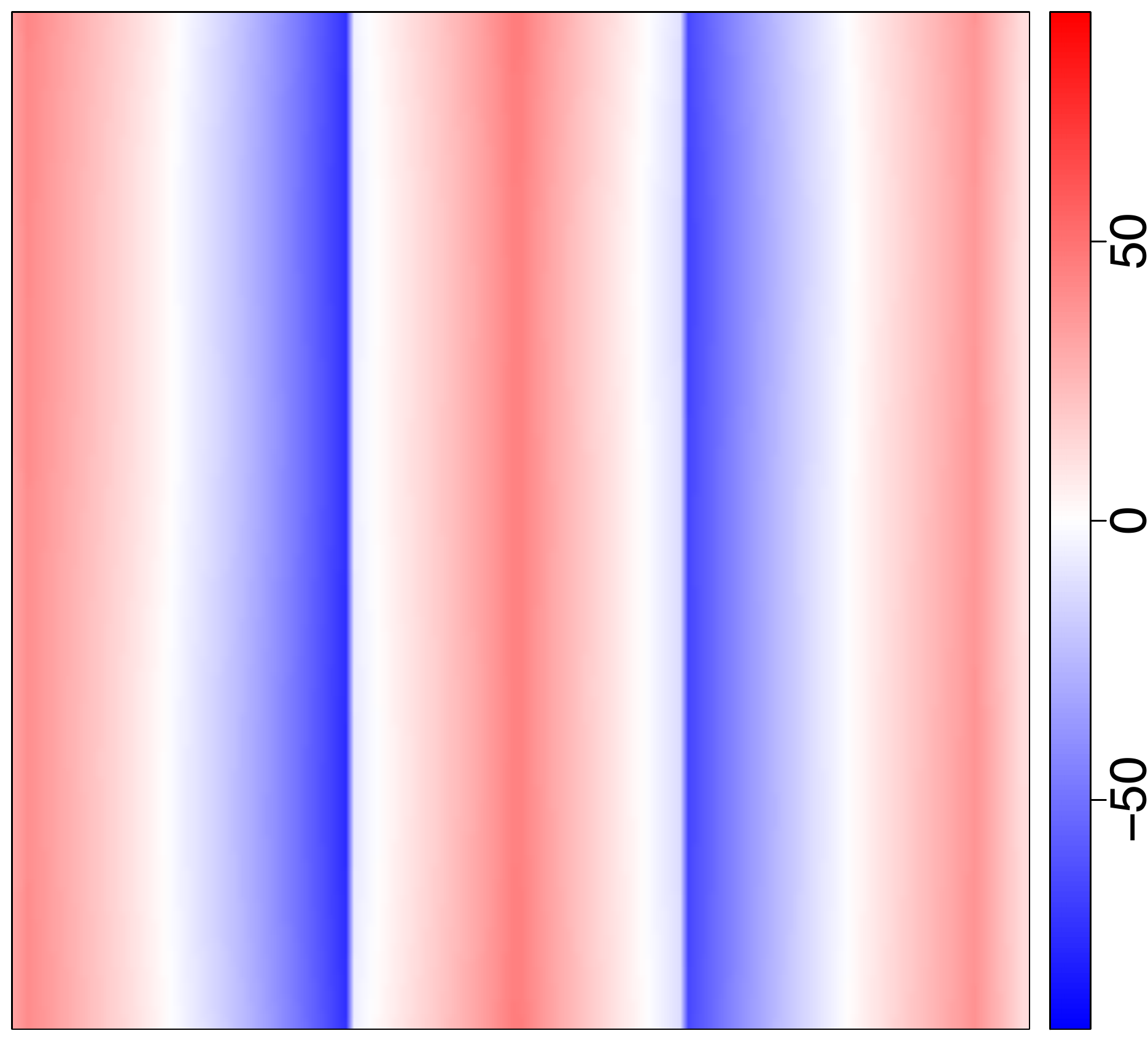}
\caption{
Estimated bias for $\widehat{\rho}_{p,m}^{V}(u)$, $u\in W=[0,1]^2$, $m=200$, and kernel estimators, based on $500$ realisations of an independently thinned simple sequential inhibition process in $W=[0,1]^2$ with intensity $\rho(x,y)=450p(x,y)$,  $p(x,y)=\1\{x<1/3\}|x-0.02| + \1\{1/3\leq x<2/3\}|x-0.5| + \1\{x\geq2/3\}|x-0.95|$, $x,y\in W$. From top-left to bottom-right: $\widehat{\rho}_{p,m}^{V}(u)$ with $p=0.1,0.3,0.5,0.7,0.9,1$; kernel estimators with bandwidths selected using Poisson likelihood cross-validation \citep{BRT15,Load99} (left) and the method of \citet{cronie2018bandwidth} (right) are on the last row.
}
\label{f:BiasTSSIR2}
\end{figure*}

\begin{figure*}[!htbp]
\centering
 \includegraphics[width=0.35\textwidth]{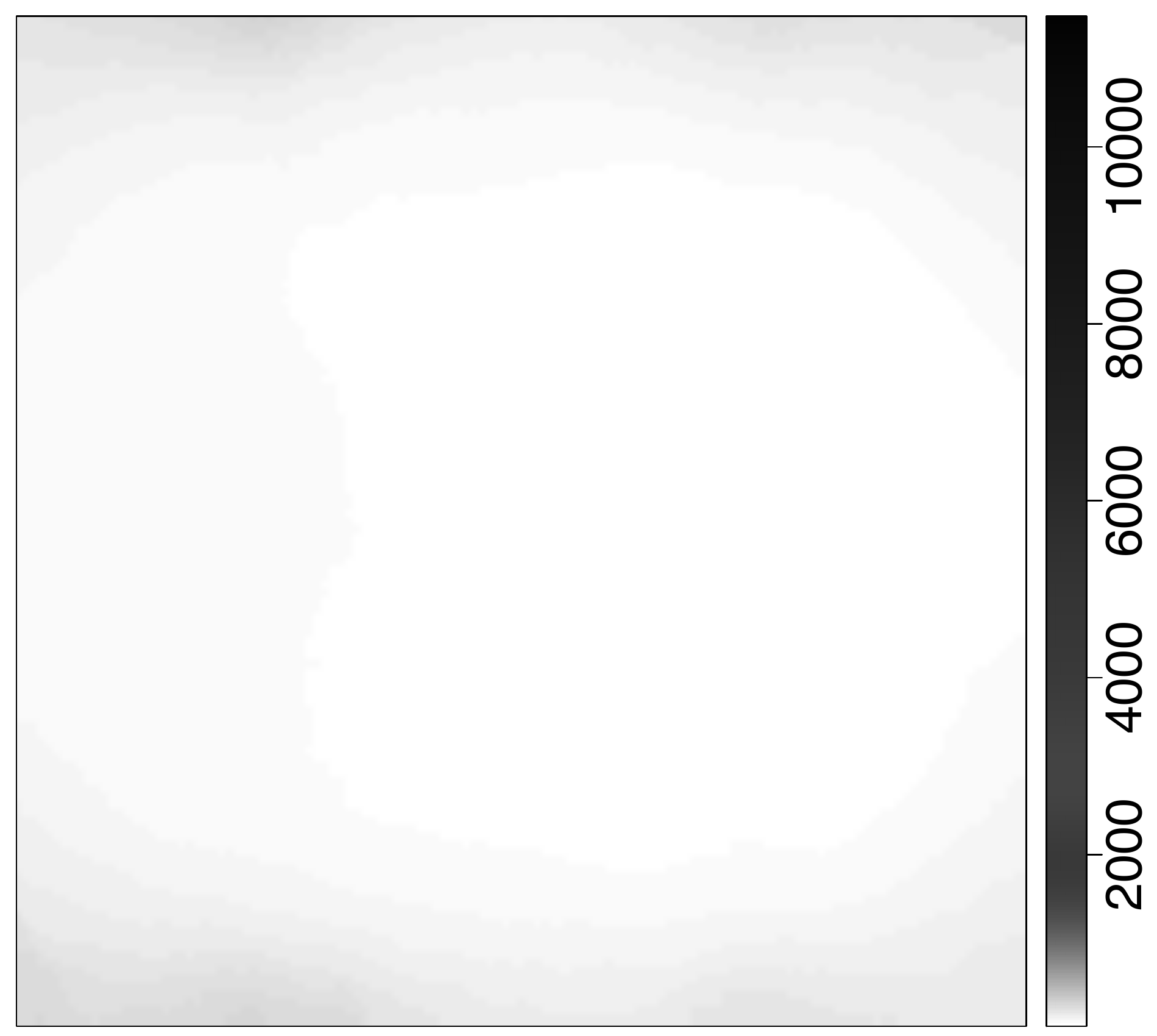}
  \qquad
 \includegraphics[width=0.35\textwidth]{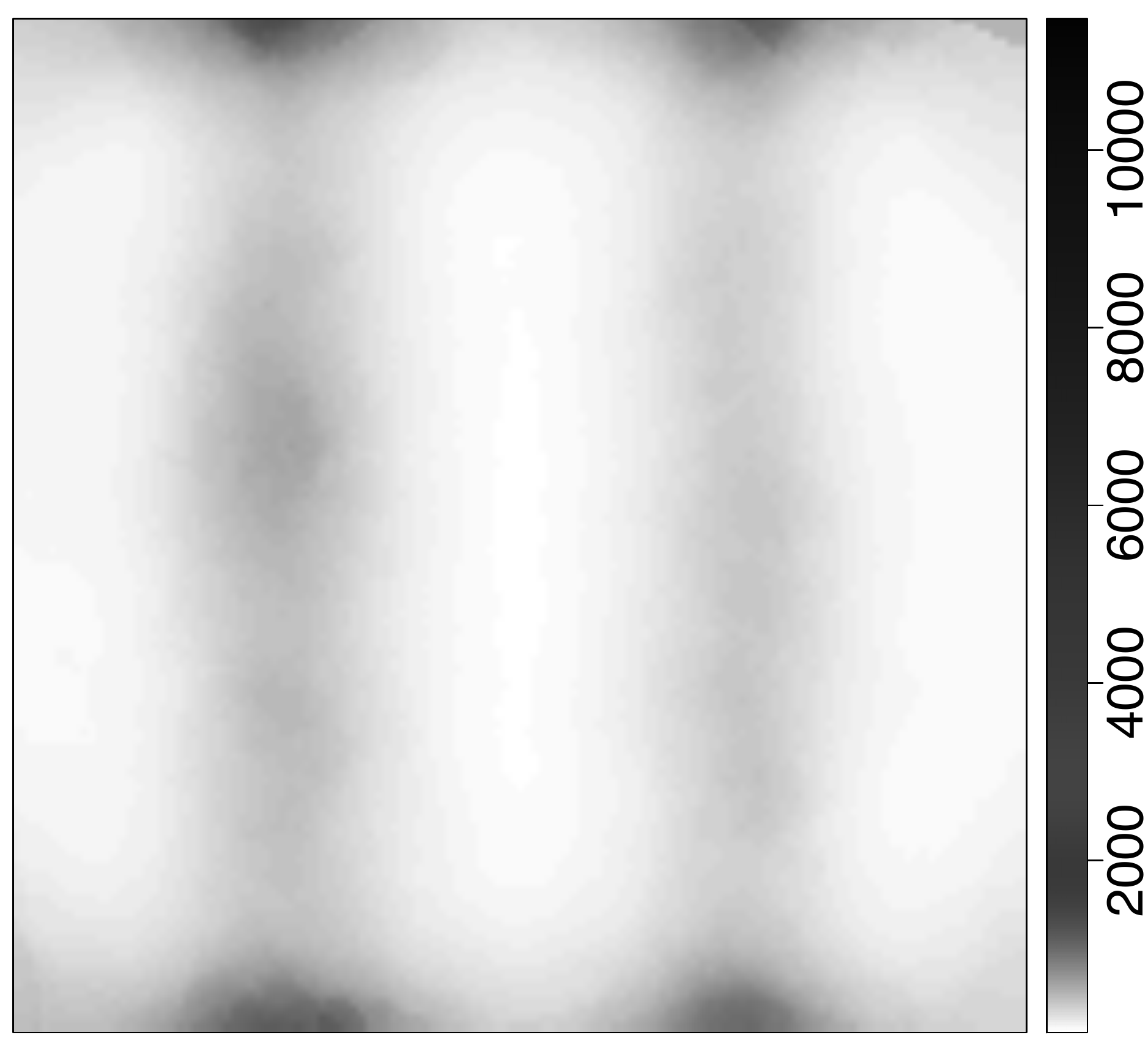}
 \qquad
  \includegraphics[width=0.35\textwidth]{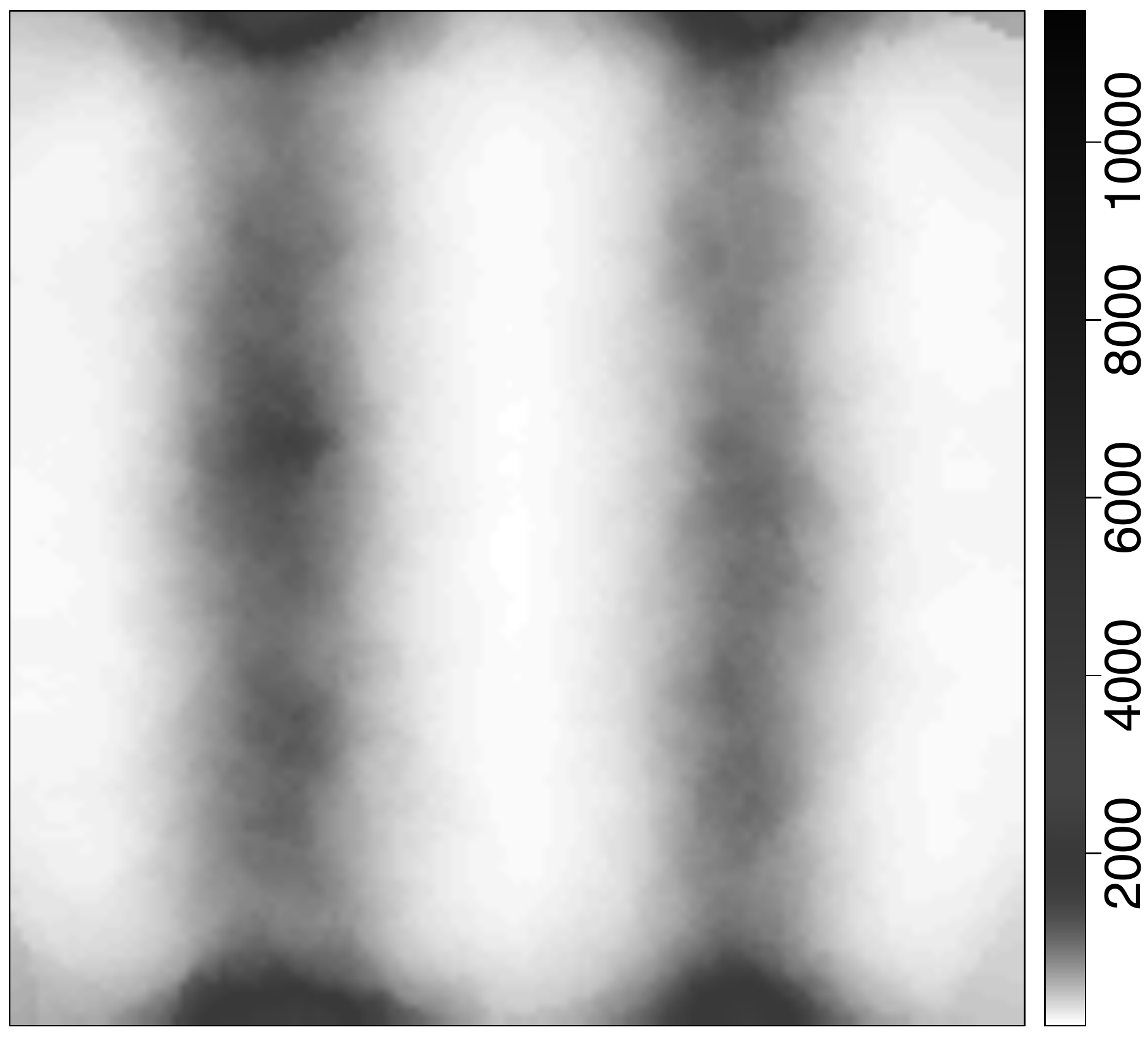}
  \qquad
 \includegraphics[width=0.35\textwidth]{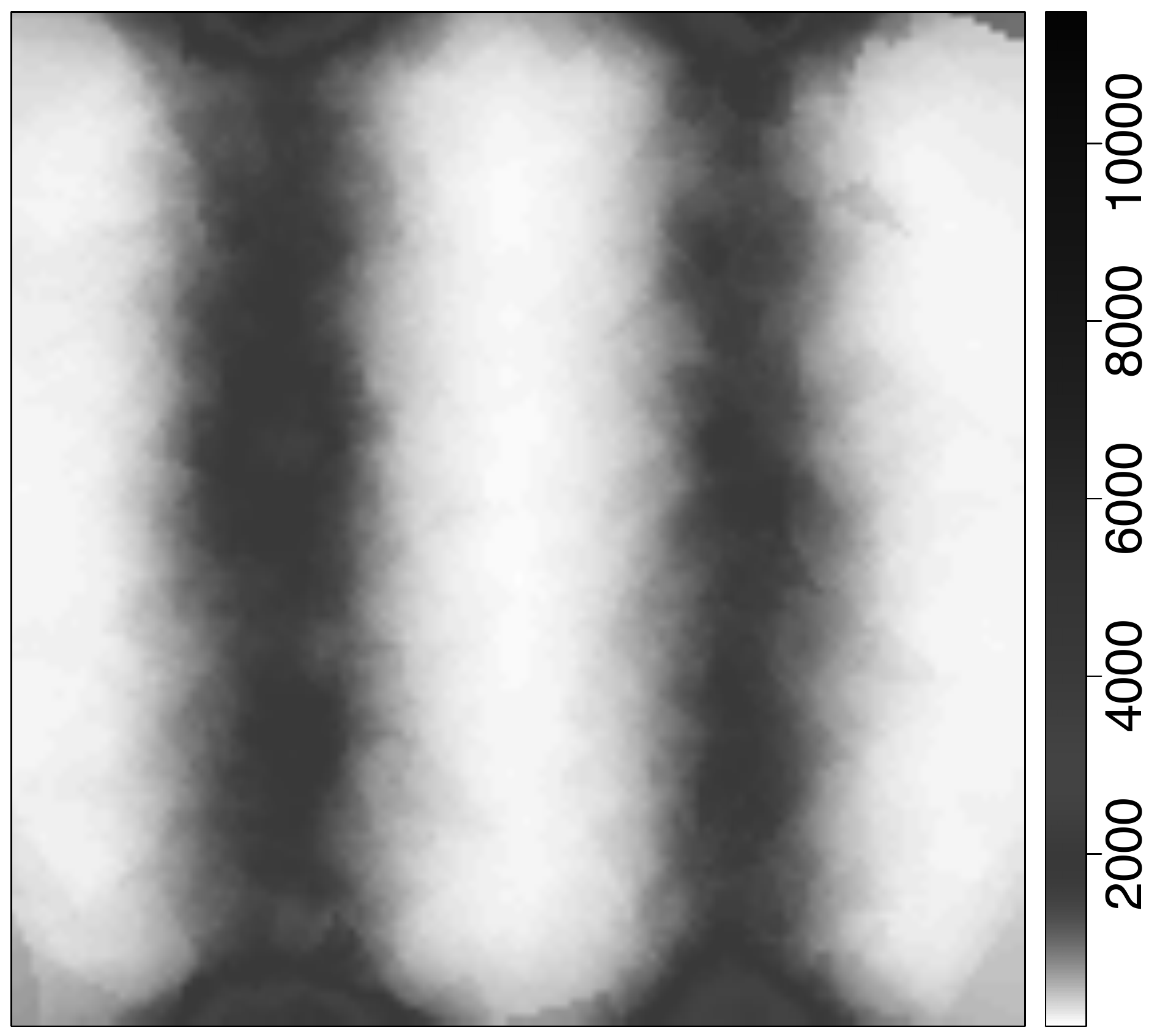}
 \qquad
  \includegraphics[width=0.35\textwidth]{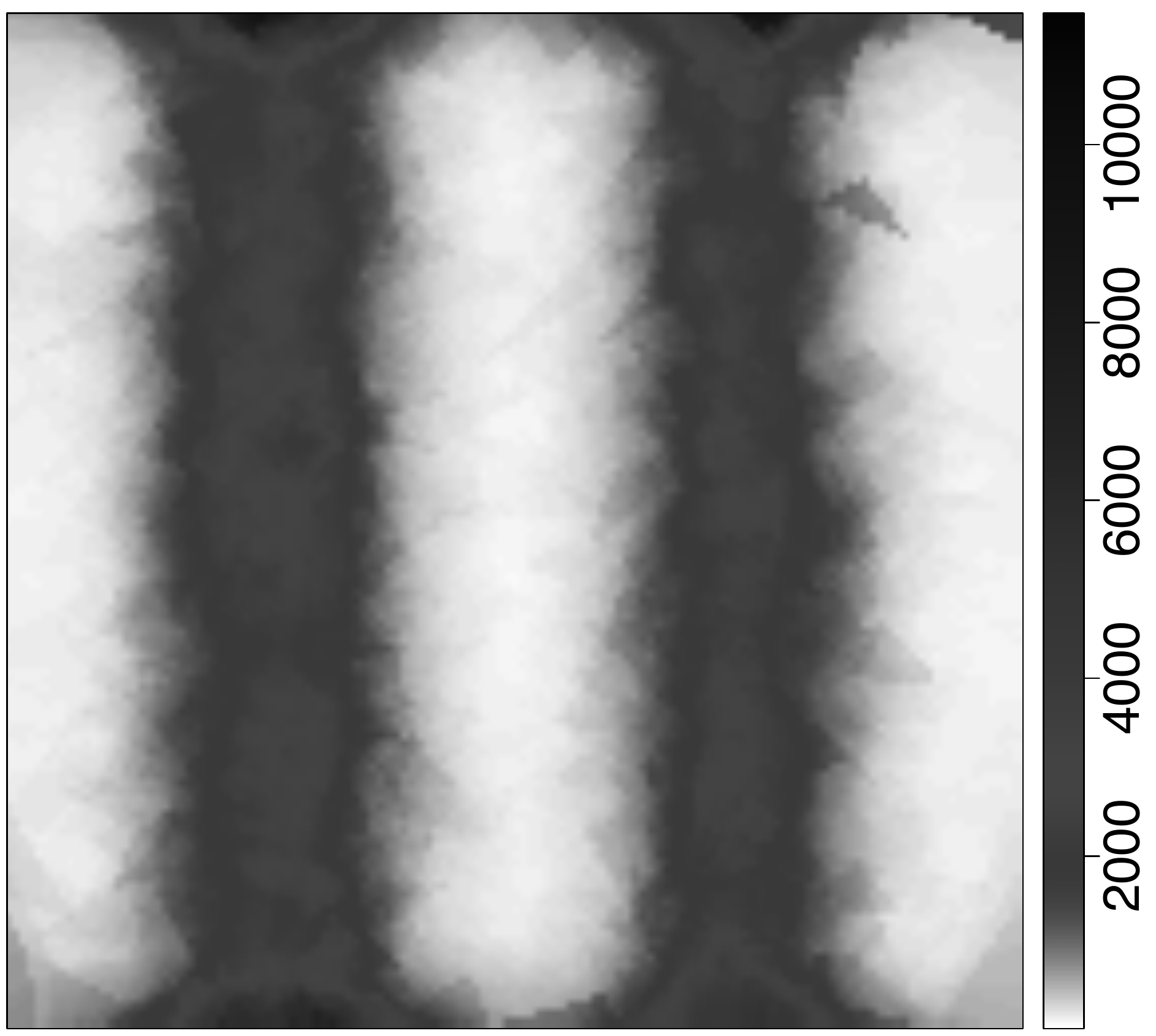}
  \qquad
  \includegraphics[width=0.35\textwidth]{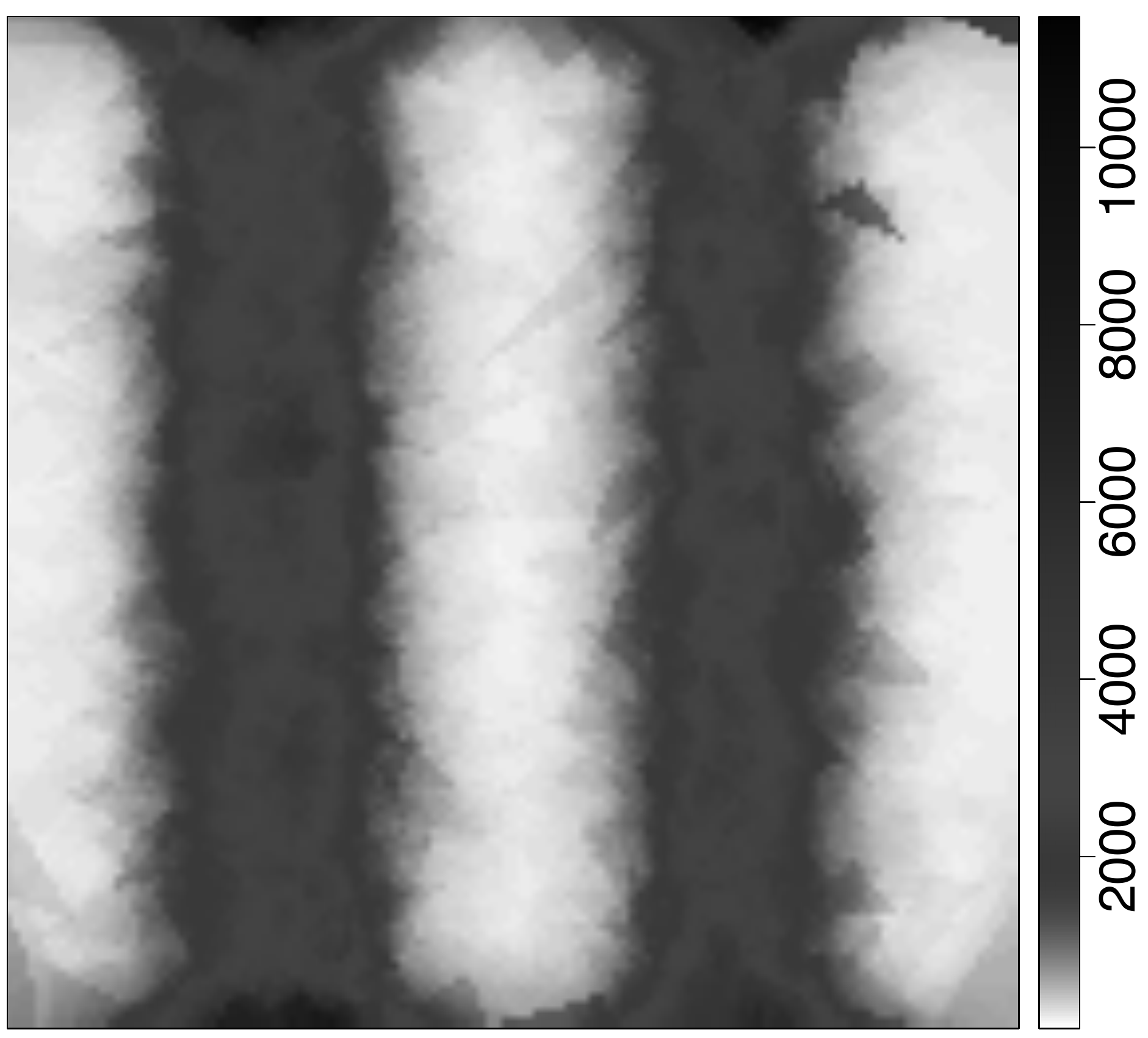}
  \qquad
  \includegraphics[width=0.35\textwidth]{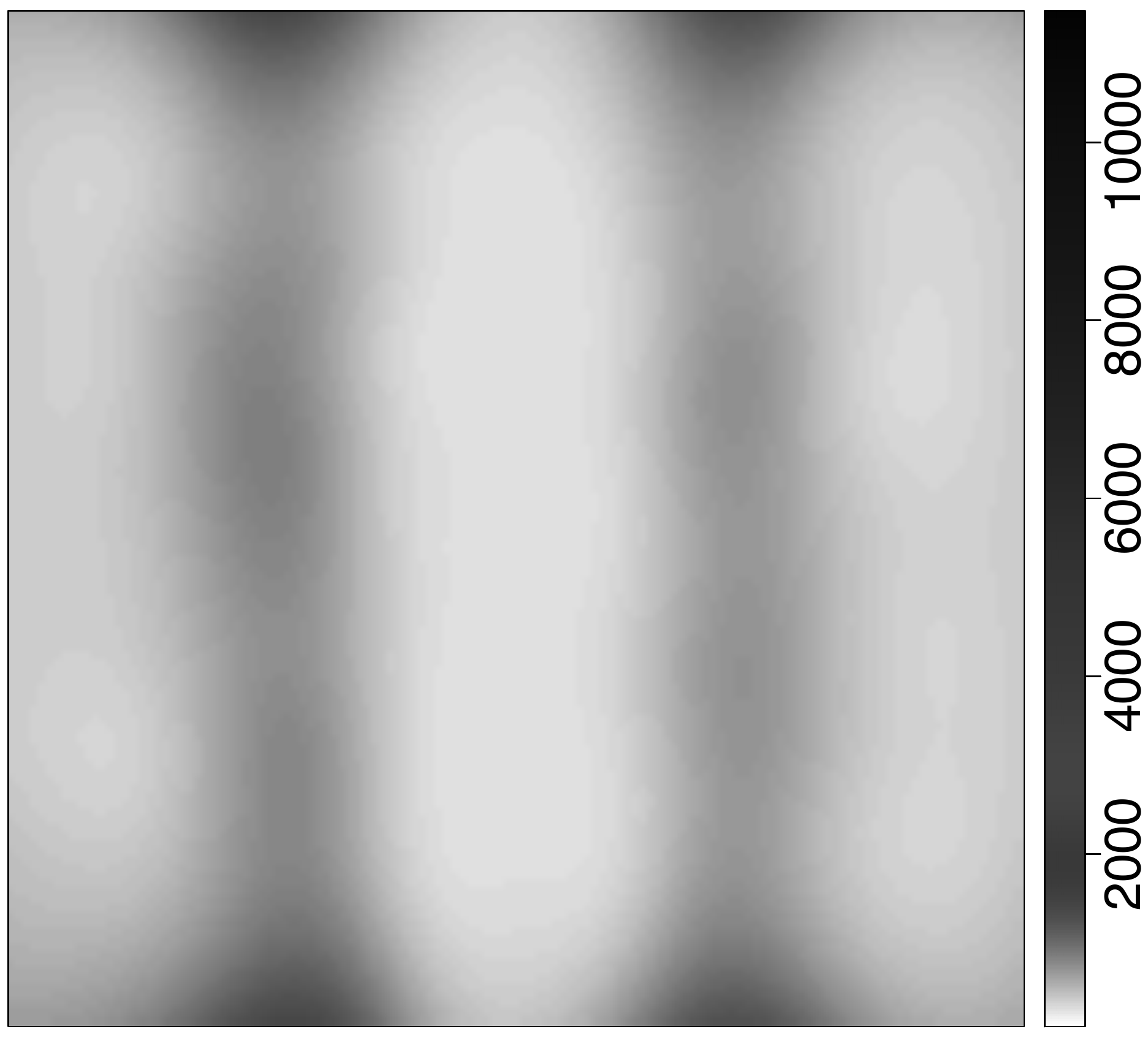}
  \qquad
  \includegraphics[width=0.35\textwidth]{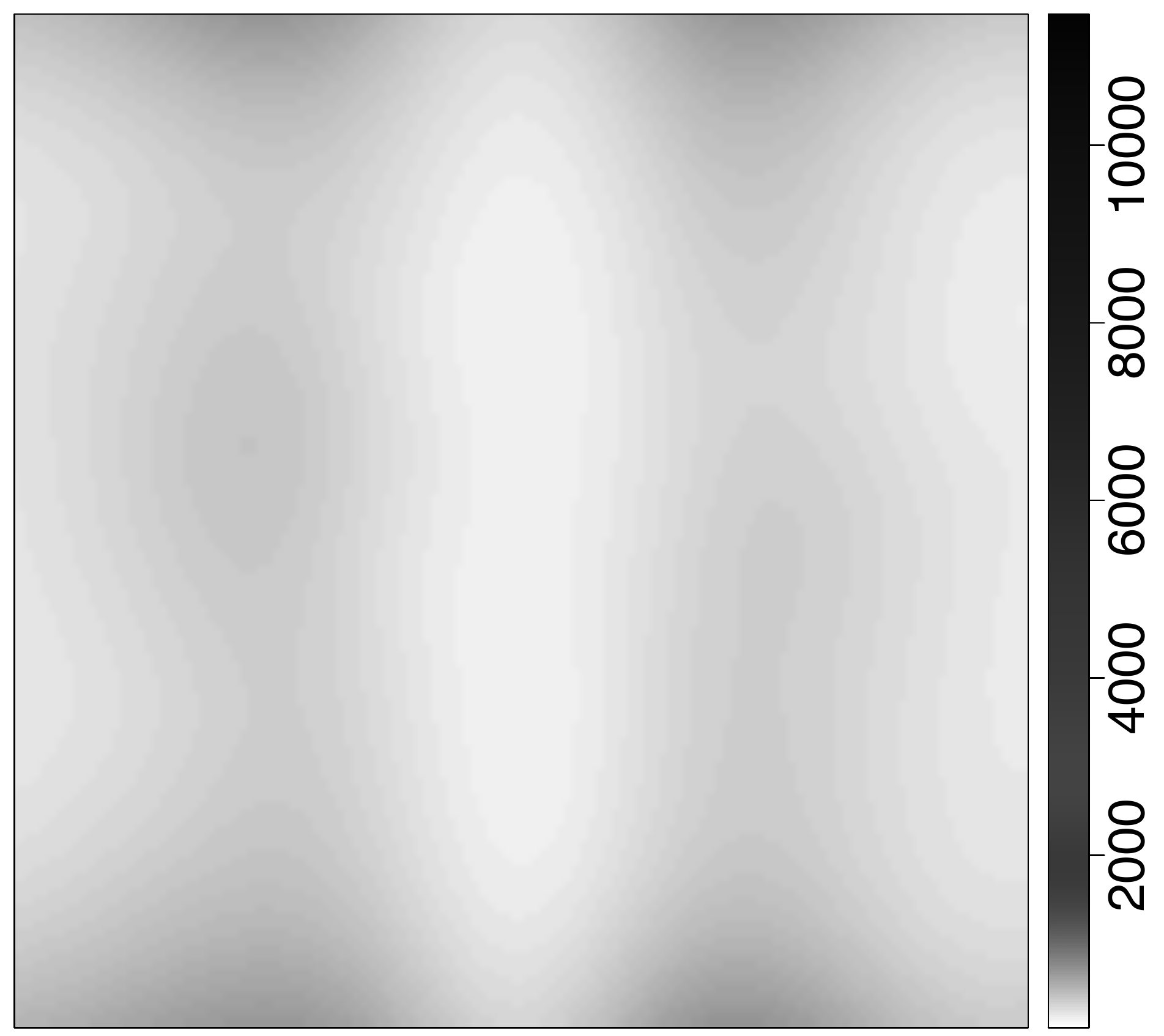}
\caption{
Estimated variance for $\widehat{\rho}_{p,m}^{V}(u)$, $u\in W=[0,1]^2$, $m=200$, and kernel estimators, based on $500$ realisations of an independently thinned simple sequential inhibition process in $W=[0,1]^2$ with intensity $\rho(x,y)=450p(x,y)$,  $p(x,y)=\1\{x<1/3\}|x-0.02| + \1\{1/3\leq x<2/3\}|x-0.5| + \1\{x\geq2/3\}|x-0.95|$, $x,y\in W$. From top-left to bottom-right: $\widehat{\rho}_{p,m}^{V}(u)$ with $p=0.1,0.3,0.5,0.7,0.9,1$; kernel estimators with bandwidths selected using Poisson likelihood cross-validation \citep{BRT15,Load99} (left) and the method of \citet{cronie2018bandwidth} (right) are on the last row.
} 
\label{f:VarTSSIR2}
\end{figure*}

\newpage






\end{document}